\documentclass[12pt]{article}

\usepackage{lmodern}
\usepackage[T2A,T1]{fontenc}
\usepackage[utf8]{inputenc}
\usepackage{geometry}
\usepackage{color}
\usepackage{bm}
\usepackage{mathtools}
\usepackage{amsmath}
\usepackage{amssymb}
\usepackage{graphicx}
\usepackage{subcaption}
\usepackage{textcomp}
\usepackage[numbers,sort&compress]{natbib}
\usepackage{hyperref}
\hypersetup{colorlinks=true,citecolor=black,linkcolor=black,urlcolor=black}
\makeatletter

\@ifundefined{textcolor}{}
{
	\definecolor{BLACK}{gray}{0}
	\definecolor{WHITE}{gray}{1}
	\definecolor{RED}{rgb}{1,0,0}
	\definecolor{GREEN}{rgb}{0,1,0}
	\definecolor{BLUE}{rgb}{0,0,1}
	\definecolor{CYAN}{cmyk}{1,0,0,0}
	\definecolor{MAGENTA}{cmyk}{0,1,0,0}
	\definecolor{YELLOW}{cmyk}{0,0,1,0}
}
\makeatletter
\def\simgt{\mathrel{\lower2.5pt\vbox{\lineskip=0pt\baselineskip=0pt
			\hbox{$>$}\hbox{$\sim$}}}}
\def\simlt{\mathrel{\lower2.5pt\vbox{\lineskip=0pt\baselineskip=0pt
			\hbox{$<$}\hbox{$\sim$}}}}

\makeatother
\numberwithin{equation}{section}

\def\sect#1{Sec.~{\ref{#1}}}

\def\tab#1{Table~{\ref{#1}}}

\def\fig#1{Fig.~\ref{#1}}
\def\Fig#1{Fig.~\ref{#1}}
\def\figs#1#2{Fig.~\ref{#1} and~{\ref{#2}}}

\newcommand{\Sec}[1]{Sec.~\ref{#1}}

\def\nn{\nonumber}
\def\spa#1.#2{\left\langle#1\,#2\right\rangle}
\def\spb#1.#2{\left[#1\,#2\right]}
\def\sand#1.#2.#3{%
\left\langle#1{\vphantom1}\right|{#2}\left|#3\right]}%
\def\sandmp#1.#2.#3{%
\left\langle#1{\vphantom1}\right|{#2}\left|#3\right]}%
\def\sandpm#1.#2.#3{%
\left[#1{\vphantom1}\right|{#2}\left|#3\right\rangle}%
\def\sandmm#1.#2.#3{%
\left\langle#1{\vphantom1}\right|{#2}\left|#3\right\rangle}%
\def\sandpp#1.#2.#3{%
\left[#1{\vphantom1}\right|{#2}\left|#3\right]}%
\def\sandmx#1.#2.#3{%
\left\langle#1{\vphantom1}\right|{#2}\left|#3\right]}%

\def\eps{\epsilon}
\def\Ord{{\mathcal O}}
\def\Operator{{\mathcal O}}
\def\Operator{{\mathbb O}}

\def\tree{{\rm tree}}
\def\id{\protect{{1 \kern-.28em {\rm l}}}}
\def\pol{\varepsilon}

\def\pol{\varepsilon}
\newcommand{\eqn}[1]{Eq.~\eqref{#1}}
\newcommand{\eqns}[2]{Eqs.~(\ref{#1}) and~(\ref{#2})}

\def\nn{\nonumber}

\def\Pf{{P^{\flat}}}
\def\Qf{{Q^{\flat}}}

\def\S{{{\mathbb S}}}
\def\pS{{\hat {\bm{S}}}}
\def\Vmom{\widehat{V}}
\def\FTLq{\bm{L}_{q}}
\def\H{{\rm H}}

\newbox\charbox
\newbox\slabox
\def\s#1{{      
        \setbox\charbox=\hbox{$#1$}
        \setbox\slabox=\hbox{$/$}
        \dimen\charbox=\ht\slabox
        \advance\dimen\charbox by -\dp\slabox
        \advance\dimen\charbox by -\ht\charbox
        \advance\dimen\charbox by \dp\charbox
        \divide\dimen\charbox by 2
        \raise-\dimen\charbox\hbox to \wd\charbox{\hss/\hss}
        \llap{$#1$} }}

\newcommand{\be}{\begin{equation}}
	\newcommand{\ee}{\end{equation}}

\usepackage{array}   
\newcolumntype{L}{>{$}l<{$}}

\allowdisplaybreaks
\begin{document}
	\interfootnotelinepenalty=10000
	\baselineskip=18pt
	\hfill
	
	\thispagestyle{empty}

	\vspace{1.2cm}

	\begin{center}
		{ \bf \Large Spinning Black Hole Binary Dynamics, \\[6pt]
  Scattering Amplitudes
 and Effective Field Theory
		}

		\bigskip\vspace{1.cm}{
			{\large 
		Zvi Bern,${}^{a}$ Andres Luna,${}^{a}$ Radu Roiban,${}^b$ 
		Chia-Hsien Shen,${}^a$ and Mao Zeng${}^c$
		}
		} \\[7mm]
		{\it  
			${}^a$Mani L. Bhaumik Institute for Theoretical Physics, \\[-1mm]
			  Department of Physics and Astronomy, UCLA, Los Angeles, CA 90095 \\ [1mm]
			${}^b$Institute for Gravitation and the Cosmos, \\[-1mm]
			  Pennsylvania State University, University Park, PA 16802, USA \\ [1mm]
			${}^c$Institute for Theoretical Physics, ETH Z\"urich, 8093 Z\"urich, Switzerland 
		}
                  \\
	\end{center}
	\bigskip
	\bigskip

	\begin{abstract} \small

We describe a systematic framework for finding the conservative
potential of compact binary systems with spin based on scattering
amplitudes of particles of arbitrary spin and effective field
theory. An arbitrary-spin formalism is generally required in the
classical limit. 
By matching the tree and one-loop amplitudes of four spinning
particles with those of a suitably-chosen effective field theory, we
obtain the spin$_1$-spin$_2$ terms of a two-body effective Hamiltonian
through $\mathcal{O}(G^2)$ and valid to all orders in velocity.
Solving Hamilton's equations yields the impulse and spin changes of the
individual bodies. We write them in a surprisingly compact form as
appropriate derivatives of the eikonal phase obtained from the
amplitude.  It seems likely this structure persists to higher orders.
We also point out various double-copy relations for general spin. 

\end{abstract}


	\setcounter{footnote}{0}
	
\renewcommand{\baselinestretch}{1}	
	\newpage
	\setcounter{tocdepth}{2}
	\tableofcontents
	
	\newpage

\section{Introduction}
\label{sec:intro}

\subsection{Overview}
The landmark detection of gravitational waves by the LIGO and Virgo
collaborations~\cite{LIGO} has opened a new window into the
universe.  The promise of major new discoveries calls for an
invigorated effort to develop new theoretical tools for
predictions of gravitational-wave signals matching the precision
of current and future observations.
Current predictions for gravitational-wave signals are based on a
variety of complementary theoretical approaches.  This includes the
effective one-body (EOB) formalism \cite{EOB}, numerical
relativity~\cite{NR}, and self-force formalisms~\cite{selfforce}.  In
the inspiral phase, we have the traditional post-Newtonian~(PN)
approximation using methods in classical
gravity~\cite{PN,JaranowskiSchafer4PN} and the nonrelativistic general
relativity (NRGR) formalism~\cite{NRGR,PN_EFT} based on effective
field theory (EFT), as well as the post-Minkowskian~(PM)
expansion~\cite{PM, DamourTwoLoop, BjerrumClassical,
  CliffIraMikhailClassical, OConnellObservables, 3PMPRL, 3PMLong,
  Antonelli:2019ytb, 2PMDiVecchia}.  The various approaches provide
important nontrivial confirmation and information in overlap regions
of the PN, PM and self-force expansions~\cite{3PMLong,6PNCrossCheck,
  SelfForceComparisions}.  For recent reviews see
Refs.~\cite{GravityReviews, PortoReview}.

In recent years the post-Minkowskian approach, which is a relativistic
weak-field expansion in Newton's constant, has risen in prominence
because, at fixed order in Newton's constant, it naturally yields the
exact velocity dependence of observable quantities.  
These properties mirror those of scattering amplitudes, which are
fundamental building blocks of observables in quantum field theory.  Combining techniques
in scattering amplitudes and EFT, effective
Hamiltonians have been derived in Refs.~\cite{RothsteinClassical,Vaidya,CliffIraMikhailClassical}
that straightforwardly determine classical dynamics of bound orbits
via their equations of motion.
The usefulness of this framework has been recently demonstrated
through the construction of the conservative two-body Hamiltonian at
the third order in Newton's constant expansion~\cite{3PMPRL, 3PMLong}.
Such Hamiltonians can be imported into the EOB
framework~\cite{DamourTwoLoop, Antonelli:2019ytb} used for
gravitational-wave template construction.
An important feature of results obtained along these lines is that
they have a much simpler analytic structure than those
obtained in other approaches because, on the one hand, the velocity
expansion is resummed and on the other, because scattering amplitudes
naturally eliminate certain gauge-redundant structures that would
generically appear.

Amplitudes-based methods leverage powerful techniques that have
been developed over the years for computing quantum scattering
amplitudes in gauge and gravity theories~(for reviews see
e.g.\ Refs.~\cite{AmplitudeReviews, BCJReview}).
The basic philosophy is to focus on gauge-invariant quantities
that can be recursively computed from simpler building blocks:
on-shell recursion relations~\cite{BCFW} allow us to build more
complex tree-level amplitudes directly from lower-point ones, and the
modern unitarity method~\cite{GeneralizedUnitarity,  RationalTerms,
  UnitarityReview,  OPP, Forde}
then assembles tree amplitudes into integral representations of loop
amplitudes.
Because there is a close link between classical physics and quantum
scattering amplitudes (see
e.g. \cite{Iwasaki:1971vb,Donoghue,HolsteinRoss, RothsteinClassical,
  BjerrumClassical, Vaidya, MoreScatteringPapers, CachazoGuevara,
  Spin2PM, DamourTwoLoop, CliffIraMikhailClassical,
  OConnellObservables, 3PMPRL, MOV, 3PMLong,
  AmplitudePotential,AmplitudePotentialBjerrum}), advanced methods for
finding the latter can also be applied to solving certain nontrivial
classical gravitational problems.
The Kawai-Lewellen-Tye (KLT)~\cite{KLT} and Bern-Carrasco-Johansson
(BCJ)~\cite{BCJ, BCJLoop, BCJReview} double-copy relations give
gravitational scattering amplitudes directly in terms of much simpler
gauge-theory ones, enabling explicit (super)gravity calculations at
remarkably high orders of perturbation theory~\cite{HighLoopSugra,
  GravityTwoloop}.
Massless and massive helicity methods~\cite{SpinorHelicity,
  MassiveSpinorHelicity} have proven to be especially effective for
calculating four-dimensional amplitudes.
These tools have already demonstrated their utility for calculations
of interest in gravitational-wave physics.

In this paper we focus on spin-dependent classical interactions of
binary systems, in the post-Minkowskian expansion.
As highlighted by the recent detection of black hole spin during
inspiral phase~\cite{SpinMeasurement}, such effects are of
considerable importance in light of astrophysical evidence that black
holes can have a variety of intrinsic angular momenta, including close
to maximally-allowed values~\cite{BlackHoleSpin}.
The presence of spin can lead to qualitative changes in the dynamics
of a binary system, such as the orbital-plane precession when the spins
are not aligned with the orbital angular momentum (see
e.g.~Ref.~\cite{OrbitalMechanicsWithSpinExamples}).  Such an effect
would lead, in particular, to a modulation of the amplitude, frequency
and phase of the observed gravitational wave signal.

\begin{figure}[tb]
	\begin{center}
		\includegraphics[scale=.7]{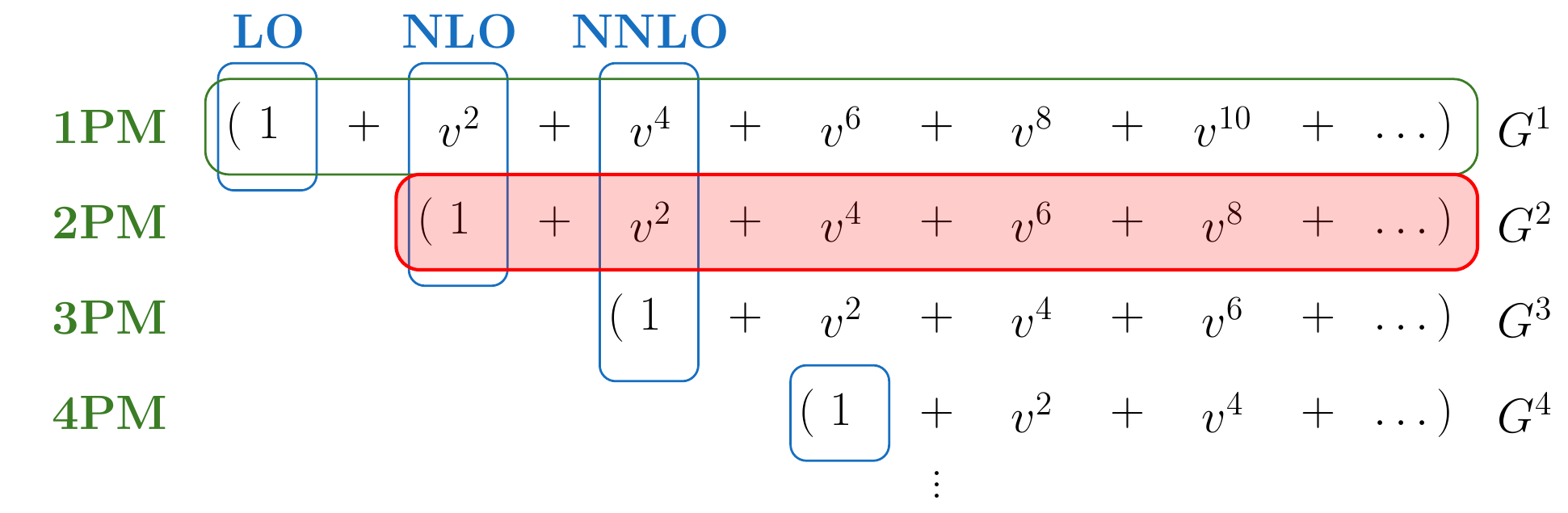}
	\end{center}
	\vskip -.3 cm
	\caption{\small The previously known results in PN and PM
          expansions of the (bilinear in spin) spin$_1$-spin$_2$
          interactions in the two-body potential, are outlined in
          horizontal (green) and vertical (blue) direction respectively. The new results in this paper
          at $\Ord(G^2)$ and all orders in velocity are correspond to the shaded (red) region.
          Each horizontal row corresponds to the same order in
          $G$, or the PM expansion.  The velocity expansion is indicated by
          $v^n$.  Each vertical column corresponds to the same PN
          order for the spin$_1$-spin$_2$ interaction, where the
          leading order (LO), next-to-leading order (NLO),
          next-to-next leading order (NNLO), and the static part at
          $G^4$ are known up to quadratic in spins.
            }
	\label{table:PNPM}
\end{figure}

Inclusion of spin effects in the post-Newtonian expansion has a long
history in a variety of frameworks~\cite{SpinStandardPN, SpinEFTEarlyWorldLine, SpinEFT, SteinhoffADMforSpin,
  SteinhoffNNLOSpin, EFTSpinHamiltonianEquiv,
  LeviSteinhoffLagrangian}.  The effect of spin in the
context of the PN approximation has also been considered using
elementary-particle scattering amplitudes~\cite{HolsteinRoss,Vaidya}.
Ref.~\cite{Vaidya} further extracts PN potentials via EFT techniques
and provides an early indication of the correspondence between minimal
coupling and Kerr black holes.
The analogous problem in the post-Minkowskian framework, where all
orders in velocity are kept, has been comparatively less explored.
For the problem of two Kerr black holes scattering,
Ref.~\cite{Vines1PM} derived a solution at linear order in Newton's
constant with the full spin dependence using traditional methods and
derived a corresponding two-body Hamiltonian.  This was later shown to
be equivalent to minimal amplitudes in massive spinor helicity
formalism~\cite{GOV,Spin2PM_1,GOV_2,SpinSmallYutin},
and these amplitudes at order~$G$ were used to derive a 
two-body effective Hamiltonian~\cite{YutinEFT}.
Physical observables with spins can also be extracted directly from scattering
amplitudes~\cite{MOV}, as demonstrated at order~$G$ in~\cite{GOV_2}.
At order $G^2$, a complete all-orders-in-velocity spin-orbit
Hamiltonian is known~\cite{Damour2PMSpinOrbit}. Beyond linear order in
spin, only partial results are available to all orders in
velocity~\cite{GOV,Spin2PM, GOV_2,Spin2PM_testBH,Spin2PM_1,
  SpinSmallYutin,DamgaardSpinAmplitude,NBI_Spin_Tree}.
The table in \fig{table:PNPM} shows the status of the spin$_1$-spin$_2$
interactions analyzed in some detail in this paper at $\Ord(G^2)$,
indicating both the previously known terms in the velocity and $G$
expansions as well the new results.

\subsection{Summary of paper}

In this paper, we aim to answer several important questions in the
amplitude-based approach.  First, conventional field theory considers
elementary spins~\cite{HolsteinRoss,Vaidya,DamgaardSpinAmplitude}. The
results are \textit{a priori} not necessarily the same as those obtained with continuous
classical spins.  Using massive-spinor-helicity formalism, the
universality of spin-dependent effects was shown at linear in $G$ order and
partially at order
$G^2$~\cite{GOV,Spin2PM_1,GOV_2,SpinSmallYutin,NBI_Spin_Tree}. 
Bootstrapping results for Kerr black holes 
from massive-spinor-helicity method however is known 
to have ambiguities beyond quartic order in spins at $G^2$~\cite{Spin2PM,Spin2PM_1,GOV,SpinSmallYutin}, 
as new spin-multipole moments are allowed.  
It would be desirable to have a complementary formulation, with arbitrary spins, that
can generate amplitudes from first principles.
Second, for applications to LIGO and VIRGO, it is crucial to 
extract quantities of interest for the bound-state problem in a
format that can be straightforwardly compared with previously known results.
In the post-Minkowskian scenario without spins, there are several methods for doing
so available \cite{DamourTwoLoop,CliffIraMikhailClassical, 3PMPRL,
  3PMLong, AmplitudePotential}.
In the presence of spins, however, the known results from EFT or
scattering angles are either limited to leading order in $G$ or to
special configurations of spins~\cite{MOV,Vaidya,YutinEFT,GOV,GOV_2}.  
The goal of this paper
is to build a systematic framework bridging the gaps between quantum
scattering amplitudes, classical gravity, and bound orbits for
spinning objects.  A key part is to identify a new direct link between the
scattering amplitudes including spin and physical observables, via the
eikonal phase~\cite{EikonalPapers,EikonalImpact}.

A central component of our paper is an amplitudes-based formalism for
incorporating spinning effects for binary systems in a
post-Minkowskian framework, i.e. fixed order in $G$ and all orders in
velocity.  Our results are new in several directions.  First we
construct a field theory with arbitrary-spin particles, smoothly
interpolating from elementary particles to classical spinning
particles.  Next, we formulate an EFT for spinning particles.  The
amplitudes from this arbitrary-spin field theory are translated,
through EFT matching, into an effective potential which can be used to
study bound-state problems. Conversely, the formalism can also turn
any classical Hamiltonians into gauge-invariant scattering amplitudes,
allowing for a straightforward comparison of the gauge-invariant
content of two Hamiltonians.  The usefulness of our setup is
demonstrated through a new result for spin-dependent effects: we
obtain a two-body Hamiltonian that describes the interactions linear
in the spin of each body, referred to as ``spin$_1$-spin$_2$'' or ``bilinear in spin'', through
order $G^2$ and to all orders in velocity.  Finally, as discussed in
more detail in Ref.~\cite{EoMEikonalPaper}, we use this Hamiltonian to
calculate physical observables---the momentum and spin transfer---via
the classical equations of motion, and organize them into an
eikonal-based formula, providing a direct link between scattering
amplitudes and classical observables.
All results presented here are for generic spin orientation.  We now
summarize each section in turn.

In \Sec{sec:basics}, we begin with a basic introduction to classical
spins, including the formulation of an arbitrary-spin field theory
following the path of Refs.~\cite{LandauQED, Khriplovich:1998ev}.  The
arbitrary-spin formalism constructed here is a natural framework for
capturing higher powers of spin interaction; this may be contrasted
with the more familiar spin 1/2 or spin 1 cases that could be used to
extract low-order spin interactions~\cite{HolsteinRoss,Vaidya,
  DamgaardSpinAmplitude}.  Non-minimal interactions at linear order
in $G$ are characterized in \Sec{sec:nonminimal} and are similar to
those in the world-line formulation~\cite{SpinEFTEarlyWorldLine} of spinning 
particles given in Ref.~\cite{LeviSteinhoffLagrangian}.  At
linear order in $G$, the Lagrangian may be interpreted as the
covariantization of the most general parity-even gravitational form
factor.  Our stress tensor reproduces the all-orders-in-spin stress
tensor at order $G$ of Ref.~\cite{Vines1PM}.  For the stress tensor,
or equivalently the on-shell two-matter-one-graviton vertex, we also
present new double-copy relations to all orders in the spin for
arbitrary nonminimal coupling, expressing the complete set of
gravitational interactions in terms of gauge-theory ones.
Double-copy properties in the context of gravitational waves have been 
discussed recently in~Refs.~\cite{DoubleCopySpin, 3PMPRL, 3PMLong,Spin2PM_1}.

In \Sec{sec:tree} and \Sec{sec:oneloop}, we compute various tree-level
and one-loop amplitudes with higher-spin particles.  At the relatively
low order considered in this paper, we will not need the full arsenal
of amplitude techniques that become important at higher orders.  We
therefore use polarization tensors to incorporate the spin
degrees of freedom. and make only modest use of the double
copy~\cite{KLT,BCJ,BCJReview} to write compact expressions for
amplitudes.  We first obtain the two-to-two scattering of higher-spin
particles at tree level, truncated to bilinear order in spins.  Then
we calculate the tree-level gravitational Compton amplitude, and find
simple KLT-like relations.  In \Sec{sec:oneloop}, we use the Compton
amplitude to extract the required contribution to the four-point
one-loop amplitude of two distinct spinning particles.  We then
reduce the integrand to a basis of scalar integrals using the massive
extension~\cite{Kilgore:2007qr} of Forde's formalism~\cite{Forde}.
This formalism clarifies the connection of basis integral coefficients and
integrands and efficiently extracts the needed contribution in the
classical limit.

In \Sec{sec:EFT}, we construct an EFT for spinning objects, following
the path of Ref.~\cite{CliffIraMikhailClassical} in the spinless case and Ref.~\cite{Vaidya} in the spin case under PN framework.
This EFT allows us to map scattering amplitudes to effective Hamiltonians,
which can then be straightforwardly applied to bound-orbit problems.  We classify spin
interactions explicitly to bilinear-in-spin order. The on-shell
matching scheme reduces the number of independent operators.  We show
how to compute scattering amplitudes in this EFT.  We also point out
the crucial role of the $SO(3)$ algebra of classical spins in order to
obtain results for generic spin orientation.  Combining with
the one-loop amplitudes obtained in \Sec{sec:oneloop}, we derive the bilinear
in spin Hamiltonian through order $G^2$ and to all-orders in velocity.

In \Sec{sec:angle}, we obtain the momentum and spin change in the
scattering regime starting from our derived classical Hamiltonian.
The three-dimensional nature of the scattering process makes the
construction of the perturbative solution of the equations of motion
somewhat more involved than for the case of spinless particles.  An
alternative approach is to directly obtain observables from the
amplitudes~\cite{OConnellObservables, MOV}, bypassing the Hamiltonian
and EFT matching.  A very interesting question is whether there exist,
in general, simple and direct relations between physical quantities and
suitably-defined finite parts of amplitudes scattering amplitudes
analogous to the one for the spinless or aligned spin case~\cite{DamourTwoLoop,
  3PMLong, AmplitudePotential,GOV}.  
The ability to do this for spin~\cite{MOV} suggests that this might be more generally possible.
We indeed find such a relation for generic spin,
generalizing the eikonal formula~\cite{EikonalPapers,EikonalImpact} to
the case with spin, obtaining not only the impulse, but also the spin
kick from appropriate derivatives of the eikonal phase.  This striking
result suggests that it should be possible to develop much more
streamlined formalisms for extracting physical observables from
scattering amplitudes at higher orders.  We leave the details to a
forthcoming paper~\cite{EoMEikonalPaper}.

In order to ensure the reliability of our results, we perform a number
of nontrivial checks for the $\Ord(G^2)$ contributions to the
interactions bilinear in spin.  This includes comparison with the
post-Newtonian NLO spin-orbit results of
Ref.~\cite{Vaidya,SteinhoffADMforSpin}, the NNLO results of
Ref.~\cite{SteinhoffNNLOSpin} in the overlapping region to $\Ord(G^2)$, and to
all-orders in velocity for the scattering angles with spins aligned to orbital angular momentum~\cite{Spin2PM_testBH,GOV}, whose spin-orbit part is in agreement with Ref.~\cite{Damour2PMSpinOrbit}.  
The latter comparison in the spin-orbit case is especially
powerful because it verifies the complete coefficient of the spin-orbit
operator.  In the test body limit, we also reproduce a simplified
isotropic gauge version\footnote{We thank Justin Vines for providing
  this form of the Hamiltonian.}  of the test-body Hamiltonian given in Appendix D of
Ref.~\cite{TestBodyQuadraticSpin} valid to quadratic order in spin and
all order in Newton's constant and velocity.
At the amplitude level and with a suitable interpretation of the
covariant spin vector, we also recover the spin-1/2 results of
Ref.~\cite{DamgaardSpinAmplitude}.  Although we work with an arbitrary
spin, this is expected because, as we argue on general grounds in
\sect{sec:basics}, for the terms linear in the spin of each particle,
spin-1/2 is sufficient as long as no special properties of the Pauli
matrices are used~\cite{HolsteinRoss,Vaidya, SpinSmallYutin,
  DamgaardSpinAmplitude}.

In this paper we use mostly-negative metric and the four-dimensional
Levi-Civita symbol is normalized as $\epsilon^{0123}=1$. Unless
otherwise specified, the boldface symbols denote spatial three vectors.  All
four momenta are outgoing.  Additional notation can be found in
\Sec{tree_and_loop_summary} and \Sec{sec:EFT_summary}, where we
summarize the results in amplitudes and in EFT.

\section{Basics for spinning particles}
\label{sec:basics}

In this section we describe the classical limit of processes involving spinning 
particles (which we identify with spinning compact astrophysical objects) 
and review basic facts on spin that we use in later sections.
We will see that the spin must be of the same classical order as the orbital angular 
momentum and therefore, from a Lagrangian perspective, the classical spinning particles should
be represented by higher-spin fields. To describe them we follow \textsection{}31 of Ref.~\cite{LandauQED}
and Ref.~\cite{Khriplovich:1998ev}.
This approach has the advantage of giving a simple relation between the 
classical spin vector and Lorentz generators in the Lagrangian, making it 
straightforward to construct a robust formalism. 
We will formulate and use a Lagrangian that captures 
the gauge-invariant completion of the most general parity-even 
spin-dependent linear response of a massive particle to a gravitational field.
We will show that the trilinear interaction of this Lagrangian is the double 
copy of similarly-general trilinear interactions of higher-spin fields with gluons,
thus extending observations of Ref.~\cite{KLT,BCJ,BCJLoop,DoubleCopySpinHenrik}.
Last but not least, we will see that, for suitably-chosen couplings, 
the gravitational stress tensor derived from our Lagrangian reproduces (in the 
classical limit)
that of the Kerr black hole, derived in \cite{Vines1PM}. Consequently, the 
scattering amplitude of two Kerr black holes discussed in that reference is 
also correctly reproduced.

\subsection{The classical limit}
\label{ClassicalLimit}

Our goal is to extract the classical potential between two
massive spinning bodies from their scattering amplitude. To define the
classical limit of an amplitude, we follow the same path
used in Refs.~\cite{CliffIraMikhailClassical, 3PMPRL, 3PMLong}.
Classical physics applies whenever the minimal inter-particle
separation $|\bm b|$ is much larger than the de~Broglie wavelength $\lambda$ of each
particle.
This macroscopic length scale $|\bm b|$ can be chosen as the impact parameter in a scattering process, or the orbital size of a bound binary system.
For incoming particles of momentum $\bm p$, we must then have
\begin{equation}
|\bm b| \gg \lambda = \frac{1}{|\bm p|} \,,
\end{equation}
where we use natural, $\hbar = 1$, units. This implies that 
for any such two-body classical system, the magnitude of orbital angular momentum $\textrm{L} =|\bm L|$ must be large
\begin{equation}
\textrm{L} \sim  |\bm p\times \bm b| \gg 1 \,.
\end{equation}
The same must hold for all other charges, such as electric charge or spin, that may be carried by classical particles.
Indeed, the difference between the classical spin and orbital angular momentum is only in the interpretation of the macroscopic 
length scale: from internal radius to the impact parameter, and rotating to translational velocities.
Thus, in the classical limit, we need the magnitude of the spin, $\textrm{S}_i =|\bm S_i|$,  
and of the orbital angular momentum, $\textrm{L} =|\bm L|$, to be commensurate,
\begin{equation}
\textrm{S}_1 \sim \textrm{S}_2 \sim \textrm{L} \ .
\end{equation}
The net effect is therefore that, classically-spinning particles should be described from a field 
theory point of view by large-spin limit of higher-spin fields.
As we will explain shortly, the details of the calculations imply that
at fixed order in Newton's constant and in the number of spin vectors,
a finite but sufficiently large spin is sufficient to capture all the
relevant contributions.

Since the impact parameter is of order of the inverse momentum
transfer in a scattering process, $|\bm b| \sim 1/|\bm q|$, the
classical limit implies the hierarchy
\begin{equation}
  m_1, m_2 , |\bm p| \sim \textrm{L}\, |\bm q|  \sim \textrm{S}_i\, |\bm q| \, \gg |\bm q|\,.
\label{eq:classical_limit}
\end{equation}
The quantum contributions enter at higher orders
in a large $\textrm{L}$ expansion or, equivalently, higher orders in a small $|\bm q|$ expansion.
This gives us the scaling
\begin{equation}
\Ord(1/\textrm{L}) \sim \Ord(1/\textrm{S}_i) \sim \Ord(|\bm q|) \sim \Ord(q) \quad \textrm{(classical expansion)} \,.
\label{eq:classicalSpin_counting}
\end{equation}
We omit the proper mass scale factors for simplicity. The italic letter denotes four-momentum components.
Unless otherwise noted, the classical expansion in this paper includes simultaneously the scaling of spins, 
orbital angular momentum, and $\bm q$.
For example, monomials in ${\bm q}\cdot {\bm S}_i$ or their covariant version 
$q\cdot S_i$ are of $\Ord(1)$ in the classical limit.

A second expansion parameter is the ratio between spin and orbital
angular momentum, which is suppressed by the internal size over impact
parameter or orbital radius if we ignore the difference in rotating
velocities.  Therefore the expansion 
in spin-induced multipole moments is
\begin{equation}
\Ord(\textrm{S}_i/\textrm{L}) \sim \Ord(\textrm{S}_i/|\bm b|) 
\sim \Ord(\textrm{S}_i |\bm q|)\sim \Ord(\textrm{S}_i q) \quad \textrm{(spin expansion)}\,.
\end{equation}
For examples, the monomials in ${\bm q}\cdot {\bm S}_i$ are classically $\Ord(1)$ but are  order by order in spin expansion.
Indeed, terms linear in the spin
correspond to a dipole moment, those quadratic in spin, $({\bm q}\cdot
{\bm S}_i)^2$, represent a quadrupole moment, {\it etc}.  While the multipole moments are not necessarily small 
when taking velocities into account, the fact that we keep the spin vector arbitrary provides a way to classify 
interactions between two particles in terms of interactions between their
respective multipole moments.\footnote{In general, multipole moments
  are symmetric and traceless combinations of spin vectors.  At
  tree-level the trace part leads to contact interactions, which are
  not of interest to us. At loop-level however, trace terms no longer
  drop out and more care is needed to relate symmetric products of
  spin vectors to multipole moment operators. Since in later sections
  we will be concerned at most with spin${}_1$-spin${}_2$ --- or
  dipole-dipole --- interactions, we will not need the complete
  identification of spin-induced multipole moments.}

The traditional PN expansion parameter relies on velocity $v \sim |\bm p|/m_i$. For bound orbits 
the virial theorem relates the scale of both $G$ and spin expansion parameters to the velocity.\footnote{
	In PN counting, spins is suppressed relative to angular momentum by $\textrm{S}_i \sim \textrm{L} v^\alpha$, 
	with $\alpha=1$ to $4$ depending on rotating speeds. See Ref.~\cite{PortoReview} for more details.}
The virial theorem however does not hold for unbound orbits and therefore the velocity expansion is 
independent from the others in scattering events.
From an amplitudes' perspective it is also more natural to keep a fully-relativistic velocity dependence.
Thus, we will not expand in velocity except to compare with results from the PN literature.
  
With this in mind, we have the following structure of classical conservative Hamiltonian expanded in the (number of) spin vectors of each particle
\begin{align}
H = \null &  H^{(0)}(r^2, p^2)
+h^{(1)}_{i}(r^2, p^2) \frac{1}{r^2} \bm L\cdot \bm S_i
+ h^{(2,1)}_{ ij}(r^2, p^2) \frac{1}{r^4}  \bm r\cdot \bm S_i \, \bm r\cdot \bm S_j + \ldots  \, ,
\label{Hamiltonian_generalA}
\end{align}
where we only keep terms up to quadratic order in spins, i.e. up to quadrupole moments.
Here ${\bm r}$ and ${\bm p}$ are center of mass distance and momentum and
the indices run over $i,j = 1,2$, $r = |{\bm r}|$, $ p = |{\bm p}|$.
$H^{(0)}(r^2, p^2)$ is the spinless Hamiltonian with the usual PM expansion in $G/r$.
At lowest order, ignoring velocity and spin dependence, the potential
is simply the Newtonian one.
The $h^{(a)}_b$ coefficient of each spin-induced moment has the same structure as $H^{(0)}(r^2, p^2)$.
Using $\Ord(1/r) \sim \Ord(|{\bm q}|)$ under Fourier transform,
we can see that each spin structure is of the same classical order $\Ord(1)$ as the spinless potential, 
but carries higher order in spins, or equivalently suppressed by the additional powers of $1/r$.
More details will be discussed in Secs.~\ref{sec:EFT} and \ref{sec:angle}.
Such Hamiltonians are a basic input into
models---such as the EOB framework~\cite{EOB}--- for building
gravitational-wave templates.  In this paper, we evaluate the
$\Ord(G^2)$ contributions to the conservative two-body potential to
all orders in the velocity and to bilinear order in the two spins.

\subsection{The spin vector and tensor}
\label{SpinVectorSubsection}

We now describe the basic field theory formalism that we use to incorporate spin
interactions into an amplitudes-based approach.  In the post-Newtonian
framework, the classical spin-orbit and spin-spin
interaction Hamiltonian of spinning particles is well-studied in the
literature~\cite{HolsteinRoss, Vaidya, SpinSmallYutin}.  A simplifying
aspect is that through $\Ord(G^2)$, spin~1/2 and spin~1 fields turn out to be sufficient 
to recover post-Newtonian results obtained via general-relativistic 
methods~\cite{SpinStandardPN, SteinhoffADMforSpin, SteinhoffNNLOSpin}.
Not surprisingly, at higher orders in spin, calculations using such low-spin fields are 
insufficient because the dimension of these representations implies that higher 
powers of Lorentz generator matrices can be expressed in terms of lower powers.
For example, the square of a Pauli matrix describing
spin 1/2 is the identity matrix, which is of course not generally true.  Thus,  to capture all multi-spin 
interactions we need a formalism that describes arbitrarily-high spins. 
Such a formalism would also provide an {\it a priori} explanation of the validity of
the low-spin observations as well as give the minimal value of the
spin that is necessary to capture some given spin-induced multipole
moment.\footnote{As known for some time in the particle
  physics phenomenology literature, spin 1/2 is sufficient to capture
  the dipole moment and spin 1 the quadrupole. See e.g. Ref.~\cite{Khriplovich:1993js}. } 
Descriptions of higher-spin particles date back to Pauli and Fierz~\cite{FierzAllSpin}.
Our amplitudes-based approach to higher spin is closely related to
the world-line approaches of Refs.~\cite{SpinEFTEarlyWorldLine,LeviSteinhoffLagrangian}.  
The formalism makes the connection between Lorentz generators in the
amplitudes, and final spin vectors relatively transparent.  
Alternative approaches based on the massive-spinor-helicity formalism
of Ref.~\cite{MassiveSpinorHelicity}, are found in
Refs.~\cite{GOV,Spin2PM, Spin2PM_testBH,Spin2PM_1,
	SpinSmallYutin,NBI_Spin_Tree}.

In quantum field theory, massive particles of integer spin $s$ are
described by symmetric traceless rank-$s$ tensor fields \cite{Singh:1974qz},
\begin{equation}
\phi_s{}^{a_1\dots a_i\dots a_j\dots a_s} = \phi_s{}^{a_1\dots a_j\dots a_i\dots a_s}  \,, \hskip 1.5 cm 
\eta_{a_1a_2}\phi_s{}^{a_1 a_2 a_3 \dots a_s} = 0\, .
\label{symmetrictraceless}
\end{equation}
Additional transversality constraints are
necessary~\cite{Singh:1974qz} to select the part corresponding to
fixed spin $s$.  The corresponding physical states are described by
polarization tensors that are symmetric traceless and transverse in
all indices,
\begin{equation}
\pol^{a_1 a_2 \cdots  a_i \cdots a_j \cdots a_m} =  \pol^{a_1 a_2 \cdots  a_j \cdots a_i \cdots a_m} \,, \hskip 1.5 cm 
\eta_{a_1 a_2} \pol^{a_1 a_2 \cdots \cdots a_m} = 0 \,, \hskip 1.5 cm 
p_{a_1} \pol^{a_1 a_2 \cdots \cdots a_m} = 0 \ .
\end{equation}
The Hermitian Lorentz generators in this representation are:
\begin{equation}
(M^{ab}){}_{c(s)}{}^{d(s)} = {2 {} i {} s}\delta^{[a}_{(c_1}\eta^{b](d_1}\delta^{d_2}_{c_2}\dots \delta^{d_s)}_{c_s)}\,,
\hskip 1.5 cm 
(M^{ab}){}_{c(s)}{}_{d(s)} = -(M^{ab}){}_{d(s)}{}_{c(s)} \, ,
\label{ExplicitLorentzGenerators}
\end{equation}
where the indices $c(s)$ and $d(s)$ stand for the symmetrized 
sets of vector indices $\{c_1,\dots, c_s\}$ and $\{d_1,\dots, d_s\}$, respectively, and they are raised and lowered with the 
appropriate symmetric product of the Minkowski metric.
The generators $M^{ab}$ satisfy the usual Lorentz algebra,
\begin{equation}
[M^{a_1 a_2},M^{a_3 a_4}]=i(
\eta^{a_3 a_1}M^{a_4 a_2}
+\eta^{a_2 a_3}M^{a_1 a_4}
-\eta^{a_4 a_1}M^{a_3 a_2}
-\eta^{a_2 a_4}M^{a_1 a_3}) \, .
\label{LorentzAlgebra}
\end{equation}

As we will explain shortly, 
apart from describing the scattering of massive spin-$s$ fields, our interest 
is to develop a formalism that avoids use of any of the special properties of fixed-spin representations of the Lorentz group. 
Thus, it suffices for our purpose to not demand that they be transverse and instead treat the fields \eqref{symmetrictraceless} as unconstrained. 
It is then convenient to follow  Refs.~\cite{LandauQED, Khriplovich:1998ev} and map them to a two-component spinor 
indices in the usual way, 
\begin{equation}
\phi_s{}_{\alpha_1\dots \alpha_s}^{{\dot\beta}_1\dots {\dot\beta}_s} = \phi_s{}^{a_1\dots a_s}
(\sigma_{a_1}){}_{(\alpha_1}{}^{({\dot\beta}_1}\dots (\sigma_{a_s}){}_{\alpha_s)}{}^{{\dot\beta}_s)} \,.
\label{spinorindex}
\end{equation}
This parametrization trivializes the tracelessness condition \eqref{symmetrictraceless}, which translates into 
symmetrization in the two-component spinor indices of the same handedness.
Half-integer spin can also be described along these lines~\cite{LandauQED,
  Khriplovich:1998ev}, as pairs of such fields with different numbers
of left-handed and right-handed indices. While we do not discuss them
in any detail (and in the classical limit they should give the same
result as the integer-spin fields), we will also describe integer-spin
fields as pairs of fields \eqref{spinorindex}:
\begin{equation}
\phi_s = \frac{1}{\sqrt 2}\begin{pmatrix} \xi^{\alpha_1\dots \alpha_u}_{{\dot\beta}_1\dots {\dot\beta}_v} \\
\chi_{{\dot \alpha}_1\dots {\dot \alpha}_u}^{\beta_1\dots \beta_v}  \end{pmatrix} .
\label{spinorindex2}
\end{equation}
For integer- and half-integer-spin particles we have
\be
u=v=s
\qquad
u=s+\frac{1}{2}
\qquad
v = s-\frac{1}{2} \, ,
\ee
respectively.\footnote{One may have more general representations, in which $u$ and $v$ differ by some finite amount, 
$u=s_L$, $v=s_R$ and $s=s_L+s_R$. 
}
For half-integer spins $\xi$ and $\chi$ are different objects; one may impose a Majorana-type condition which identifies 
one with the conjugate of the other.  In the remaining part of this paper we use only integer spin, since that is sufficient
for describing large spin.

When taking the classical limit of quantum-mechanical expectation values it is necessary to choose states that minimize the standard 
deviation of observables being considered\footnote{E.g. for the harmonic oscillator, classical physics is recovered of one chooses 
it to be in a coherent state.}. For a spin system in the rest frame, the relevant states are 
the so-called ``spin coherent states''~ \cite{CoherentStates}. Their defining property is that
\begin{align}
\label{CoherentState}
\langle \bm n | \bm n\rangle = 1  \ , \qquad  
\langle \bm n | \hat {\bm S} | \bm n \rangle = \bm S \equiv  |\bm S| \bm n \ , \qquad    
\frac{\Delta \hat {\bm S}}{|\bm S|}\rightarrow 0 \ ,
\end{align}
where $ \hat {\bm S}$ is the rest frame spin operator, related to the rotation generator $M_{jk} $ in the usual way, 
$ \hat S {}^i = \frac{1}{2}\epsilon^{ijk} M_{jk} $,  and $\bm n$ is the unit vector along the classical spin. The state 
$ | \bm n \rangle$ localizes the spin along the unit vector $\bm n$ as much as it is allowed by quantum mechanics.

We define the covariant spin vector and spin tensor of a particle by boosting their rest-frame three-dimensional
counterparts $S^i$ and $S^{ij}$, which are related in the standard way
\begin{equation}
S^i = \frac{1}{2}\epsilon^{ijk} S_{jk}\, .
\end{equation}
The boost from the particle's rest frame gives,
\begin{align}
{ S}(p, {\bm S})^\mu &= \big(\, \frac{{\bm p}  \cdot {\bm S}}{m}, {\bm S} + \frac{{\bm p} \cdot {\bm S}}{m(E+m)} {\bm p} \, \big)
\, , \nn \\
{ S}(p, {\bm S})^{i0} &= -{ S}(p, {\bm S})^{0i} = \frac{1}{m}S^{il} {p_l} = \epsilon^{iln} \frac{p_l}{m} S_n \,,
\label{eq:spin_covToNR_map} \\
{ S}(p, {\bm S})^{ij} &= S^{ij}  - 2 \frac{p^{[i} S^{j] l} p_l}{m(m+E)}  
= \epsilon^{ijk}\left[\frac{E}{m} S_k - \frac{{\bm p} \cdot {\bm S}}{m+E} \frac{p_k}{m}\right] .
\nn 
\end{align}
where Roman letters from the middle of the alphabet indicate spatial indices. We raise and lower the indices of
the three-dimensional rest-frame spin vector with the Euclidean 3d metric, so $S^k = S_k$ (which should not to 
be confused with the spatial part of $S^\mu$).
These expressions can be summarized in a covariant format:
\begin{equation}
S^{\alpha\beta}(p)= -\frac{1}{m}\epsilon^{\alpha\beta\gamma\delta}{p}_{\gamma}{S}_{\delta}(p)
\,, \hskip 2 cm 
S^\alpha(p) = -\frac{1}{2m} \epsilon^{\alpha\beta\gamma\delta}{p}_{\beta}{S}_{\gamma\delta}(p)
\,,
\label{SpinTensorVectorRelation}
\end{equation}
where the four-dimensional Levi-Civita symbol is normalized as
$\epsilon^{0123}=1$. We will later denote the covariant spin vector of
particle $a$ as $S_a \equiv { S}(p_a, \bm S_a)$.
Our definition of the covariant spin vector implies that it obeys the so-called covariant spin supplementary condition,
\begin{equation}
p_\mu { S}(p, {\bm S})^\mu = 0  \, .
\label{CovariantSpinSupplementary}
\end{equation}

Boosting the relations \eqref{CoherentState} to an arbitrary frame
will not change the scalar product of coherent states and will yield the covariant spin
vector on the right-hand side of the second equation. Boosting the ket
and the bra states to momenta differing by some momentum transfer $q$
is less trivial. It is lengthy but straightforward to show that (see
also Appendix~E of Ref.~\cite{Spin2PM_1} for some details on the
derivation of $\pol({\bm s},p_1) \cdot \pol({\bm s},p_2)$)
\begin{align}
\pol({\bm s},p_1) \cdot \pol({\bm s},p_2)  &= \left(1  - i  
\frac{\epsilon_{rs k }p_1^r p_2^s S^k}{m(m +E({\bm p}_1))}
+ {\cal O}(\textrm{S}^2\bm q^2 )\right) + {\cal O}(q) \,, \nn\\[5pt]
\pol({\bm s},p_1) M^{ab} \pol({\bm s},p_2) &= { S}(p_1, {\bm S})^{ab} \; \pol({\bm s},p_1) \cdot \pol({\bm s},p_2) 
+{\cal O}(q^0) \,, 
\nn \\[5pt]
 \pol({\bm s},p_1)\frac{1}{2}\{ M^{ab} , M^{cd} \}\pol({\bm s},p_2) &= { S}(p_1, {\bm S})^{ab}\; { S}(p_1, {\bm S})^{cd} \; 
\pol({\bm s},p_1) \cdot \pol({\bm s},p_2) +{\cal O}(q^{-1})\, ,
\label{SpinEval}
\end{align}
where $\pol({\bm s},p_1)$ and $\pol({\bm s},p_2)$ are the incoming and
outgoing polarization tensors of a particle in a convention where both
momenta are taken to be outgoing, so that the momentum transfer is $q=
-{p}_1- {p}_2$. We denote the spin label by ${\bm s}$ to emphasize that, in general, 
it can be a quantum property of the particle and to distinguish it from the rest-frame classical spin vector $\bm S$.
The rest frame spin is assumed to be large in the
classical limit, with ${\bm q} \cdot {\bm S}/m\sim {\cal O}(1)$ as discussed in Sec.~\ref{ClassicalLimit}, 
and that the number of left-handed and right-handed indices in Eq.~\eqref{spinorindex2}, $u$ and $v$, 
are commensurate, i.e. $u-v\ll u, v$.
The terms in the parenthesis of first line are classical, and we only exhibit them to linear order in spin.
The rest of ${\cal O}(q^\alpha)$ denote classical expansion, where we show only the leading term. Recall that classical expansion count both $\bm q$ and spins.
The momentum of the spin
tensor can be chosen to be any combination of $p_1$ and $p_2$ since
all such combinations differ by terms proportional to the momentum
transfer $q$.
In tree amplitudes the momentum dependence of vertices
makes all these contributions subleading in the classical limit. We
will revisit the subleading terms in the second of Eq.~\eqref{SpinEval}
in our discussion of the effective field theory and the comparison of
its amplitudes with those of the higher-spin Lagrangian we discuss
next. As we will see, they do not affect observables.  Another
subtlety is that, as indicated in the final formula in \eqn{SpinEval},
only the symmetric product of Lorentz generators is interpreted directly
as products of spin tensors.  As will be discussed in
\sect{sec:nonminimal}, this is sufficient for obtaining the classical
limit of a product of two Lorentz generators after accounting for the
standard commutation relation~\eqref{LorentzAlgebra}.

\subsection{Higher-spin Lagrangians}
\label{HigherSpinLagrangianSubsection}

Theories of massive higher-spin fields have long history. 
A free action was constructed in Ref.~\cite{Singh:1974qz}. 
Spin-$s$ fields are described by rank-$s$ symmetric tensors. Fields
transforming in the $(s+1, s+1)$ representation of the $SO(3,1)\simeq
SU(2)\times SU(2)$ Lorentz group\footnote{We denote representations of
  the Lorentz group by $(d_L, d_R)$, where the two entries are the
  dimensions of the two $SU(2)$ representations.} contain many
representation of the rotation group.  To eliminate all but the
spin-$s$ representation (i.e. the $2s+1$-dimensional representation of
the $SO(3)$ rotation group), the tensor field is usually constrained
to be transverse. Implementing this in a Lorentz-invariant Lagrangian can be done~\cite{Singh:1974qz}
with the aid of $s$ rank-$k$ auxiliary fields, with $k=0,\dots, s-1$.
Preservation of tree-level unitarity when coupling this free action with gravity turns out to 
require introduction of dimension-four terms involving both the higher-spin fields and the 
Riemann curvature tensor~\cite{Deser:2001dt, HigherSpin}.
Aspects of an interacting Lagrangian constructed along these lines for all spin-induced 
multipole moments were discussed in Ref.~\cite{SpinSmallYutin}.

While we use a Lagrangian to organize the interactions of
higher-spin fields with gravity, we take a different approach
than earlier ones, which is tailored to our needs for
constructing classical limits of amplitudes with explicit 
dependence on the spin vector. It may be interpreted as a
relativistic effective theory that captures all spin-induced multipole
moments and thus all linear-response of spinning objects to gravity.
This approach provides a minimal completion of any desired
three-particle interactions which is invariant under the nonlinear
diffeomorphism transformations and offers a convenient way to align
our derivations with earlier ones.
Here we will use this approach to understand aspects of low-point
interactions of higher-spin fields and gravitons in the classical
limit and derive the effective interaction potential of two
higher-spin fields due to exchange of gravitons through $\Ord(G^2)$.

For our purpose it is not important that the matter fields transform in an
irreducible representation of the rotation group. It is, however, 
important that all irreducible components be treated uniformly.
To this end we take our fields to be traceless
rank-$s$ tensors in their spinor formulation \eqref{spinorindex} and
not require that they be transverse.  The validity of this approach
can be verified {\it a posteriori}, through the independence of the
result on the number of components of the tensor field.
One may intuitively expect that this will be the case as the number of
components of matter fields can arise only from loops containing them
and such graphs do not contribute in the classical limit. 
As we will see, this framework provides a minimal value for the spin needed
to capture the complete spin dependence of an $L$-loop four-point amplitude
in the classical limit. Such lower bonds are similar in spirit with 
the observation~\cite{HolsteinRoss, Vaidya, SpinSmallYutin} that
calculations at fixed and low spin can be used to reproduce the part
of the spin-dependent Hamiltonian that is available in the literature
and was originally derived though general-relativistic techniques.

We describe the gravitational field in the vielbein rather than the
metric formulation because it exposes the tangent-space Lorentz
generators, making it easier to identify the (classical) spin
vector. Since we are not interested in matter contact interactions of
higher-spin fields (because they do not contribute to the long-range 
potential), we will focus on a single higher-spin field, $\phi_s$.

Our higher-spin Lagrangian has two parts:
\begin{equation}
{\cal L}={\cal L}_{\rm min} + {\cal L}_{\rm nonmin} \ , \qquad S = \int d^4 x \sqrt{-g} {\cal L} \ .
\end{equation}
The minimal Lagrangian, i.e. the Lagrangian with the minimal number of
derivatives, including the terms needed to preserve tree-level
unitarity~\cite{Deser:2001dt, HigherSpin, HigherSpinString},
is\footnote{The sign of the $\H$ term follows form the one in
  \cite{Deser:2001dt, HigherSpin, HigherSpinString} by changing the
  signature of the metric to mostly-minus and converting to Hermitian
  Lorentz generators.}
\begin{align}
{\cal L}_{\rm min} &=-R(e, \omega) + \frac{1}{2} g^{\mu\nu} \nabla(\omega)_\mu\phi_s\nabla(\omega)_\nu\phi_{s} 
- \frac{1}{2}m^2\phi_s\phi_{s}
+\frac{\H}{8}R(e, \omega)_{efgh} \, \phi_s M^{ef} M^{gh} \phi_{s} +\dots \,,
\label{Ls}
\end{align}
where $\H$ is an adjustable parameter,  we take the higher-spin field $\phi_s$ to be real and the
ellipsis stand for terms that vanish on shell.  The $M^{ab}$ are the
Hermitian Lorentz generators in the $(s+1, s+1)$
representation~\eqref{spinorindex}-\eqref{spinorindex2}.  The
covariant derivative is
\begin{equation}
 \nabla(\omega)_{\mu} \phi_s \equiv  \partial_{\mu} \phi_s +\frac{i}{2} \omega{}_\mu{}_{ef} M^{ef}\phi_s \, ,
\end{equation}
where $\omega$ is the spin connection. To shorten the expression we do
not display the many tangent-space indices of $\phi_s$.  They are
understood as contracted via matrix multiplication.  The spinor
notation we use for the higher-spin field emphasizes that they are
assumed to carry {\em only} tangent-space indices; similarly, the
curvature tensor and the Lorentz generators also carry tangent-space
indices.
We postpone describing nonminimal higher-spin Lagrangians to the next section.

The last term displayed in \eqn{Ls} is a gravitational quadrupole
interaction; its coefficient may be set to $\H=1$ by requiring
that amplitudes have an improved high-energy behavior, delaying
violations of partial-wave unitarity~\cite{HigherSpin, HigherSpinString}. 
String theory predicts a different value for this coefficient~\cite{HigherSpinString}. 
This may be interpreted as being due to the other higher-spin 
fields of string theory further contributing to the unitarity constraint.
Here we keep $\H$ as a free parameter. This term does not affect any
interaction linear in the particle's spin,
but it is important at higher order in spin and, as we will see in
\Sec{onshellvertex}, plays an important role in giving a
field-theory description of the stress tensor of the Kerr black
hole~\cite{Vines1PM}. The value of $\textrm{H}$ found by matching to a Kerr
black hole reproduces the one required by improved partial-wave
unitarity.

At tree level there is no physical difference between the scattering
of higher-spin fields described by the Lagrangian \eqref{Ls} and by
one that enforces transversality of the higher-spin polarization
vectors. This is because four-point tree level scattering amplitudes
of higher-spin fields contain no Feynman graphs with propagators for
these fields.

Let us now examine the relation between calculations carried out with 
low-spin fields and with arbitrary-spin fields, beyond tree level and
for the case where each vertex contains no more than one Lorentz
generator.  The Lagrangians for massive vectors and massive spin-2
fields without curvature couplings are both of the form of \eqn{Ls}. 
Theories of such low-spin fields will yield the same amplitudes as the 
Lagrangian~\eqref{Ls} {\em as long as} special relations obeyed by symmetric 
products of generators of four-dimensional Lorentz group 
in representations $(2, 2)$ and $(3, 3)$ are {\em not} used.
These relations stem from the fact that, for a spin-$s$ representation 
of $SU(2)$, with generators $J_s$,
\begin{align}
(\Xi\cdot J_s)^{k\ge 2s+1} = \sum_{n=1}^{2s} a_n(\Xi, k, s) (\Xi \cdot J_s)^n
\label{Mrelations}
\end{align}
for some coefficients $a_n(\Xi, k, s)$.  Here $\Xi$ is an arbitrary three-component vector;
differentiating Eq.~\eqref{Mrelations} $k$ times with respect to $\Xi$ yields the decomposition
of a symmetric product of $k$ generators of $SO(3)$ into a linear combinations of symmetric 
products of at most $s$ generators.
If each matter-graviton vertex of a Feynman diagram of an $L$-loop four-point matter 
amplitude contains at most one Lorentz generator, it is easy to see that each matter line of 
this diagram contains a (symmetric) product of at most $L+1$ generators, multiplied 
from the left and the right with polarization tensors. 
For \eqref{Mrelations} not to operate if the fields are in a chiral representation of 
the Lorentz group (i.e. they transform only under one of the two $SU(2)$ factors), it is 
therefore necessary that they be in a representation of dimension ${\rm dim}\ge L+2$.

This counting suggests that one-loop calculations carried out with fields with the Lorentz 
representations $(3, 1)$ or/and $(1, 3)$ should be sufficient within our formalism, because they 
yield products of at most two Lorentz generators (in e.g. box or triangle graphs) for a matter 
line. At two loops, where we may get products of three Lorentz generators,  Lorentz 
representations of the type $(4, 1)$ or/and $(1, 4)$ are needed.
We interpret these bounds as being sufficient to capture the complete spin dependence within our formalism.
Similarly, on a case-by-case basis it may be possible to evade them and use e.g. spin-1 fields in our formalism 
and capture the complete spin dependence at one loop. One may reach this conclusion by e.g. 
constructing a relation analogous to \eqref{Mrelations} for the $SO(3,1)$ generators in the $(2,2)$ 
representation and demanding that the decomposition be manifestly covariant. 

We stress that the counting above refers specifically to our formalism and does not necessarily apply 
to actions that e.g. use properties of special representations of $SO(3,1)$.
The actual bound might be even lowered after the classical limit is applied.
In general, perhaps the most straightforward approach to using low values of spin is to use 
a formulation of the low-spin Lagrangians that does not implicitly employ relations between Lorentz 
generators that use their four dimensional nature. Moreover, such relations should not be used at 
any step in the calculation of amplitudes.

Trilinear couplings containing more than one Lorentz generator, such
as the $\H$-dependent term in \eqref{Ls} and the higher-derivative
terms discussed below change the counting argument above and
suggest a need for larger representations at lower-loop orders.

\subsection{Expansion of the minimal Lagrangian}
\label{LargragianExpansionSubsection}

The spin connection $\omega$ is an auxiliary field, which can be
eliminated via its equation of motion, as usually done in
supergravity theories. This expresses $\omega$ in terms of the
vielbein and matter fields. Once replaced in the original Lagrangian,
the matter-field dependence yields only matter contact terms and is
thus irrelevant for long-range interactions of matter fields.  The
remainder matter-independent solution of the $\omega$ equation of
motion is equivalent to the solution to the vielbein postulate,
$\nabla_\mu(\omega) e{}_\nu{}^a = 0$. We will denote it by
$\omega(e)$.

Following standard methods we define the graviton field as the fluctuation of the metric around Minkowski background.  
Local Lorentz symmetry can be used to choose the fluctuations of the vielbein to be symmetric, $h{}_\mu{}_a=h{}_a{}_\mu$.
We take,
\begin{align}
& g_{\mu\nu}  = \eta_{\mu\nu}+h_{\mu\nu}\,, \hskip 1. cm 
e{}_\mu{}^a =\delta_\mu^a + \frac{1}{2}h{}_\mu{}^a-\frac{1}{8}h_{\mu\rho}h^{a\rho}+\Ord(h^3) \, \nn \\
& \omega(e){}_\mu{}_{cb} = -\partial_{[c}h_{b]\mu}
 - \frac{1}{4} h{}^\rho{}_{[c} \partial_\mu  h{}_{b]\rho}
+ \frac{1}{2} h{}^\rho{}_{[c} \partial_\rho  h{}_{b]\mu}
 - \frac{1}{2} h{}^\rho{}_{[c} \partial_{b]}   h_{\mu\rho} + \Ord(h^3) \, ,
\end{align}
where the antisymmetrization includes division by the number of terms and we take $\eta_{\mu\nu}$ to be in the mostly-minus
convention.
One may make different choices for the metric fluctuations to e.g. make the expansion of the vielbein simpler; 
while this has no effect on scattering amplitudes, it makes gravitational vertices depart from their standard form. 
With the choice above, the expansion of the Riemann tensor to second order in fluctuations is:
\begin{align}
\label{Riemann_expansion}
 R(e, \omega(e))_{e f g h} = & -2\partial_{[e|}\partial_{[g}h_{h]|f]}
+ (h{}^\mu{}_{[e}\delta^\nu_{f]} + \delta{}^\mu_{[e} h^\nu{}_{f]})\partial_{\mu}\partial_{[g}h_{h]\nu}
\nn \\
&\null 
 -\frac{1}{2} \partial_{[e|} h{}^\rho{}_{[g} \partial_{|f]}  h{}_{h]\rho}
 + \partial_{[e|} h{}^\rho{}_{[g} \partial_\rho  h{}_{h]|f]}
 -  \partial_{[e|}  h{}^\rho{}_{[g} \partial_{h]}   h_{|f]\rho}
 \nn \\
& +2 \partial_{[g}h_{c][e|} \partial_{[d}h_{h]|f]}\eta^{cd} + \Ord(h^3) \,.
\end{align}

\begin{figure}
\begin{center}
\includegraphics[scale=.75]{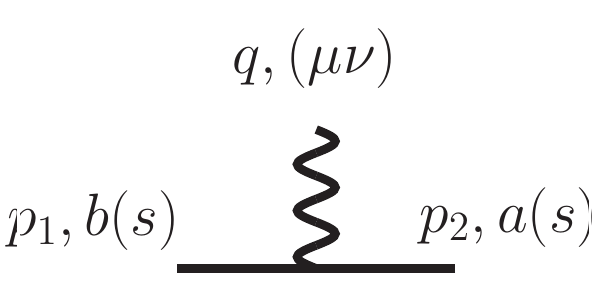}
\end{center}
\vskip -.2 cm
\caption{\small The three vertex labels.  All momenta are outgoing.  }
\label{VwertexFigure}
\end{figure}

Following the usual procedure we can extract Feynman vertices.  Consider the 
three-point vertex in \fig{VwertexFigure}.
The contribution from the three vertex from the minimal Lagrangian \eqref{Ls} is then
\begin{align}
-iV{}^{\mu\nu}_\text{min} {}_{a(s)}{}^{b(s)}(q, p_1, p_2) = &  p_1^{(\mu} p_2^{\nu)}\delta_{a(s)}^{\; b(s)} 
- \frac{1}{2}\eta^{\mu\nu} (p_1\cdot p_2 +m^2)\,\delta_{a(s)}^{\; b(s)} 
- \frac{i}{2} q_{\rho}(p_2-p_1)^{(\mu} (M^{\nu)\rho}){}_{a(s)}{}^{b(s)} 
\nn \\
& + \frac{\H}{2} q_{\rho}q_{\sigma}(M^{\rho(\mu|}M^{\sigma|\nu)}){}_{a(s)}{}^{b(s)} \,,
\label{ThreeVertexOffShell}
\end{align}
where the legs carrying momenta $p_1$ and $p_2$ are spin-$s$ fields with sets of tangent-space 
Lorentz indices $b(s)$ and $a(s)$, respectively, and the symmetrization of the two graviton indices 
has unit strength (i.e. it includes division by the number of terms).  
A useful property of the vertex, following from diffeomorphism invariance of the action, 
is the on shell analog of stress tensor conservation. That is, when its external legs are placed on shell, the 
three-point vertex is  transverse with respect to the graviton momentum,
\begin{equation}
q_\mu V{}^{\mu\nu}_\text{min} {}_{a(s)}{}^{b(s)}(q, p_1, p_2) = 0\,, \hskip 2 cm 
q_\nu V{}^{\mu\nu}_\text{min} {}_{a(s)}{}^{b(s)}(q, p_1, p_2) = 0 \, .
\label{ThreeVertexTransverse}
\end{equation}
This does not require any special properties, such as transversality, for the higher-spin polarization tensors. 
This is consistent with our setup, in which higher-spin fields are not required to be transverse. 
As for low-spin particles, this property guarantees the gauge-choice independence of the tree-level four-point matter amplitude.

\section{Nonminimal Interactions}
\label{sec:nonminimal}

The general form of the stress tensor of an arbitrary-spin particle in
a parity-invariant theory is described in
Ref.~\cite{Khriplovich:1998ev} in terms of four independent form
factors. When used as vertices in a scattering amplitude, two of them
contribute only contact terms.
In this section we describe a nonminimal part ${\cal
  L}_\text{non-min}$, corresponding to the remaining two form factors
of the Lagrangian~\eqref{Ls}.

\subsection{Nonminimal higher-spin Lagrangians and cubic vertices}
\label{HigherSpinNonMinimalLagrangianSubsection}

To construct manifestly-covariant spin-dependent Lagrangian
interactions it is convenient to define an off-shell
manifestly-covariant analog of the Pauli-Lubanski vector: \be
\S^{a}\equiv \frac{-i}{2m}\epsilon^{a b cd}M_{cd}\nabla(\omega)_{b}
\,.
\label{Shat_operator}
\ee It carries an explicit tangent-space vector index and two implicit
labels for the spin representation which we will choose to be $(s+1, s+1)$.
Then, all the terms linear in the graviton and bilinear in higher-spin fields are 
\begin{align}
\label{Lnonmin}
{\cal L}_\text{non-min}&=\sum_{n=1}^{\infty} \frac{\left(-1\right)^n}{\left(2n\right)!}
\frac{C_{ES^{2n}}}{m^{2n}} \nabla(\omega)_{f_{2n}}\cdots \nabla(\omega)_{f_3} R_{f_1 a f_2 b} 
\nabla(\omega)^a{\phi}_s\,  \S^{(f_1} \dots \S^{f_{2n})} \nabla(\omega)^b \phi_s
\\
&-\sum_{n=1}^{\infty} \frac{\left(-1\right)^n}{\left(2n+1\right)!}
\frac{C_{BS^{2n}}}{m^{2n+1}} \nabla(\omega)_{f_{2n+1}}\cdots \nabla(\omega)_{f_3}
\frac{1}{2}\epsilon_{a b (c| f_1} R^{a b}{}_{|d) \, f_2} \nabla(\omega)^c  {\phi}_s  \S^{(f_1} \dots \S^{f_{2n+1})}  \nabla(\omega)^d\phi_s
\,.
\nonumber
\end{align}
where, as in \eqn{Ls}, the indices on $\phi_s$ are implicit.  
The operators included here are in one-to-one correspondence to the non-minimal couplings in the worldline spinning-particle 
action of~Ref.~\cite{LeviSteinhoffLagrangian}.
As in the minimal Lagrangian, here all indices are flat and we assume that the fields are real.

We note that, through cubic order in fields, the matrix elements of the quadrupole term in \eqref{Ls} are indistinguishable 
from the matrix elements of the $n=1$ term on the first line of \eqn{Lnonmin}. Thus, when combining the minimal and non-minimal Lagrangians and their contributions to vertices, we will drop the $\H$ term in favor of $C_{ES^2}$.

One can include further terms, with two or more Riemann tensors. At
${\cal O}(G)$ in a scattering process, they necessarily imply emission of gravitational
radiation. At higher orders in Newton's
constant they contribute also to the conservative part of the
two-particle spin-dependent Hamiltonian.
Some of them can be identified as bilinears in the single-graviton
operators included in \eqn{Lnonmin}. We will not attempt to classify
here all such operators to all orders in spin.


To find the non-minimal vertices we expand \eqref{Lnonmin} around
Minkowski space.  Following the same reasoning as in
\sect{LargragianExpansionSubsection}, we may take the spin
connection to be given by the solution to the vielbein postulate. This
implies that the expansion of the Riemann tensor is given by
\eqn{Riemann_expansion}. Since all non-minimal terms contain a Riemann
tensor, none of the other vielbein or spin connections in
\eqn{Lnonmin} contribute to the three-point vertex.
It is convenient to consider separately the contribution of the terms depending on the Riemann tensor and its dual: 
\begin{align}
-iV_3{} ^{\mu\nu}_\text{non-min}&=-iV_3{} ^{\mu\nu}_{\text{non-min}, E}-iV_3{} ^{\mu\nu}_{\text{non-min}, B} \, .
\end{align}
Each of them is given by 
\begin{align}
\label{nonminimal_vertices}
-iV_3{} ^{\mu \nu }_{\text{non-min}, E}&=
p_1^{(\mu }p_2^{\nu )}{\rm Sym}[(q\cdot \S_0(p_1)), (q\cdot \S_0(p_1)), 
{\hat E}(q\cdot \S_0(p_1)) 
]+{\cal O}(q, p_i^2-m^2) \,,
\\
-iV_3{} ^{\mu \nu }_{\text{non-min}, B}&=
 - \frac{i}{2} m^2 q_{\rho}  (p_2-p_1)^{(\mu } {\rm Sym}[M^{\nu ) \rho}, q\cdot \S_0(p_1) , 
 {\hat B}(q\cdot \S_0(p_1)) 
 ] 
 +{\cal O}(q, p_i^2-m^2, p_i^\mu M_{\mu\nu}) \, ,
 \nonumber
\end{align}
where $\S_0(p)^\mu$ is the 
Fourier-transform of the linearization of the operator $\S^\mu$ in \eqn{Shat_operator} (i.e. the 
part that is independent of the metric fluctuations), 
\begin{equation}
\label{S0hat}
 \S_0(p)^\mu \equiv \frac{1}{2m}\epsilon^{\mu \nu\rho\sigma} p_{\nu}M_{\rho\sigma} \,.
\end{equation}
In the second Eq.~\eqref{nonminimal_vertices} we neglected terms containing $p_i^\mu M_{\mu\nu}$ because
such terms vanish in the classical limit up to contributions subleading in $q$.
The operators  ${\hat E}(X)$ and ${\hat B}(X)$ are defined as
\begin{align}
{\hat E}(X) = \sum_{n=1}^{\infty} \frac{1}{\left(2n\right)!}
\frac{C_{ES^{2n}}}{m^{2n}}  X^{2n-2}  
\, , \hskip 1.5 cm
{\hat B}(X) =  \sum_{n=1}^{\infty} \frac{1}{\left(2n+1\right)!}
\frac{C_{BS^{2n+1}}}{m^{2n+2}} 
X^{2n-1}   \,,
\end{align}
and the operator ${\rm Sym}[\dots]$ symmetrizes with unit strength in
all of its arguments. Last, if an argument is the $n$th power of an
operator, then it is interpreted as $n$ distinct entries.

\subsection{The higher-spin and the Kerr black hole stress tensor}
\label{onshellvertex}
\label{matchKerr}

To construct the on-shell stress tensor we contract the three-point vertex  with polarization tensors for the higher-spin fields, 
use the mass-shell conditions and transversality of the graviton polarization tensor,
\begin{equation}
q^2 = 0\,, \hskip 1.5 cm q^\mu \pol(q)_{\mu\nu}  = q^\nu \pol(q)_{\mu\nu} = 0\,,
\hskip 1.5 cm p_1^2 = p_2^2 = m^2 \,,
\end{equation}
and evaluate
\begin{eqnarray}
T^{\mu\nu}=\frac{i}{m} \frac{\pol(\bm s, p_2)  (V_{3, \text{min}}^{\mu\nu}+V_{3, \text{non-min}}^{\mu\nu}) \pol(\bm s, p_1)}
{\pol(\bm s, p_2)\cdot\pol(\bm s, p_1)} \,.
\label{HSstresstensor}
\end{eqnarray}
Following our original setup, we do not assume that the higher-spin polarization tensors are transverse. 
The division by ${\pol(\bm s, p_2)\cdot\pol(\bm s, p_1)}$ can be understood as a choice of position-space coordinate conjugate to 
the graviton momentum.
With this normalization this coordinate is covariant and thus include a certain 
shift proportional to the rest-frame spin \cite{Khriplovich:1998ev}.

To see this it is useful to recall that, as it was understood long ago by Foldy and Wouthuysen~\cite{Foldy:1949wa} in the context 
of the free Dirac theory, the operator $\bm x_\text{cov}$ whose expectation value is the position of a particle is in fact a particular combination of the canonical position operator ${\bm  x}_\text{}$ and the spin operator.
For a particle with momentum $p$ and rest-frame spin $\bm S$ it is
\begin{align}
\bm x_\text{cov} =  \bm x_\text{} -  \frac{\bm p \times \bm S}{m(E(\bm p)+m)} \, .
\label{FW}
\end{align}
A similar relation was shown in Ref.~\cite{Vines1PM} to be a
consequence of switching between the covariant and canonical spin
supplementary conditions. Using the all-orders-in-spin generalization
of Eqs.~\eqref{SpinEval},
\begin{align}
\pol(\bm s,p_1) \cdot \pol(\bm s,p_2) 
 = \exp\left[ - i   \frac{\epsilon_{rs k }p_1^r p_2^s S^k}{m(E+m)} \right] +{\cal O}(q) \, ,
 \label{full_epsdoteps}
\end{align}
which may be proven directly, by writing the polarization tensors as boosts of rest-frame coherent
states, as in Sec.~\ref{sec:basics}, it is straightforward to see that  
\begin{align}
\int d^2 \bm q  \, e^{-i \bm r_\text{}  \cdot \bm q}
{\pol(\bm s, p_2)  (V_{3, \text{min}}^{\mu\nu}+V_{3, \text{non-min}}^{\mu\nu}) \pol(\bm s, p_1)}
=\int d^2 \bm q  \, e^{-i \bm r_\text{cov} \cdot \bm q} \frac{\pol(\bm s, p_2)  (V_{3, \text{min}}^{\mu\nu}+V_{3, \text{non-min}}^{\mu\nu}) \pol(\bm s, p_1)}{\pol(\bm s, p_2)\cdot\pol(\bm s, p_1)} \, .
\label{FWAmplitudeRelation}
\end{align}
This choice facilitates comparisons with Ref.~\cite{Vines1PM}, which uses the covariant coordinate and covariant spin tensor 
in the derivation of the Kerr black hole stress tensor. 
To express our results in terms of the rest-frame spin, it is necessary to restore the spin dependence contained
in the product of polarization tensors; we will do so in later sections.

The classical limit of Eq.~\eqref{HSstresstensor} can be taken by using a generalization of \eqn{SpinEval} to the symmetric 
product of an arbitrary number of Lorentz generators. By boosting from the rest frame, where such products can be computed using 
the properties of the coherent states and the explicit forms of Lorentz generators, it is not difficult to find that
\begin{align}
\label{genericSymMMM}
\pol({\bm s},p_2){\rm Sym}[M^{\mu_1\nu_1},\dots,M^{\mu_n\nu_n}]\pol({\bm s},p_1)
=S(p_1, {\bm S})^{\mu_1\nu_1} \dots S(p_1, {\bm S})^{\mu_n\nu_n}\pol({\bm s},p_2)\cdot \pol({\bm s},p_1) +\Ord(q^{-(n-1)}) \, .
\end{align}
This relation can be used to evaluate the expectation value of a generic product of Lorentz generators. Indeed, using the Lorentz 
algebra one can rewrite an arbitrary monomial in Lorentz generators as a sum of completely symmetric products or generators, with coefficients given by the structure constants of the algebra. For example,
\begin{align}
M^{\mu_1\nu_1}M^{\mu_2\nu_2}&=\frac{1}{2}\{M^{\mu_1\nu_1}, M^{\mu_2\nu_2}\}+\frac{1}{2}[M^{\mu_1\nu_1},M^{\mu_2\nu_2}]
\label{eq:cov_symmetrization} \\
&=\frac{1}{2}\{M^{\mu_1\nu_1}, M^{\mu_2\nu_2}\} +
\frac{i}{2}(
\eta^{\mu_3 \mu_1}M^{\mu_4 \mu_2}
+\eta^{\mu_2 \mu_3}M^{\mu_1 \mu_4}
-\eta^{\mu_4 \mu_1}M^{\mu_3 \mu_2}
-\eta^{\mu_2 \mu_4}M^{\mu_1 \mu_3}) \,.
\nonumber
\end{align}
Then, the expectation value of each factor can be evaluated using
Eq.~\eqref{genericSymMMM}. Each time Lorentz algebra is used, the
number of generators decreases by one; the expectation value of the
resulting monomials is subleading compared to that of the symmetric
product of the original number of generators. Such subleading terms
will be crucial at one loop in Sec.~\ref{sec:oneloop} to obtain the
correct classical terms. By the same reasoning, this will true at
higher loops as well.

At tree level however, the maximal number of generators already gives
a classical contribution, so all comparatively subleading terms can be
ignored.
Upon using  \eqn{genericSymMMM}, all Lorentz generators $M^{\mu\nu}$ become spin tensors. Moreover, using the contraction of 
Eq.~\eqref{SpinTensorVectorRelation} with the momentum transfer $q$,
\begin{equation}
\frac{1}{2}\epsilon_{\mu\nu\rho\sigma} q^{\mu}p_1^\nu S(p_1)^{\rho\sigma} = -m \, q_\mu S(p_1)^\mu \equiv -m\, q\cdot S(p_1)\,,
\label{identity}
\end{equation}
ignoring terms subleading in the small-$q$ expansion and defining $C_{ES^{0}}=1$ and $C_{BS^{0}}=1$, the stress tensor becomes
\begin{align}
\label{Lstresstensor}
T^{\mu\nu}(p_1, q)=\frac{p_1^{\mu}p_1^{\nu}}{m}
\sum_{n=0}^{\infty} \frac{C_{ES^{2n}}}{\left(2n\right)!}
\left(\frac{q\cdot S(p_1)}{m}\right)^{2n}  \!\!
-\frac{i}{m} q_\rho p_1^{(\mu}S(p_1)^{\nu)\rho}
\sum_{n=1}^{\infty} \frac{C_{BS^{2n+1}}}{\left(2n+1\right)!}
\left(\frac{q\cdot S(p_1)}{m}\right)^{2n} \! .
\end{align}
As expected, it has a form consistent with the general stress tensor that contributes to long-range interactions~\cite{Khriplovich:1998ev}.

Eq.~\eqref{identity} also implies that, as stated previously, the
coefficient $C_{ES^{2}}$ is equivalent to the quadrupole $\H$ term in
Eq.~\eqref{Ls}. Indeed, using the relation between the covariant spin
vector and tensor it is not difficult to show that
\begin{eqnarray}
S(p_1)^{\mu\rho} S(p_1)^{\nu \sigma} q_\rho q_\sigma =  
-\frac{1}{m^2}p_1^\mu p_1^\nu  (q\cdot S(p_1))^2+\dots 
=+ \frac{1}{m^2}p_1^{(\mu} p_2^{\nu)}  (q\cdot S(p_1))^2+\dots 
\label{SSidentity}
\end{eqnarray}
where ellipsis stands for terms that vanish when the free indices are
contracted with an on-shell graviton polarization tensor.  Thus, as
noted earlier, we are justified to ignore the quadrupole term in \eqn{Ls} when the
non-minimal interaction Lagrangian \eqref{Lnonmin} is included. Comparing  Eq.~\eqref{ThreeVertexOffShell} with the 
$n=1$ term in the first Eq.~\eqref{nonminimal_vertices} in the classical limit and using \eqref{SSidentity}, it is easy to see that 
the coefficient $\H$ is related to $C_{ES^2}$ as
\begin{equation}
\H = C_{ES^2} \,.
\label{CvsH}
\end{equation}
Thus, the value of $\H$ for the Kerr black hole can be found by comparing Eq.~\eqref{Lstresstensor} 
to the stress tensor of the Kerr black hole constructed in Ref.~\cite{Vines1PM}.

To carry out this comparison we first organize the result of Ref.~\cite{Vines1PM} in our notation.
It is found there by casting the linearized Kerr metric in the form of an operator acting on a free-particle Green's function:
\be 
h_{\rho\sigma} = 4G{\cal P}^\text{de Donder}_{\rho\sigma\mu\nu} {\hat T}^{\mu\nu}
\frac{1}{r} 
\ , \qquad
{\cal P}^\text{de Donder}_{\rho\sigma\mu\nu} = 
\frac{1}{2}\eta_{\mu\alpha}\eta_{\nu\beta}+\frac{1}{2}\eta_{\nu\alpha}\eta_{\mu\beta}
-\frac{1}{D-2}\eta_{\mu\nu}\eta_{\alpha\beta}
\, .  
\ee 
Here ${\cal P}^\text{de Donder}_{\rho\sigma\mu\nu}$ is the tensor structure of the
graviton propagator in de~Donder gauge and $r$ is the flat-space
four-dimensional coordinate distance, $r^2=\eta_{\mu\nu} x^\mu x^\nu $.
The stress-tensor operator ${\hat T}^{\mu\nu}$ is given by (cf.~Eq.~(32a) of Ref.~\cite{Vines1PM}) 
\begin{equation}
\label{VinesStressTensor}
{\hat T}^{\mu\nu} = m \exp(a*\partial){}^{(\mu}{}_\rho u^{\nu)}u^\rho \, ,
\end{equation}
where
\begin{equation}
a^\mu = \frac{1}{2p^2} \epsilon{}^\mu{}_{\nu\rho\sigma}p^\nu S^{\rho\sigma} \,,
\hskip 1.5 cm 
(a*\partial)^\mu{}_{\nu} \equiv \epsilon^{\mu}{}_{\nu\rho\sigma}a^{\rho}\partial^\sigma \,,
\end{equation}
and $u$ is the four-velocity of the black hole, $u^\mu=p^\mu/m$.  Identities such as
\be
(a*\partial)^\mu{}_{\nu} (a*\partial)^\nu{}_{\rho} \frac{u^\rho}{r} = -(a\cdot \partial)^2\, \delta^\mu_\rho \frac{u^\rho}{r} \,,
\qquad
(a*\partial)^\nu{}_{\rho} \frac{u^\rho}{r} = S(p)^\mu{}_\rho \partial^\rho \frac{1}{r} \,,
\ee
may be used reorganize the exponential factor.

To compare with our trilinear vertex we need to Fourier-transform ${\hat T}^{\mu\nu}$ to momentum space, which is
easily done via the substitution $\partial_\mu \rightarrow i q_\mu$. It leads to
\begin{align}
{\hat T}^{\mu\nu} = m \exp(ia*q){}^{(\mu}{}_\rho u^{\nu)}u^\rho &= m (\cos a*q+ i \sin a*q){}^{(\mu}{}_\rho u^{\nu)}u^\rho 
\cr
&= m \biggl(\cosh (a\cdot q) \delta^{(\mu}_\rho+ i \frac{(a*q){}^{(\mu}{}_\rho}{a\cdot q} \sinh (a\cdot q)\biggr) u^{\nu)}u^\rho 
\cr
&= m \cosh(a\cdot q) u^{\mu}u^\nu -  \frac{i}{a\cdot q} \sinh(a\cdot q)\, q^\rho S(p){}_\rho{}^{(\mu} u^{\nu)} \,.
\label{FT1pm}
\end{align}
It is not difficult to see, 
using Eq.~\eqref{identity}, that the building block of this expression, $a\cdot q$,  is the same as the building block 
of Eq.~\eqref{Lstresstensor}:
\begin{equation}
a\cdot q = \frac{1}{2m^2}\epsilon_{\mu\nu\rho\sigma} q^{\mu}p^\nu S(p)^{\rho\sigma} = -\frac{q\cdot S(p)}{m} \,.
\end{equation}
Further using the relation between momentum and velocity,
Eqs.~\eqref{FT1pm} and~\eqref{Lstresstensor} become identical if we choose
\begin{equation}
C_{ES^{2n}} = 1 \qquad C_{BS^{2n}} = 1 \,.
\end{equation}
Eq.~\eqref{CvsH} then implies that the $\H$ parameter of the Kerr black hole is
\begin{equation}
\H=1 \,.
\end{equation}
As mentioned earlier, this value is the one required~\cite{HigherSpin, HigherSpinString} for amplitudes of higher-spin fields  to
have an improved high-energy behavior delaying violations of partial-wave unitarity.

The relation between the Kerr stress tensor \cite{Vines1PM} and the
one following from the Lagrangian \eqref{Ls} and \eqref{Lnonmin}
implies that the tree-level scattering amplitude of two higher-spin
fields---and consequently the 1PM effective Hamiltonian---will also
reproduce the scattering amplitude of two Kerr black holes found in
Ref.~\cite{Vines1PM}.
In subsequent sections we will be interested in Hamiltonian terms that
contain at most one spin vector for each of the two particles, so we
will focus on the minimal Lagrangian \eqref{Ls} and ignore the
higher-derivative terms in the nonminimal Lagrangian~\eqref{Lnonmin}.

\subsection{The double-copy properties of general three-point vertex}
\label{doublecopyonshellvertex}

For computations beyond leading order in Newton's constant it can be
quite useful to exploit the double-copy structure of gravitational
theories. This property played a useful role in the computation of the
two-body Hamiltonians for spinless particles at
$\Ord(G^3)$~\cite{3PMPRL, 3PMLong}.
The double-copy properties of amplitudes of massive spin-1/2 and
spin-1 massive particles have been studied in some detail
in~\cite{DoubleCopySpinHenrik}, using the massive-spinor-helicity
formalism.
Here we make a few observations on the properties of the trilinear
vertex for graviton-coupled arbitrary spin particles and in
\sect{sec:tree} we derive a double-copy formula for the tree-level
gravitational Compton amplitude of the minimal Lagrangian. This
tree-level amplitude, together with the three- and four-point
amplitudes of higher-spin particles, are the building blocks of the
one-loop amplitude we construct in~\sect{sec:oneloop}.

As we now show, for generic values of the  parameters $C_{ES^{2n}}$ and $C_{BS^{2n}}$, the complete on-shell
trilinear graviton-higher-spin vertex,
\begin{equation}
V_3^{\mu\nu} = V_{3, \text{min}}^{\mu\nu}+V_{3, \text{non-min}}^{\mu\nu} \, ,
\label{full3vertex}
\end{equation}
can be expressed as the double-copy of trilinear vertices coupling higher-spin fields with vector fields. The double copy is usually 
formulated in terms of non-abelian vector fields; for three-point interactions, the non-abelian structure is not essential, so one may 
equally well describe \eqref{full3vertex} as the double-copy of trilinear vertices coupling higher-spin fields with a Maxwell field. 
Extension of the double-copy property for four- and higher-point amplitudes is an interesting open question. 

To see explicitly these properties, consider the general trilinear vector-higher-spin vertex \footnote{As in their coupling to
gravity, here too higher-spin fields are not required to be transverse.} arising from the Lagrangian
\begin{align}
\label{Lguess}
{\cal L} &= 
\frac{1}{4} F^a_{\mu\nu}F^{a,\mu\nu}
\\
&-\frac{1}{2}\sum_{n=0}^{\infty}\frac{C_n \eta_{\mu_0\nu_0} }{2^{2n}m^{4n}} 
\epsilon_{\mu_1\nu_1\rho_1\sigma_1}\dots  \epsilon_{\mu_{2n}\nu_{2n}\rho_{2n}\sigma_{2n}}
\cr
&\qquad\qquad\qquad
D^{(\mu_0}  D^{\mu_1}\dots D^{\mu_{2n})}{\varphi}_{s_g} {\rm Sym}[M^{\rho_1 \sigma_1},\dots, M^{\rho_{2n}\sigma_{2n}}] D^{(\nu_0}  D^{\nu_{1}}\dots D^{\nu_{2n})}  \varphi_{s_g}
\cr
&+\frac{1}{2}m^2 \sum_{n=0}^{\infty}\frac{C_n}{2^{2n} m^{4n}} 
\epsilon_{\mu_1\nu_1\rho_1\sigma_1}\dots  \epsilon_{\mu_{2n}\nu_{2n}\rho_{2n}\sigma_{2n}}
\cr
&\qquad\qquad\qquad
D^{(\mu_1}\dots D^{\mu_{2n})} \varphi_{s_g} {\rm Sym}[M^{\rho_1 \sigma_1},\dots, M^{\rho_{2n} \sigma_{2n}}] D^{(\nu_{1}}\dots D^{\nu_{2n})}  \varphi_{s_g}
\cr
&+\frac{i}{2} \sum_{n=0}^{\infty}\frac{E_n}{2^{2n} m^{4n}} \epsilon_{\mu_1\nu_1\rho_1\sigma_1}\dots  \epsilon_{\mu_{2n}\nu_{2n}\rho_{2n}\sigma_{2n}}
\cr
&\qquad\qquad\qquad
D^{(\mu_1}\dots D^{\mu_{2n})} {\varphi}_{s_g} {\rm Sym}[M_{\mu_0 \nu_0} F^{\mu_0\nu_0} ,M^{\rho_1 \sigma_1},\dots, M^{\rho_{2n} \sigma_{2n}} ]D^{(\nu_{2n}}\dots D^{\mu_{1})}  \varphi_{s_g} \,.
\nonumber
\end{align}
Here we assume that the real higher-spin field $\varphi_{s_g}$ is in a real non-adjoint representation of some gauge group $\cal G$,  
$D^{\nu}$ is the corresponding covariant derivative and  $F_{\mu_0 \nu_0} \equiv F^{a}_{\mu_0 \nu_0} T^a$ is its field strength. 
Eq.~\eqref{Lguess} may be given in a (slightly) more compact form in terms of an operator obtained from $\S$ defined in 
Eq.~\eqref{Shat_operator} by replacing the gravitational covariant derivative with a gauge-covariant derivative.
Both Lorentz and gauge group indices are contracted via matrix multiplication.
The scalar coefficients $C_n$ and $E_n$ are arbitrary except for $C_0=-1$ defining the 
quadratic term of the higher-spin field; all quadratic terms with more than two derivatives
cancel out upon integration by parts. 
Last, Lorentz generators $M$ are in the $({s_g}+1, s_g+1)$ representation; for the purpose of this action one 
may think of them only as the Clebsch-Gordan coefficients for projection 
$(s_g+1, s_g+1)\times (s_g+1, s_g+1)\rightarrow (3,1)\oplus(1,3)$.

The color-stripped three-point vertex can be easily read from the action:
\begin{align}
\label{vectorvertex}
 -i V_3^{\mu_0}{}_{a(s_g)}{}^{b(s_g)}  = & \frac{i}{2}(p_2-p_1)^{\mu_0} \sum_{n=0}^\infty 
 {\widehat C}_n^{\mu_1\nu_1,\dots,\mu_{2n}\nu_{2n}} 
  (M^{\mu_1 \nu_1}\dots M^{\mu_{2n} \nu_{2n}})  {}_{a(s_g)}{}^{b(s_g)}
 \cr
 & - \sum_{n=0}^\infty {\widehat E}_n^{\mu_1\nu_1,\dots,\mu_{2n}\nu_{2n}} 
  {\rm Sym}[M^{\mu_0\nu_0}q_{\nu_0},   M^{\mu_1 \nu_1},\dots , M^{\mu_{2n} \nu_{2n}}]{}_{a(s_g)}{}^{b(s_g)} \, . 
\end{align}
Here $p_{1}$ and $p_{2}$ are the momenta of the tensor fields with Lorentz indices $b(s_g)$ and $a(s_g)$, respectively, 
and $q=-p_1-p_2$ is the gluon momentum. 
The tensors $\widehat C_n$ and $\widehat E_n$ are symmetric under the interchange of pairs of 
$(\mu_i\nu_i)$ indices and read: 
\begin{align}
{\widehat C}_n^{\mu_1\nu_1,\dots,\mu_{2n}\nu_{2n}} 
&= \frac{C_n}{2^{2n} m^{4n}}\epsilon^{\rho_1\sigma_1\mu_1\nu_1}q_{\rho_1}p_{1\sigma_1} \dots \epsilon^{\rho_{2n}\sigma_{2n}\mu_{2n}\nu_{2n}}q_{\rho_{2n}}p_{1\sigma_{2n}} \,,
\nn \\
{\widehat E}_n^{\mu_1\nu_1,\dots,\mu_{2n}\nu_{2n}} 
&= \frac{E_n}{2^{2n} m^{4n}}\epsilon^{\rho_1\sigma_1\mu_1\nu_1}q_{\rho_1}p_{1\sigma_1} \dots \epsilon^{\rho_{2n}\sigma_{2n}\mu_{2n}\nu_{2n}}q_{\rho_{2n}}p_{1\sigma_{2n}} \,.
\end{align}
In deriving them we ignored terms proportional to the free equations of motion and 
terms that contain more powers of the gluon momentum than the ones shown. In the contraction 
of the Levi-Civita symbols, Lorentz generators and momenta we may recognize the repeated appearance of the operator $\S_0$
defined in Eq.~\eqref{S0hat}.
Up to numerical coefficients, the tensors ${\widehat C}_n$ and ${\widehat E}_n$ are proportional, so their product also is totally 
symmetric in all $(\mu_i\nu_i)$ pairs of indices.

To construct the double copy of two such vertices, one with $s_g=s_L$ and the other with $s_g=s_R$, 
we need the projection of the product $(s_L+1, s_L+1)\times (s_R+1, s_R+1)$ 
onto $(s+1, s+1)$ with $s=s_L+s_R$. Using the fact that the double-copy vertex is contracted with a polarization 
tensor in the $(s+1, s +1)$ representation, the relevant projection (denoted by the vertical bar and realized e.g.
by contracting all two-component spinor indices with identical commuting spinors) is
\begin{align}
\label{general_product}
&(M^{\mu_1 \nu_1}\dots M^{\mu_n \nu_n}) {}_{a({s_L})} {}^{b({s_L})}  \otimes (M^{\rho_1 \sigma_1}\dots M^{\rho_m \sigma_m})_{a({s_R})} {}^{b({s_R})} \Big| 
\nn \\
&\qquad\qquad\qquad\qquad\qquad\qquad
= {\cal C}(n,m, s_L, s_R)
(M^{\mu_1 \nu_1}\dots M^{\mu_n \nu_n} M^{\rho_1 \sigma_1}\dots M^{\rho_m \sigma_m})_{a({s})} {}^{b({s})} \,,
\end{align}
where 
\begin{equation}
{\cal C}(n,m, s_L, s_R) = \frac{s_L!}{(s_L-n)!}\frac{s_R!}{(s_R-m)!}\frac{(s-n-m)!}{s!} \,.
\end{equation}
Evaluating 
\begin{equation}
-i\pol(\bm s, p_2)V^{ \mu_0 \nu_0} _\text{3,DC}\pol(\bm s, p_1) =  
\left[-i\pol(\bm s_L, p_2)V^{ \mu_0 \nu_0} _\text{3,L}\pol(\bm s_L, p_1) \right]
\left[-i\pol(\bm s_R, p_2)V^{ \mu_0 \nu_0} _\text{3,R}\pol(\bm s_R, p_1) \right] \,,
\label{doublecopy3vertex}
\end{equation}
in the classical limit by using the relation between polarization tensors 
$\pol(\bm s, p) = \pol(\bm s_L, p)\otimes \pol(\bm s_R, p)|$, the identity~\eqref{SSidentity} and ignoring terms that vanish for 
on-shell gravitons, we find
\begin{align}
\label{three_point_general_2copy_amplitude}
-i\pol(\bm s, p_2)V^{ \mu_0 \nu_0} _\text{3,DC}\pol(\bm s,  p_1) =& \null p_1^{\mu_0}p_1^{\nu_0} 
\sum_{n, m=0}^\infty  {\cal C}(n,m,s_L, s_R) C_nC_m \,
\pol(\bm s, p_2) \left(\frac{q\cdot \S_0}{m}\right)^{2m+2n} \pol(\bm s, p_1)
\\
&- p_1^{\mu_0}p_1^{\nu_0} \sum_{n, m=0}^\infty   {\cal C}(n,m,s_L, s_R)  
E_nE_m \,  \pol(\bm s, p_2)\left(\frac{q\cdot \S_0}{m}\right)^{2n+2m+2}	\pol(\bm s, p_1)
\nn\\                
&- i\sum_{n, m=0}^\infty  C_nE_m \,\pol(\bm s, p_2){\rm Sym} \biggl[ \big({\cal C}(n,m,s_L, s_R) p_1^{\mu_0} M^{\nu_0\sigma_0}q_{\sigma_0} 
\nn\\                                
 & \qquad\qquad\qquad 
 + {\cal C}(m,n,s_L, s_R)
                 p_1^{\nu_0} M^{\mu_0\sigma_0}q_{\sigma_0}\big) ,
                \left(\frac{q\cdot \S_0}{m}\right)^{2m+2n}\, \biggr] \pol(\bm s, p_1) \, .
\nn
\end{align}
If $s_L \ne s_R$ the expression above contains an antisymmetric part
which is identified with a coupling with the Neveu-Schwarz $B$-field. It is not difficult to see that the
symmetric part has the same tensor structure as the
graviton-higher-spin vertex in Eq.~\eqref{Lstresstensor}; moreover,
any choice of coefficients $C_{ES^{2n}}$ and $C_{BS^{2n}}$ can be
reproduced by adjusting the coefficients $C_n$ and $E_n$.

As a special case of this general relation, we show that the double-copy of two minimal couplings of a higher-spin field with a 
vector yields the gravitational minimal coupling \eqref{ThreeVertexOffShell}, including the  quadrupole contribution. The former
may be found by truncating \eqref{vectorvertex} to terms with at most one Lorentz generator (i.e. setting to zero all $C_{n\ne 0}$ 
and $E_{n \ne 0}$):
\begin{equation}
-iV^\mu{}_{a(s_L)}{}^{\! b(s_L)}(q, p_1, p_2) = i p_1^\mu \delta_{a(s_L)}^{b(s_L)}
         - q_{\rho}(M^{\mu \rho})_{a(s_L)}{}^{b(s_L)}    \,.
\end{equation}
Constructing the double-copy vertex \eqref{doublecopy3vertex} and projecting onto the $(s+1, s+1)$ representation 
gives
\begin{align}
-iV^{\;\; \mu\nu}_{3,{\rm DC}\, a(s)}{}^{\! b(s)}(q, p_1, p_2)  
&=
-p_1^\mu p_1^\nu \delta_{a(s)}^{b(s)}
    -\frac{i}{s_L+s_R} \left( s_R \, q_\rho  p_1^{\mu} (M^{\nu \rho})_{a(s)}{}^{b(s)}+s_L \, q_\rho  p_1^{\nu} (M^{\mu \rho})_{a(s)}{}^{b(s)}\right)
    \cr
&    +  \frac{1}{2}\left(2 \frac{s_L s_R}{s (s-1)}\right) q_\rho q_\sigma (M^{(\mu| \rho}M^{|\nu) \sigma})_{a(s)}{}^{b(s)}  \, .
\label{HigherSpinVertexDoubleCopy}
\end{align}
For $s=s_L+s_R=1$ as well as for $s_L=0$ or $s_R=0$ the quadrupole term is not generated.
As in the general case, for $s_L\ne s_R$ the on-shell double-copy three-point vertex $V^{\;\; \mu\nu}_{3,{\rm DC}}$ contains an 
antisymmetric part, representing the coupling of the Neveu-Schwarz $B$-field with the higher-spin tensor.
The term linear in Lorentz generators in the symmetric part of $V^{\;\; \mu\nu}_{3,{\rm DC}}$ is universal, independent on 
$s_L$ and $s_R$; this is a reflection of the universality of gravity.
It is easy to see that the symmetric part of Eq.~\eqref{HigherSpinVertexDoubleCopy} is the on-shell value of the graviton-higher-spin 
vertex \eqref{ThreeVertexOffShell} derived from the minimal Lagrangian \eqref{Ls}, with a gravitational quadrupole coefficient given by
\begin{equation}
\H = 2 \frac{s_L s_R}{s (s-1)} \,.
\label{HConsistency}
\end{equation}
This value is the same as the one found in string theory~\cite{HigherSpinString}.  
Eq.~\eqref{HConsistency} implies that, in the classical limit, where all spins are large, $\H<1/2$. For low values 
of the spins, $\H$ can reach unity.

\section{Tree amplitudes}
\label{sec:tree}

\begin{figure}[tb]
\begin{center}
  \includegraphics[scale=.65]{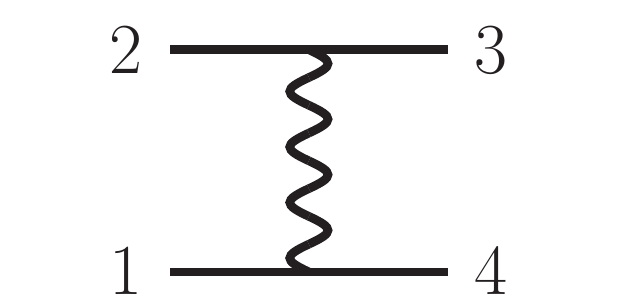}
\end{center}
\vskip -.4 cm
\caption{\small The tree-level Feynman diagram containing the $\Ord(G)$ spin interactions. 
Because we are interested only in long range interactions contact terms where the graviton 
propagator cancel can be ignored.}
\label{FourPtTreeFigure}
\end{figure}

One of our goals is to obtain new results for the terms bilinear in
spin in the two-body Hamiltonian, through $\Ord(G^2)$ and to all orders in
velocity.  The key input that we need is the one-loop 2-to-2
scattering amplitudes for spinning particles.  In turn, the
generalized unitarity method~\cite{GeneralizedUnitarity, RationalTerms,
  UnitarityReview, OPP, Forde} for constructing loop
amplitudes relies on suitable tree-amplitude building blocks.  In this
section, we describe the ones that will be required in subsequent 
sections to obtain the desired gravitationally-induced interactions of spinning
particles.
To construct these tree amplitudes we use the arbitrary-spin formalism 
set up in \sect{sec:basics}.  Here we are interested only in
terms linear in the spin of each particle, so the minimal
Lagrangian in \eqn{Ls} with $\H=0$ is sufficient; the nonminimal interactions
discussed in Sec.~\ref{sec:nonminimal} are all quadratic or of higher
order in spin and will be useful for future studies.

\begin{figure}[tb]
\begin{center}
  \includegraphics[scale=.6]{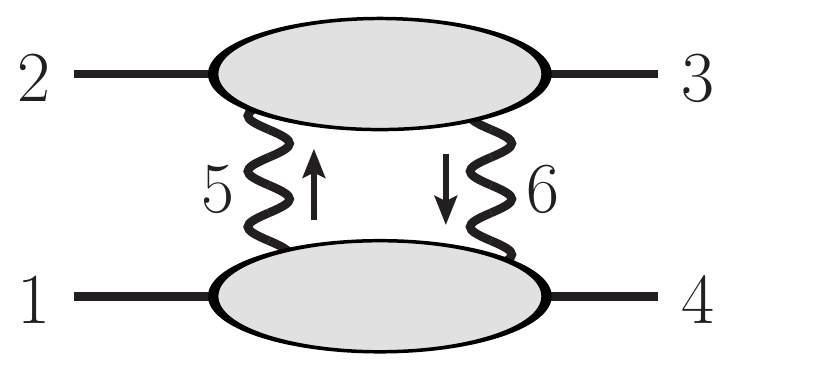}
\end{center}
\vskip -.3 cm
\caption{\small The two-particle cut needed for extracting classical
  dynamics.  The blobs represent on-shell tree amplitudes and the
  exposed lines indicate that the propagators are replaced with
  on-shell conditions.  }
\label{TwoParticleFigure}
\end{figure}

\begin{figure}[tb]
\begin{center}
  \includegraphics[scale=.6]{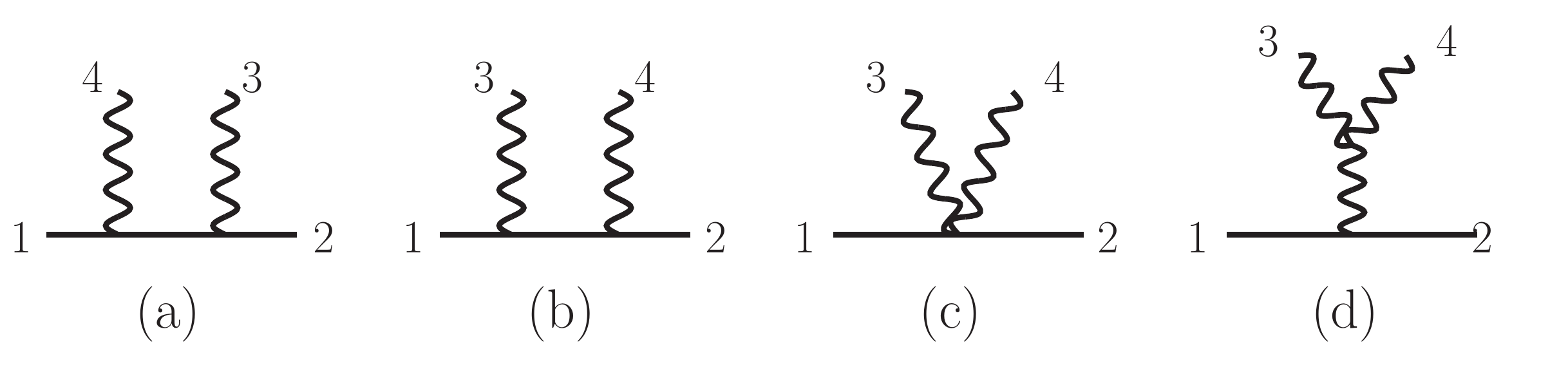}
\end{center}
\vskip -.4 cm
\caption{\small The tree-level Feynman diagram for gravitational Compton scattering.
 For integer-spin electrodynamics diagram (d) is absent. Here the internal lines represent
 Feynman propagators.}
\label{ComptonFigure}
\end{figure}

First, we obtain the tree-level amplitude with four external particles
of arbitrary spin.  The sole contributing diagram is shown in
\fig{FourPtTreeFigure}.  It gives us the necessary information to
determine the $O(G)$ two-body Hamiltonian.  Then, we proceed to obtain
the tree-level amplitude which will be used in \sect{sec:oneloop} to construct 
the one-loop amplitude that encodes the spin-orbit and spin-spin interactions at $\Ord(G^2)$.
As noted in Refs.~\cite{RothsteinClassical, BjerrumClassical, CliffIraMikhailClassical, 
2PMDiVecchia} and reviewed in Sec.~\ref{ClassicalLimit}, only a limited number of 
terms in the one-loop amplitude contribute to the long-range classical potential. 
They are captured by the unitarity cuts that separates the two matter
lines, as illustrated at one loop in \fig{TwoParticleFigure}.
Thus, to build the relevant parts of the one-loop amplitude,
we only need the tree amplitudes contributing to these cuts, 
i.e. the gravitational analog of the tree-level Compton amplitude, whose
diagrams are shown \fig{ComptonFigure}. 
In this section, we will present these tree amplitudes, constructed
using our arbitrary-spin formalism. We also comment on some of their important properties,
including generalized on-shell Ward identities and double-copy properties.
  
\subsection{Tree-level 2-to-2 scattering of spinning particles}

Consider the 2-to-2 tree-level scattering amplitude encoded in the diagram in~\fig{FourPtTreeFigure} and 
obtained by sewing together two three-point vertices~\eqref{ThreeVertexOffShell} with $\H=0$.  
We can drop any term that cancels the graviton propagator, 
since we are only interested in long range interactions.
This effectively places the
graviton on-shell with the result that each vertex is automatically
transverse and thus their contribution to the amplitude is independent
of the gauge choice.  We may therefore use the relatively simple
de~Donder gauge propagator,
\begin{equation}
P^{\mu\nu\alpha\beta}_{\textrm{\tiny{de Donder}}}(q)
=\frac{i}{q^2}
\left(\frac{1}{2}\eta^{\mu\alpha}\eta^{\nu\beta}+\frac{1}{2}\eta^{\nu\alpha}\eta^{\mu\beta}
-\frac{1}{D-2}\eta^{\mu\nu}\eta^{\alpha\beta}\right),
\label{DeDonder}
\end{equation}
since any longitudinal terms in the graviton physical-state
projector are automatically set to zero by the two vertices.

Using the vertex \eqref{ThreeVertexOffShell} and the propagator
\eqref{DeDonder} it is then straightforward to obtain the desired
tree-level amplitude from the diagram in \fig{FourPtTreeFigure}:
\begin{align}
& i\mathcal{M}^{\textrm{tree}}(1^{ s_1}, 2^{ s_2}, 3^{s_2}, 4^{s_1}) 
\nn\\
& \hskip 2 cm \null =  
{\pol_4}^{a(s)}
V{}^{\mu\nu}_\text{min} {}_{a(s)}{}^{b(s)}(q, p_1, p_4)
{\pol_1}_{b(s)}
P^{\mu\nu\alpha\beta}_{\textrm{\tiny{de Donder}}}(q)
{\pol_3}^{c(s)}
V{}^{\mu\nu}_\text{min} {}_{c(s)}{}^{d(s)}(-q, p_2, p_3)
{\pol_2}_{d(s)} \,,
\label{Formal4PtTree}
\end{align}
where $\pol_i\equiv \pol({\bm s}_i, p_i)$ and  we can drop any terms that cancel the graviton propagator.
Starting with the tree-level vertex given in \eqn{ThreeVertexOffShell} and defining
\begin{equation}
M_{ij}(a,b) \equiv \pol(p_i) M^{\mu\nu} \pol(p_j) a_\mu b_\nu
\ , \qquad
M_{ij}(e^\mu, a)\equiv \pol_i M^{\mu\nu} \pol_j a_\nu \,,
\label{defMxx}
\end{equation}
where $e^\mu$ is a unit vector signifying that the $\mu$ index is uncontracted, 
from \eqn{Formal4PtTree} we obtain,
\begin{align}
 i\mathcal{M}^{\textrm{tree}}(1^{s_1}, 2^{ s_2}, 3^{s_2}, 4^{s_1}) = &
- \frac{16\pi iG m_1^2m_2^2}{q^2}\biggl\{ 
 (2\sigma^2-1) \pol_1\cdot \pol_4 \,\pol_2\cdot \pol_3
 + \frac{2i\sigma}{m_1m_2} M_{14}(p_2, q) \pol_2\cdot \pol_3 
\nn\\
& \hskip 1 cm \null 
 - \frac{i}{m_1^2} M_{14}(p_1, q) \pol_2\cdot \pol_3 
 - \frac{2i\sigma}{m_1m_2} \pol_1\cdot \pol_4M_{23}(p_1, q) +  \frac{i}{m_2^2}   \pol_1\cdot \pol_4 M_{23}(p_2, q)
\nn\\
& \hskip 1 cm \null 
 + \frac{1}{m_1^2m_2^2} \left(     M_{14}(p_2, q) M_{23}(p_1, q) - M_{14}(p_1, q) M_{23}(p_2, q)     \right)
\cr
&\hskip 1 cm \null
 -\frac{\sigma}{m_1 m_2}  M_{14}(e^\mu, q) M_{23}(e^\nu, q)\eta_{\mu\nu}\biggr\} \,,
\label{treeS1S2minimal}
\end{align}
where we use the dimensionless kinematic variable, 
\begin{equation}
\sigma \equiv \frac{p_1\cdot p_2}{m_1 m_2} \,.
\end{equation}
In the classical limit, the products of polarization tensors and Lorentz generators \eqref{defMxx} are related to the spin tensors
of the two particles through Eqs.~\eqref{SpinEval}; thus, the tree amplitude can be expressed solely in terms of them as well 
as the products $\pol_i\cdot \pol_j$. Further using  \eqn{SpinTensorVectorRelation} and the covariant spin supplementary condition 
\eqref{CovariantSpinSupplementary}, the amplitude becomes
\begin{align}
i\mathcal{M}^{\textrm{tree}}(1^{ s_1}, 2^{ s_2}, 3^{s_2}, 4^{s_1})  = 
& -\frac{16\pi iG m_1 m_2}{q^2} \,
\pol_1\cdot \pol_4\, \pol_2\cdot \pol_3  \null \,
\biggl\{
m_1m_2(2\sigma^2-1)- 2\sigma \left(iS_2(p_1,q) - iS_1(p_2,q)\right) \nn  
\label{treeS1S2Alt}\\
& \null 
- 
\left(\frac{1}{m_1 m_2} S_1(p_2,q) S_2(p_1,q)+ \sigma S_1(e^\mu,q) S_2(e^\nu,q)  \eta_{\mu\nu}\right)
\biggr\}
+ \Ord(q, S_i^2) \,,
\end{align}
where, in close analogy with Eq.~\eqref{defMxx}, we defined
\begin{equation}
S_{i}(a,b)\equiv a_\mu b_\nu 
S^{\mu\nu}(p_{i}),\qquad 
S_{i}(e^\mu,b)\equiv b_\nu 
S^{\mu\nu}(p_{i})
\,.
\label{Stensori}
\end{equation}
The parametrization of the classical amplitude in terms of the spin tensor emphasizes its close relation to 
its complete quantum origin. As we will see in Sec.~\ref{sec:oneloop}, this persists at loop level and 
we will organize the amplitude in a similar form, which will have a structure close to that of 
the unitarity cuts.

Using Eq.~(\ref{SpinTensorVectorRelation}), and the identities for
products of Levi-Civita tensors, we can express the result in terms of the 
spin vector. During this transformation, terms with more than two powers
of the momentum transfer $q$ appearing in the bilinears in spin tensor as 
well as terms canceling the graviton propagator
are discarded. We find
\begin{align}
\label{treeS1S2}
i\mathcal{M}^{\textrm{tree}}(1^{ s_1}, 2^{ s_2}, 3^{s_2}, 4^{s_1}) = 
& -\frac{16\pi iG m_1 m_2}{q^2} \,
\pol_1\cdot \pol_4 \,\pol_2\cdot \pol_3 \, 
 \biggl\{
m_1m_2(2\sigma^2-1) 
 \\
& \null 
- 2i\sigma {} \epsilon^{\mu\nu\rho\sigma}{p_1}_\mu {p_2}_\nu q_\rho 
\left(\frac{S_1{}_\sigma}{m_1}+ \frac{S_2{}_\sigma}{m_2} \right)
+ 
(2\sigma^2-1) q\cdot S_1 q\cdot S_2
\biggr\}
+ \Ord(q, S_i^2) \, .
\nn
\end{align}
It is straightforward to further write the amplitude in terms of the rest-frame spin vectors of the two particles;
we will do this in Sec.~\ref{tree_and_loop_summary} and will be an important input in the 
construction of the EFT in Sec.~\ref{sec:EFT}.

\subsection{Tree-level gravitational Compton amplitude}
\label{sec:Comptons}

To obtain the tree-level gravitational Compton amplitude needed to
construct the one-loop four-point matter amplitude, we follow the same basic
procedure.
It is obtained by straightforwardly evaluating the four Feynman diagrams in \fig{ComptonFigure}, 
with two external arbitrary-spin particles and two gravitons.  
We need to include the contribution of the four-point vertex arising from the Lagrangian~\eqref{Ls}
and shown in \fig{ComptonFigure}(c) as it contributes, together with other contact terms 
from collapsing internal propagators, to the classical part of the loop amplitudes we construct 
from this tree amplitude.

Once we have obtained the Compton amplitude we put it in a
factorized form inspired by the KLT relations~\cite{KLT}, which
express gravitational amplitudes in terms of gauge-theory ones.  First
we note that the case of spinless external matter has a simple
factorization into a product of amplitudes in scalar
electrodynamics,
\begin{equation}
i {\mathcal M}(1^0,2^0,3^h,4^h)= -4 \pi i G \,
\frac{p_1\cdot p_3 p_1\cdot p_4}{p_3\cdot p_4} 
\bigl[A (1^0, 2^0, 3^A, 4^A)\bigr]^2 \,,
\label{ComptonGravitySpin0}
\end{equation}
where $ A (1^0, 2^{{0}},3^A, 4^A)$ is the scalar electrodynamics
Compton amplitude, the $0$ superscript indicated that the matter leg
is spinless and the $h$ and $A$ superscripts indicate the leg is a
graviton and photon, respectively. The factors of $i$ are due to our choices for
normalizing the amplitudes.
Similarly, the arbitrary-spin amplitude also
factorizes into electrodynamics amplitudes,
\begin{equation}
i {\mathcal M}(1^{ {s}},2^s,3^h,4^h)= -4 \pi i G \,
\frac{p_1\cdot p_3 \, p_1\cdot p_4}{p_3\cdot p_4} A (1^0, 2^{{0}},
3^A, 4^A) \, A(1^{{s}},2^s,3^A,4^A) \,,
\label{ComptonGravitySpinS}
\end{equation}
where the second amplitude $A(1^{{s}},2^s,3^A,4^A)$ is
an electrodynamics Compton amplitude for an arbitrary-spin particle
as indicated by the superscript $s$.
In \eqn{ComptonGravitySpinS} the graviton polarization tensor is
identified in terms of a product of two photon polarizations,
\begin{eqnarray}
\pol^{\gamma_1\gamma_2}(p_3) = \pol^{\gamma_1}(p_3)  \pol^{\gamma_2}(p_3)\,.
\end{eqnarray}
While inspired by KLT factorization, \eqn{ComptonGravitySpinS} differs
somewhat from the usual field theory KLT relation in two way: it holds
for arbitrary-spin massive particles and the factorization involves
abelian rather than nonabelian gauge-theory amplitudes.

The scalar electrodynamics amplitude $ A (1^0, 2^{{0}},3^A, 4^A)$  is derived 
from the standard Lagrangian,
\begin{equation}
\mathcal{L}_{s=0,\textrm{EM}}= - \frac{1}{4}F^{\mu \nu} F_{\mu \nu} 
+ D_\mu^{\dagger} \, \bar\phi  D^\mu \phi -m^2 \bar\phi \phi \,,
\label{LagrangianSpin0EM}
\end{equation}
where $F_{\mu \nu}$ is the usual Maxwell field strength and 
$D^\mu$ the corresponding covariant derivative.
Similarly, the arbitrary-spin electromagnetic Compton
amplitude in \eqn{ComptonGravitySpinS}, arises from the Lagrangian
with a gyromagnetic ratio, $g=2$,
\begin{eqnarray}
\mathcal{L}_{s,\textrm{EM}}= - \frac{1}{4}F^{\mu \nu} F_{\mu \nu} 
+ D_\mu^{\dagger}\bar{\phi}_s D^\mu \phi_s
- m^2 \bar{\phi}_s \phi_s + e(g-1)F_{\mu\nu} 
\bar{\phi}_s M^{\mu\nu}  \phi_s
 \,,
\label{LagrangianSpinEM}
\end{eqnarray}
where $M$ is a Lorentz generator.  We suppress the Lorentz indices of the higher-spin fields, 
as in \eqn{Ls}. Because we are coupling to a $U(1)$ gauge field we take the higher-spin field
$\phi_s$ to be complex here and \eqref{LagrangianSpinEM} is the complex version of the 
two-derivative truncation of the Lagrangian~\eqref{Lguess} used in Sec.~\ref{doublecopyonshellvertex}
to show that the minimal higher-spin-graviton vertex has a double-copy structure\footnote{Eq.~\eqref{LagrangianSpinEM} 
is also a rewriting of Eq.~\eqref{Lguess} for an $SO(2)$ gauge group and higher-spin matter fields in its vector representation.}. 
As discussed there, the $\H$ (quadrupole) term is not generated if one of the two spins vanishes, which is consistent with the 
left-hand side of Eq.~\eqref{ComptonGravitySpinS} arising from the Lagrangian \eqref{Ls} with $\H=0$. 

To present the explicit form of the amplitudes we strip them of their external polarization vectors and tensors,
\begin{equation}
A(1^{{s}},2^s,3^A,4^A) =
\pol_\alpha(p_3)
\pol_\beta(p_4)
{\pol}_s(p_1) \cdot {A^{\alpha\beta}}(1^{{s}},2^s,3^A,4^A)
\cdot {\pol_s}(p_2) \,,
\end{equation}
where as usual we suppress the higher-spin indices for 
legs 1 and 2.  The dot products refer to the contraction of these indices.
For spinless matter fields we have,
\begin{equation}
A(1^0, 2^{{0}}, 3^A, 4^A) \equiv
 \pol_\alpha(p_3)\pol_\beta(p_4) {A}^{\alpha\beta}(1^0, 2^{{0}}, 3^A, 4^A) \,.
\end{equation}
The explicit polarization-stripped amplitude is, 
\begin{align}
\nonumber {{A}^{\alpha\beta}}(1^{{s}},2^s,3^A,4^A) =&
 {A}^{\alpha\beta} (1^0, 2^{{0}}, 3^A, 4^A)  \id
\\
\nonumber &-2i\bigg\lbrace \left(
\frac{p_1^\beta}{p_1\cdot p_4}
-\frac{p_2^\beta}{p_2\cdot p_4}
\right){p_3}_\gamma i{M^{\gamma\alpha}}
+\left(
\frac{p_1^\alpha}{p_1\cdot p_3}
-\frac{p_2^\alpha}{p_2\cdot p_3}
\right){p_4}_\delta i{M^{\delta\beta}}\\
& \qquad\qquad -{p_3}_\gamma{p_4}_\delta
\left(
\frac{1}{p_1\cdot p_4}{M^{\alpha\gamma}}{M^{\beta\delta}} 
+\frac{1}{p_1\cdot p_3} {M^{\beta\delta}}{M^{\alpha\gamma}} 
\right)\bigg\rbrace,
\label{AmplitudeEMspin}
\end{align}
where we have suppressed the indices of the higher spin fields, 
$M$ is a Lorentz generator and $\id$ is the identity matrix of the $(s_1+1, s_2+1)$ representation with $s=s_1+s_2$,
$g = 2$, and we have dropped the electromagnetic coupling constant $e$.
The scalar part is,
\begin{eqnarray}
A^{\alpha\beta}(1^0, 2^{{0}}, 3^A, 4^A) = 2i
\left(
\frac{p_1^\beta p_2^\alpha}{p_1\cdot p_4}+
\frac{p_1^\alpha p_2^\beta}{p_1\cdot p_3}+
\frac{p_3^\alpha p_3^\beta+p_4^\alpha p_4^\beta}{p_3\cdot p_4}+
\eta^{\alpha\beta}
\right).
\label{AmplitudeEMscalar}
\end{eqnarray}

From \eqn{ComptonGravitySpinS} the arbitrary-spin polarization-stripped
gravitational Compton amplitude (derived from the Lagrangian \eqref{Ls} with $\H=0$) is
\begin{equation}
i \mathcal M^{\gamma_1\gamma_2, \delta_1\delta_2}(1^s, 2^s, 3^h, 4^h) 
= -4 \pi i G \, \frac{p_1\cdot p_3\, p_1\cdot p_4}{p_3\cdot p_4} \,
A^{\gamma_1\delta_1}(1^0,2^0, 3^A, 4^A) \,
 A^{\gamma_2\delta_2}(1^{{s}}, 2^s, 3^A, 4^A)\,.
\label{AmplitudeQGoffshell}
\end{equation}
Similarly, the case of spinless matter can be written as,
\begin{equation}
i \mathcal M^{\gamma_1\gamma_2, \delta_1\delta_2}(1^{0},2^{0},3^h,4^h) = 
-4 \pi i G\, \frac{p_1\cdot p_3 \, p_1\cdot p_4}{p_3\cdot p_4}\,
A^{\gamma_1\delta_1}(1^0,2^{0},3^A,4^A)  \,
   A^{\gamma_2\delta_2}(1^0,2^{0},3^A,4^A) \,.
\label{AMplitudeQGscalar}
\end{equation}
In practice, the spinless limit follows simply by setting the Lorentz generators $M$ to zero, and 
the scalar products of massive-particle polarization tensors $\pol({\bm s}, p)$ to be unity.

The KLT-inspired form of the gravitational amplitudes 
inherits useful properties directly from the photon
amplitudes.  Specifically, the spin-0 and spin-$s$ electromagnetic Compton amplitudes 
are automatically transverse on each photon leg, without the need contracting the other legs with polarization tensors.  
From \eqns{AmplitudeEMspin}{AmplitudeEMscalar} it is straightforward to verify that, for any $s$ including $s=0$,
\begin{align}
& {p_3}_\gamma A^{\gamma\delta}(1^{ {s}},2^{s}, 3^A, 4^A) = 0 \,, \hskip 1.5 cm 
  {p_4}_\delta A^{\gamma\delta}(1^{ {s}},2^{s}, 3^A, 4^A) = 0 \,,
\label{WardQED}
\end{align}
using only the antisymmetry of the Lorentz matrices and the on-shell conditions for all external momenta.  
The net effect is that, when sewing photon lines of amplitudes with this property, physical-state projectors 
are {\em not} required. 
Note that the terms depending on $p_3$ and $p_4$ in Eq.~(\ref{AmplitudeEMscalar}), which
vanish when contracted with physical gluon polarization vectors, are crucial to ensure this property.  

The gravity amplitudes stripped of polarization tensors automatically
inherit similar Ward identities via the KLT-like relation \eqref{ComptonGravitySpinS}. 
Together with \eqn{WardQED}, this implies that 
\begin{align}
& {p_3}_{\gamma_1} \mathcal M^{\gamma_1\gamma_2,\delta_1\delta_2}(1^{ s}, 2^s, 3^h, 4^h) =0 \,, \hskip 2 cm 
{p_3}_{\gamma_2} \mathcal M^{\gamma_1\gamma_2,\delta_1\delta_2}(1^{ s}, 2^s, 3^h, 4^h) =0 \,, \nn\\
& {p_4}_{\delta_1} \mathcal M^{\gamma_1\gamma_2,\delta_1\delta_2}(1^{ s}, 2^s, 3^h, 4^h) =0 \,, \hskip 2 cm 
{p_4}_{\delta_2} \mathcal M^{\gamma_1\gamma_2,\delta_1\delta_2}(1^{ s}, 2^s, 3^h, 4^h) =0 \,.
\label{TransverseProperty}
\end{align}
Polarization-stripped gravitational amplitudes constructed through standard methods will not automatically satisfy
these identities. Typically, on-shell Ward identities hold only after transversality is imposed on all other 
legs by contracting them with physical state polarization.  
The difference between such generic forms of polarization-stripped amplitudes and those obeying 
the generalized on-shell Ward identities \eqref{TransverseProperty} are terms that vanish upon contraction 
with the physical polarization tensors. 
An advantage of using amplitudes obeying the generalized form of Ward identities is that the
graviton physical state projectors used to sew tree amplitudes into loops reduces to the simple de Donder 
gauge one \eqref{DeDonder}, without requiring ghosts.  This in turn simplifies $D$-dimensional cut constructions 
of loop amplitudes, recently exploited in Refs.~\cite{3PMPRL, 2PMDiVecchia,3PMLong}.

\subsubsection{Connection to field theory KLT relations}

Although the original form of the relations was given for all string theory states, including 
the massive arbitrary-spin ones, the field theory KLT relations are usually formulated in 
terms of massless states of spin $s\le 2$ on the gravitational side. The amplitudes appearing 
in the two factors are those of nonabelian Yang-Mills theories (perhaps coupled to additional matter)
Here, \eqn{ComptonGravitySpinS} holds for a single massive particle of arbitrary spin and involves 
abelian amplitude factors. 
Thus, while there is a strong similarity between the double-copy relation
\eqref{ComptonGravitySpinS} and the celebrated KLT relations~\cite{KLT}, they are not identical. 
It is therefore worth commenting on the precise connection. 

\begin{figure}
\begin{center}
  \includegraphics[scale=.6]{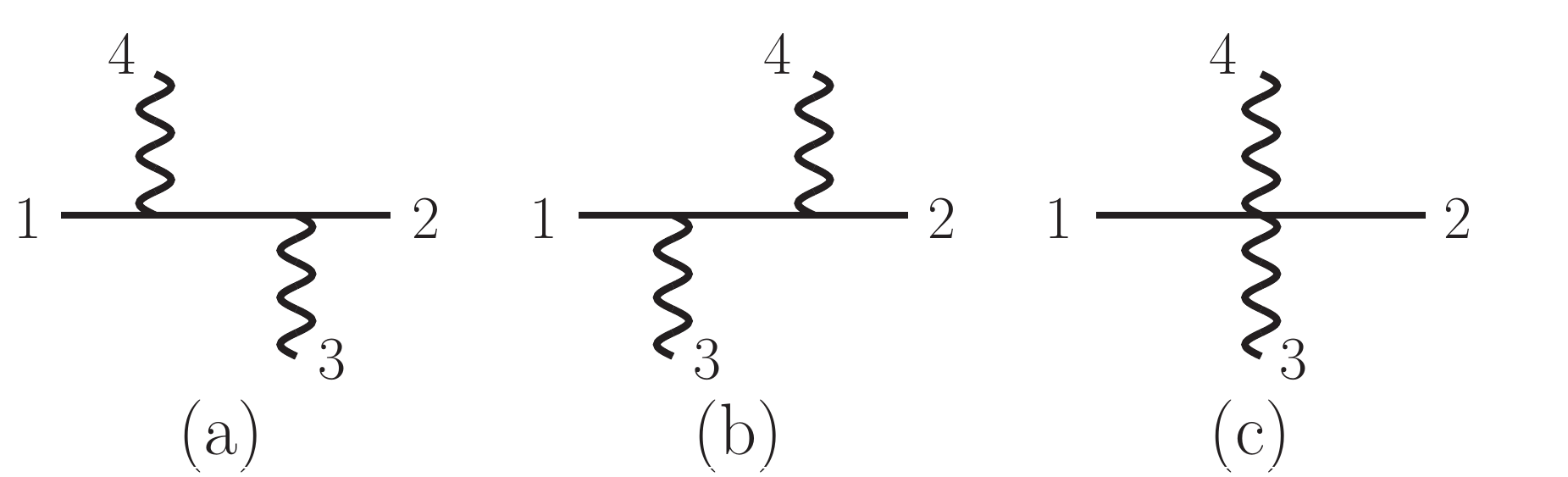}
\end{center}
\vskip -.3 cm
\caption{\small The color ordered tree-level Yang-Mills Compton Feynman diagrams, with 
ordering 1,3,2,4, where legs 1 and 2 are bosonic particles of spin $s$.}
\label{OrderYMFigure}
\end{figure}

To understand this, we will construct the gravitational Compton amplitude for massive scalars, by 
dimensionally-reducing the standard massless $D$-dimensional KLT relations to four dimensions.
This will yield the amplitude in a form that will allow us to connect its factors to the amplitudes of 
massive scalar electrodynamics that enter Eq.~\eqref{ComptonGravitySpin0}.

The $D$ dimensional four-graviton tree-level amplitude in KLT form is~\cite{KLT},
\begin{equation}
i \mathcal M(1^h,2^h,3^h,4^h)= -  16 \pi i G\, p_1 \cdot p_4 
A^{\rm YM}(1^g,2^g,3^g,4^g)A^{\rm YM}(1^g,3^g,2^g,4^g) \,.
\end{equation}
where the YM label indicates that these are color-ordered~\cite{AmplitudeReviews}
amplitudes of Yang-Mills theory.  The superscript $g$ indicates that the leg is a gluon, while the superscript 
$h$ indicates that it is a graviton. Using the four-point BCJ amplitude relation~\cite{GSW,BCJ}
between partial amplitudes,
\begin{equation}
A^{\rm YM}(1^g,2^g,3^g,4^g)=\frac{p_1\cdot p_3}{p_3\cdot p_4} A^{\rm YM}(1^g,3^g,2^g,4^g) \, ,
\end{equation}
allows us to rewrite the four-graviton amplitude as
\begin{equation}
i \mathcal M(1^h,2^h,3^h,4^h)= -16 \pi i G \, \frac{p_1 \cdot p_3\, p_1\cdot p_4}{p_3 \cdot p_4}
[A^{\rm YM}(1^g,3^g,2^g,4^g)]^2\,.
\label{KLTsquare}
\end{equation}

The components of the vector fields in the extra $(D-4)$ dimensions appear as scalars in four dimensions. 
Moreover, the components of the momentum in the extra dimensions acts as a mass for the four dimensional 
particles. Thus,  in both gauge theory amplitudes, we will choose gluons 3 and 4 to be vectors in four dimensions, 
with no momenta in the extra dimensions and particles 1 and 2 to be scalars -- i.e. vectors pointing in the extra 
dimensions. 
We will also assume that they have momentum components in the extra dimensions, so they are massive from 
a four-dimensional standpoint. (See, for example, Eqs.~(3.3) and (3.4) of Ref.~\cite{3PMLong} for more details.) 
Thus, the KLT relation for two massive scalars and two-graviton amplitude is
\begin{equation}
i \mathcal M(1^{0},2^{0},3^h,4^h) = - 16 \pi i G \,
 \frac{p_1 \cdot p_3\, p_1\cdot p_4}{p_3 \cdot p_4}
  [A^{\rm YM}(1^{0},3^g,2^{0},4^g)]^2\,.
\label{SpinlessMassiveKLT}
\end{equation}
This now of a similar form as \eqn{ComptonGravitySpin0} except that it
is in terms of nonabelian gauge-theory amplitudes instead of
electrodynamics.  This difference is inconsequential because
the color-ordered diagrams that contribute to the particular
color ordering in Eq.~\eqref{SpinlessMassiveKLT} and collected in \fig{OrderYMFigure} 
do not contain a three-gluon interaction.  
They are therefore the same as Maxwell amplitudes, after accounting for different 
normalizations and signs from reordering the diagrams. With standard  normalization 
of color generators used to define the color-order gauge-theory amplitudes in the KLT relation,
one must divide by a factor of $\sqrt{2}$ for each factor of the electric charge and
account for color ordering signs,
\begin{equation}
A^{\rm YM}(1^{0},3^g,2^{0},4^g) = -\frac{1}{2} A^{\rm EM}(1^{0},2^{0},3^A,4^A) \,,
\end{equation}
where the legs on the left side are ordered and on the right hand side
unordered.  Thus, we obtain,
\begin{equation}
i \mathcal M(1^{0},2^{0},3^h,4^h)=
-4 \pi i G \,
\frac{p_1\cdot p_3\, p_1\cdot p_4}{p_3\cdot p_4}\, 
[A^{\rm EM}(1^{0},2^{0},3^A,4^A)]^2\,,
\end{equation}
in agreement with \eqn{ComptonGravitySpinS}. 


The same discussion extends straightforwardly to the tree-level 
scattering amplitude of two spin 1 particles 
and two gravitons, offering a simple proof of \eqn{ComptonGravitySpinS} for $s=1$.
We note that, as verified by explicitly computing both the gravitational 
and electromagnetic Compton amplitudes arising respectively from the 
Lagrangians in \eqns{Ls}{LagrangianSpinEM}, 
the factorization \eqref{ComptonGravitySpinS} requires that the electromagnetic 
amplitudes include a magnetic moment coupling, as indicated in 
\eqn{LagrangianSpinEM}. This mirrors the situation for the three-point vertex 
discussed in Sec.~\ref{doublecopyonshellvertex}, where such a coupling was 
necessary to generate the complete higher-spin-graviton three-point vertex. 
It is an interesting problem to explore such relations in general, 
especially at arbitrary orders of the spin and in the presence of 
higher-dimension operators as in Eq.~\eqref{Lguess}.  They should
become important at higher orders, where they will help simplify calculations 
and expose new structures.

\section{One-loop amplitudes}
\label{sec:oneloop}

\begin{figure}
\begin{center}
\includegraphics[scale=.6]{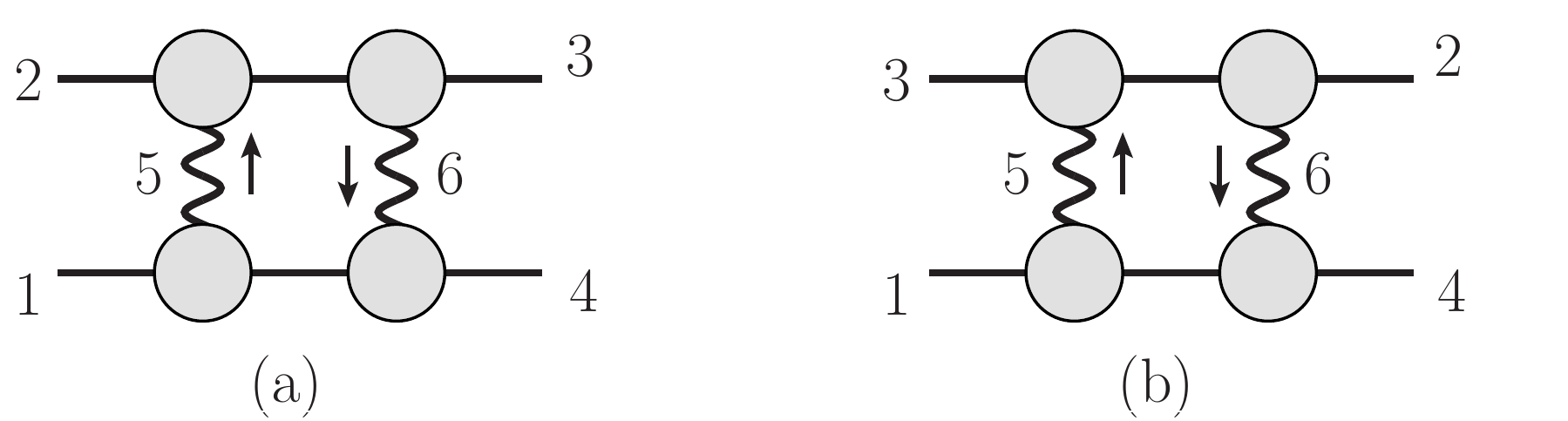}
\end{center}
\vskip -.5 cm
\caption{\small The quadruple cut from which the coefficients of the
  two box integrals in \fig{OneLoopIntegralsFigure} are extracted. All
  four external lines are placed on shell.  }
\label{QuadCutFigure}
\end{figure}

\begin{figure}
\begin{center}
\includegraphics[scale=.6]{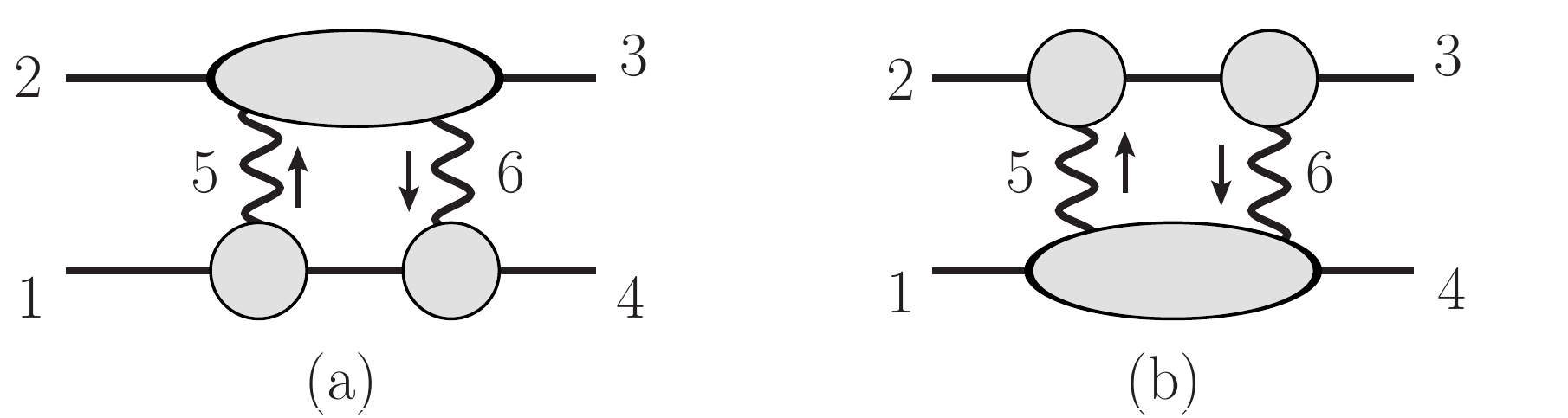}
\end{center}
\vskip -.5 cm
\caption{\small The triple cuts from which the coefficient of the
  triangle integrals are extracted.  In each case the three exposed
  lines are placed on shell.  }
\label{TripleCutFigure}
\end{figure}

Using the generalized unitarity method~\cite{GeneralizedUnitarity}, we
now construct the parts of the one-loop amplitude needed to extract
the classical interaction potential between spinning particles. As
reviewed at length in Ref.~\cite{3PMLong}, not all generalized cuts
contain useful information about the classical limit of the
amplitude. (See also Refs.~\cite{RothsteinClassical, BjerrumClassical,
  CliffIraMikhailClassical,2PMDiVecchia}.) Since we are interested
only in long-range interactions, four-point matter contact
interactions can be dropped; this implies that whenever graviton
propagators that connect the two matter lines are cancelled by
numerator factors they can be set to zero.  In the generalized unitarity language
this implies that the contributing terms must have two cut graviton legs
separating the two matter lines, as illustrated in \fig{TwoParticleFigure}.
Moreover, the fact that we are scattering classical particles requires
that each loop must contain at least one matter line; thus, the cuts
that contribute to the classical limit of the amplitude must contain
at least one cut matter line per loop.  This implies an integrand
containing the contributions that we are interested in can
be obtained from the quadruple and triple cuts in
\figs{QuadCutFigure}{TripleCutFigure}.  As we 
describe below, the quadruple-cut contributions correspond to 
iteration contributions.  The triple cuts contain
the classical pieces we wish to obtain.

\begin{figure}
\begin{center}
\includegraphics[scale=.5]{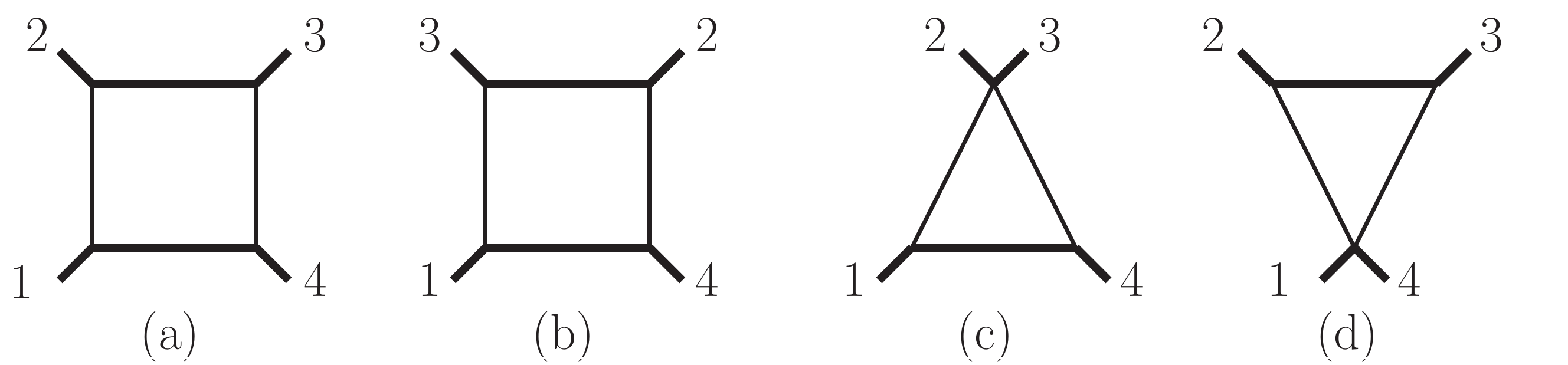}
\end{center}
\vskip -.5 cm
\caption{\small The one-loop scalar box integrals (a) and (b) that 
contribute to iterations and the triangle integrals (c) and (d) that 
contribute to $\mathcal{O} (G^2)$ terms in the potential.  }
\label{OneLoopIntegralsFigure}
\end{figure}

After an integrand is constructed from the unitarity constraints, we
apply standard integral reduction methods to express it as a linear
combination of scalar integrals. Using these methods any one-loop
amplitude can be organized into linear combination of scalar box,
triangle, bubble and tadpole integrals~\cite{PassarinoVeltman}.  We
immediately drop any bubble and tadpole integrals that result from the
reduction because they are not relevant in the classical limit,
leaving only the box and triangle integrals in
\fig{OneLoopIntegralsFigure}.  Not only do bubble and tadpoles not
have the required unitarity cuts, but a direct inspection of the
explicit values of the integrals~\cite{ExplicitIntegrals} reveals that
their dependence on the transferred momentum $q$ is inconsistent with
classical dependence while the rational prefactor cannot compensate.

Thus, the classically-relevant part of the four-point amplitude is a
linear combination of box, $I_{{\rm B}}$ and $I_{\overline{\rm B}}$,
and triangle, $I_{\bigtriangleup}$ and $I_{\bigtriangledown}$,
integrals, shown in \fig{OneLoopIntegralsFigure}(a) and~(b), and (c) and (d),
respectively:
\begin{equation}
i\mathcal{M}^\text{1 loop}_4 = d_{\rm B}\, I_{\rm B} + d_{\overline{\rm B}}\, I_{\overline{\rm B}}
+ c_{\bigtriangleup} \, I_{\bigtriangleup} + c_{\bigtriangledown}\,  I_{\bigtriangledown} \, .
\label{Org_Scalar_Integrals}
\end{equation}
The coefficients $d_{\rm B}$, $d_{\overline{\rm B}}$,
$c_{\bigtriangleup} $ and $c_{\bigtriangledown} $ are rational
functions of external momenta, polarization tensors and Lorentz
generators. In the classical limit of the latter can be converted to
spin tensors and vectors through an appropriate choice of polarization
tensors, using the relations in \sect{SpinVectorSubsection}.  The
evaluation of scalar integrals in the classical limit is
straightforward~\cite{Donoghue, BjerrumClassical,CliffIraMikhailClassical,2PMDiVecchia}. 
In any case, since
the one-loop Feynman integrals are known~\cite{ExplicitIntegrals}, we can 
also simply extract the classical limit directly from these from these.

One issue that we encounter is that because the box integrals have a
stronger-than-classical behavior, subleading terms in the relation
between polarization and spin tensors are required for a consistent
construction of the classical limit and extraction of the classical
interaction potential. Whenever this issue arises we postpone
introducing the classical spin tensors until \sect{sec:EFT}. It turns
out that simply by matching the infrared divergences between the
amplitudes of the full theory and those of the effective field theory
for spinning particles, we find that the physics is insensitive
to these subtleties at this order in $G$.

Well-developed methods for extracting the coefficient of the basics scalar 
integrals exist for extracting them~\cite{OPP,Forde} directly
from generalized unitarity cuts~\cite{GeneralizedUnitarity,
  UnitarityReview}. We use the method due to Forde~\cite{Forde}, as
extended to massive particles in
Ref.~\cite{Kilgore:2007qr}.
In this method the coefficient of scalar box integrals are computed
from quadruple cuts, i.e. cuts in which four internal propagators are
replaced with one-shell conditions, as illustrated in
\fig{QuadCutFigure}.  The coefficients of triangle integrals are then
extracted from the triple cuts, as illustrated in \fig{TripleCutFigure},
in which three internal propagators are replaced with on-shell
conditions.
While there are alternative ways to carry out the integral reduction
to the basis of scalar integrals, this method naturally maintains an
organized structure for the coefficients of the integrals even for
high-rank numerator tensors of the type that are encountered in
multi-spin interactions of higher-spin fields described in
Secs.~\ref{HigherSpinNonMinimalLagrangianSubsection} and
\ref{matchKerr}.

\subsection{Sewing tree amplitudes}

The triple and quadruple cuts, shown in \fig{TripleCutFigure} and
\fig{QuadCutFigure} respectively, that are necessary for constructing
the coefficients of the scalar integrals in
Eq.~\eqref{Org_Scalar_Integrals} can all be obtained from a regular
$t$-channel two-particle cut, shown in \fig{TwoParticleFigure}.  Given
that the required gravitational Compton amplitudes are relatively
simple, this two-particle cut is a convenient starting point for
obtaining the quadruple and triple cuts. To do so one simple replaces
additional propagators with on shell conditions.

We carry out the calculation in $D$ dimensions. Although it is
generally more efficient to use four-dimensional tree amplitudes in
the unitarity cuts, $D$-dimensional cuts make it straightforward to
implement dimensional regularization and thus identifying all infrared
singularities.  This is particularly useful at higher loops, as one
needs to ensure that no terms are missed due to subtleties or
incomplete handling dimensional regularization~\cite{3PMLong}.
At one loop it is not difficult to show that the difference between
four- and $D$-dimensional methods for construction of the integrand
amounts to certain rational terms in the
amplitude~\cite{RationalTerms} that do not have the correct scaling at
small transferred momentum to contribute to the classical limit.  In
particular, they do not have the characteristic $1/\sqrt{-q^2}$
behavior that arises from triangle integrals~\cite{Donoghue,
  BjerrumClassical}.

The two-particle cut corresponding to \fig{TwoParticleFigure} is given by
\begin{align}
C_2 & =\sum_{\lambda\lambda'}
\pol(k_5){}_{\lambda}^{\gamma_1\gamma_2}\pol(-k_6){}_{-\lambda'}^{ \delta_1\delta_2}
M_{\gamma_1\gamma_2\delta_1\delta_2} 
\bigl(4^{s_1}, 1^{ s_1}, 5^h, -6^h\bigr)\nn \\
& \hskip 1 cm \null \times 
\pol(-k_5){}_{-\lambda}^{{\gamma_1}'{\gamma_2}'}\pol(k_6){}_{\lambda'}^{{\delta_1}'{\delta_2}'}
M_{{\gamma_1}'{\gamma_2}'{\delta_1}'{\delta_2}'} 
\bigl(2^{ s_2}, 3^{s_2}, -5^h, 6^h\bigr) \,,
\label{TwoGravitonCut}
\end{align}
where the two tree amplitudes correspond to the two blobs in that figure and the sum runs over
the physical polarization of the cut gravitons. They may be expressed naturally in terms of the sum 
$\mathcal{P}^{\mu\nu}(k)$ over the physical polarizations of a vector field, as
\begin{equation}
\sum_{\lambda}\pol(k)^{\mu\nu}_\lambda\pol(-k)^{\alpha\beta}_{-\lambda}
= \frac{1}{2}\mathcal{P}^{\mu\alpha}\mathcal{P}^{\nu\beta}
+ \frac{1}{2}\mathcal{P}^{\nu\alpha}\mathcal{P}^{\mu\beta}
- \frac{1}{D-2}\mathcal{P}^{\mu\nu}\mathcal{P}^{\alpha\beta} \,,
\label{CompletenessRelationGravity}
\end{equation}
where 
\begin{equation}
\mathcal{P}^{\mu\nu}(k)=\eta^{\mu\nu}-\frac{r^\mu k^\nu+r^\nu k^\mu}{r\cdot k} \, .
\label{PhysicalStateProjector}
\end{equation}
and $r^\mu$ is an arbitrary null reference vector which should drop
out of physical expressions.

The appearance of terms dependent on this reference vector in
intermediate expressions complicates the evaluation of the generalized
unitarity cut, especially at higher loops.  Even at one loop it is
best to eliminate them as early as possible.  Because our tree-level
amplitudes satisfy, by construction, the on-shell generalized Ward
identities~\eqref{TransverseProperty}, these terms automatically drop
out from the physical-state projectors because in every such term the
graviton momentum contracts with a manifestly-transverse amplitude.
Thus the completeness relation \eqref{CompletenessRelationGravity}
reduces to the numerator of the graviton propagator in de Donder
gauge,
\begin{equation}
\sum_{\lambda}\pol(k)^{\mu\nu}_{\lambda}\pol(-k)^{\alpha\beta}_{-\lambda}
=\frac{1}{2}\eta^{\mu\alpha}\eta^{\nu\beta}+\frac{1}{2}\eta^{\nu\alpha}\eta^{\mu\beta}
-\frac{1}{D-2}\eta^{\mu\nu}\eta^{\alpha\beta} 
\equiv \mathcal{P}^{\mu\nu\alpha\beta}_{\textrm{\tiny{de Donder}}} \, .
\label{CompletenessDeDonder}
\end{equation}
A key difference between our construction and the usual de Donder
gauge Feynman diagram approach is that here, despite the appearance of
the same projector, only physical states propagate so ghosts are not
necessary to remove unphysical degrees of freedom.
Combining everything, the two-graviton cut in \eqn{TwoGravitonCut} becomes,
\begin{align}
C_{2}= M_{\gamma_1\gamma_2\delta_1\delta_2}
   \bigl(4^{s_1}, 1^{  s_1}, 5^{h}, -6^{h}\bigr)  \,
\mathcal{P}_{\textrm{\tiny{de Donder}}}^{\gamma_1\gamma_2{\gamma_1}'{\gamma_2}'}\,
\mathcal{P}_{\textrm{\tiny{de Donder}}}^{\delta_1\delta_2{\delta_1}'{\delta_2}'}\,
M_{{\gamma_1}'{\gamma_2}'{\delta_1}'{\delta_2}'} 
\bigl(2^{  s_2}, 3^{s_2}, -5^{h}, 6^{h}\bigr)
\, .
\label{TwoGraviton}
\end{align}
where the superscripts $s_1$ and $s_2$ indicates the spin of the massive particles and 
the superscript $h$ indicates that the legs are gravitons.
The sewing of tree amplitudes with the de~Donder projector~\eqref{CompletenessDeDonder} substantially simplifies the
evaluation of generalized cuts at both one- and higher loops..

The on shell conditions for the external and cut legs,
\begin{equation}
k_5^2 = 0
\, , \qquad
k_6^2\equiv (k_5+q)^2=0 \, ,
\end{equation}
alter the naive scaling in the limit of small momentum transfer, defined as $q\equiv p_2+p_3$. We use the momentum 
assignment in \fig{TripleCutFigure}.  Indeed, they imply that
\begin{equation}
p_2\cdot q=q^2/2,\qquad p_1\cdot q=-q^2/2,\qquad k_5\cdot q=-q^2/2 \, .
\end{equation}
This improved scaling further simplifies the generalized cuts in the 
classical limit and, consequently, also the box and triangle coefficients in Eq.~\eqref{Org_Scalar_Integrals}.

\subsection{Quadruple and triple cuts}

The two-graviton cut \eqref{TwoGraviton}, obtained by sewing two
tree-level gravitational Compton amplitudes, leads to a rational
function of momentum invariants and polarization tensors, whose
numerator depends explicitly on loop momenta.  
We use the entirely algebraic formalism of
Refs.~\cite{Forde,Kilgore:2007qr}, which extracts the coefficients of
box and triangle scalar integrals from quadruple and triple cuts,
respectively, shown in \fig{QuadCutFigure} and
\fig{TripleCutFigure}. They in turn are straightforwardly obtained by
imposing two and one additional cut conditions on the two-particle
cut~\eqref{TwoGraviton}, where the input tree amplitudes for capturing
the terms bilinear in spin are given in \eqn{treeS1S2minimal}.

\subsubsection{Quadruple cuts}

The quadruple cut corresponding to Fig.~\ref{QuadCutFigure}(a), which
determine the coefficient $d_{\rm B}$ in
Eq.~\eqref{Org_Scalar_Integrals}, is obtained from
Eq.~\eqref{TwoGraviton} by cutting the two matter
lines carrying momenta $p_1+k_5$ and $p_2-k_5$. It is therefore
defined as
\begin{equation}
C^{{\rm (a)}}_{4}\equiv  (-2ip_1 \cdot k_5)(2ip_2 \cdot k_5) \, C_2 \bigr|_{p_1\cdot k_5\rightarrow 0, p_2\cdot k_5\rightarrow 0} \,.
\label{QuadCuts}
\end{equation}
Because the scalar box integral is more singular in the classical
limit than the expected classical terms, it is necessary that the
classical limit be taken carefully, by keeping subleading terms in the
relation between Lorentz generators and spin tensors.  Their details
are correlated to the choice of effective field theory that we use to
construct the effective interaction Hamiltonian.  We therefore
temporarily postpone the classical limit and list here the
spin-independent and the terms linear and quadratic in Lorentz
generators.

To shorten the expressions we anticipate that each Lorentz generator
yields a factor of the spin tensor and therefore has a leading
$|q|^{-1}$ scaling in the classical limit. The same expectation allows
us to use the covariant spin supplementary condition in the form
$M_{14}(p_1, e^\mu) = M_{23}(p_2, e^\mu) = 0$. This clearly holds to
leading order in the classical limit, cf.~Eq.~\eqref{SpinEval}; we
also verify that the needed subleading terms do not spoil this
relation. Last but not least, we also take the loop
momentum to scale as $k_5^2 \sim q^2$ at small $|q|$. We will verify all
these expectations in Sec.~\eqref{GetCoefficients}, and emphasize that
they are not a necessary step in the construction of the (classical
limit of the) amplitude.  At one-loop it is simple enough to confirm 
any assumption, by direct computation.

Thus, the part of the quadruple cut is independent of Lorentz
generators is,
\begin{align}
\frac{C_4^{{\rm (a)},S_i=0}}{64 \pi^2 G^2 m_1 m_2} &= 
  4 m_1^3 m_2^3 (2\sigma^2-1)^2\pol_1\cdot \pol_4 \pol_2\cdot \pol_3 \, .
\label{Cut4S=0}
\end{align}
The momentum dependence reproduces that of the classical limit of the quadruple cuts of
one-loop four-point scattering amplitude of scalar fields~\cite{DamourTwoLoop, CachazoGuevara, BjerrumClassical, CliffIraMikhailClassical, 2PMDiVecchia}.

The part of the quadruple cut that is linear in Lorentz generators is
\begin{align}
\frac{C_4^{{\rm (a)},{\rm SO}}}{64 \pi^2 G^2 m_1 m_2} = 
&-8 i m_1^2 m_2^2 \sigma {} (2\sigma^2-1) M_{14}(p_2, q) \pol_2\cdot \pol_3
+8 i m_1^2 m_2^2 \sigma{} (2\sigma^2-1) M_{23}(p_1, q) \pol_1\cdot \pol_4
\cr
&+2 i {} \left(4 m_1^2 m_2^2 \sigma {} (2\sigma^2-1) + m_1 m_2^3 (4\sigma^2-1)\right)M_{14}(k_5, q) \pol_2\cdot \pol_3
\cr
&+2 i {} \left(4 m_1^2 m_2^2 \sigma {} (2\sigma^2-1) + m_1^3 m_2 (4\sigma^2-1)\right)M_{23}(k_5, q) \pol_1\cdot \pol_4 
\, ,
\label{Cut4SO}
\end{align}
where the notation  $M_{ij}(a, b)$ is defined in Eq.~\eqref{defMxx}. 
These terms will yield the spin-orbit terms in the coefficients of scalar box integrals.


Finally,  the part of the quadruple cut is quadratic in Lorentz generators with each matter line carrying at least one of them is 
\begin{align}
\frac{C_4^{S_1S_2}}{64 \pi^2 G^2 m_1 m_2} =&
-8 m_1 m_2 (2\sigma^2+1) M_{14}(k_5, p_2)M_{23}(k_5, p_1)
\nn \\&
+4 m_1 m_2 (2\sigma^2-1)  M_{14}(p_2, q)M_{23}(p_1, q)
\cr&  
+\left(8 m_2^2 \sigma  +4 m_1 m_2 (2\sigma^2+1) 
\right)M_{14}(k_5, q)M_{23}(k_5, p_1)
\cr&  
-\left(8 m_1^2 \sigma  +4 m_1 m_2 (2\sigma^2+1) 
\right)M_{14}(k_5, p_2)M_{23}(k_5, q)
\cr& 
-4 m_1 m_2 (2\sigma^2+1)  M_{14}(p_2, q)M_{23}(k_5, p_1)
-4 m_1 m_2 (2\sigma^2+1)  M_{14}(k_5, p_2)M_{23}(p_1, q)
\cr&  
+ 4m_1 m_2 (3\sigma^2-1) M_{14}(p_2, q)M_{23}(k_5, q)
- 4m_1 m_2 (3\sigma^2-1)  M_{14}(k_5, q)M_{23}(p_1, q)
\cr&  
+4m_1^2 m_2^2 \sigma {} (2\sigma^2-1) M_{14}(e^\mu, q)M_{23}(e^\nu, q)\eta_{\mu\nu}
\cr&
+ 8 m_1^2 m_2^2  \sigma {} (2\sigma^2-1)  M_{14}(k_5, e^\mu)M_{23}(k_5, e^\nu)\eta_{\mu\nu}
\cr&
+4 q^2 m_1 m_2 \sigma^2 M_{14}(k_5, e^\mu)M_{23}(p_1, e^\nu)\eta_{\mu\nu}
- 4 q^2 m_1 m_2 \sigma^2 M_{14}(p_2, e^\mu)M_{23}(k_5, e^\nu)\eta_{\mu\nu}
\cr&
+2 q^2 m_1 m_2 \sigma^2 M_{14}(e^\mu, q)M_{23}(p_1, e^\nu)\eta_{\mu\nu}
- 2 q^2 m_1 m_2 \sigma^2 M_{14}(p_2, e^\mu)M_{23}(e^\nu, q)\eta_{\mu\nu}
\cr&
+ 4 m_1^2 m_2^2\sigma {} (2\sigma^2-1) M_{14}(e^\mu, q)M_{23}(k_5, e^\nu)\eta_{\mu\nu}
\cr&
+4 m_1^2 m_2^2\sigma {} (2\sigma^2-1) M_{14}(k_5, e^\mu)M_{23}(e^\nu, q)\eta_{\mu\nu} 
\, ,
\label{Cut4S1S2}  
\end{align}
where $M_{ij}(e_\mu, a)$ is defined Eq.~\eqref{defMxx}.  The complete
quadruple cut corresponding to \fig{QuadCutFigure}(b) and determining
the coefficient $d_{\rm B}$ in Eq.~\eqref{Org_Scalar_Integrals} is
\begin{align}
C^{\rm {\rm (b)}}_4 = C_4^{\rm (b),S_i=0} +C_4^{\rm (b),{\rm SO}} +C_4^{{\rm (b)},S_1S_2} \,. 
\label{fullquad}
\end{align}
The quadruple cut $C^{{\rm (b)}}_4$ corresponding to \fig{QuadCutFigure}(b) and determining the coefficient 
$d_{\overline{\rm B}}$  in Eq.~\eqref{Org_Scalar_Integrals} is obtained by interchanging the external momenta 
$p_2$ and $p_3$ in Eq.~\eqref{fullquad}.


\subsubsection{Triple cuts}

The triple cuts, which will be used to determine the coefficients $c_{\bigtriangleup}$ and $c_{\bigtriangledown}$ 
in Eq.~\eqref{Org_Scalar_Integrals}, are shown in \fig{TripleCutFigure}. They may be obtained by sewing together 
one Compton and two 
three-point gravitational amplitudes or by imposing an  additional cut condition on one of the mater propagators in the 
two-particle cut \eqref{TwoGraviton}. We follow this second approach:
\begin{eqnarray}
C_{3}^{\rm {{\rm (a)}}} \equiv 2ip_2 \cdot k_5 \, C_2 \bigr|_{p_2\cdot k_5\rightarrow 0} \, ,
\hskip 1.5 cm 
C_{3}^{\rm {(b)}} \equiv  -2ip_1 \cdot k_5 \, C_2 \bigr|_{p_1\cdot k_5\rightarrow 0} \, .
\label{TripleCuts}
\end{eqnarray}
They are related by the relabeling $(m_1, m_2, p_1, p_2, p_3, p_4, k_5)\leftrightarrow (m_2, m_1, p_2, p_1, p_4, p_3, -k_5)$, so we 
need to evaluate only one of them. Each of the two cuts may be further separated into two parts related by the interchanges $p_1\leftrightarrow p_4$ and $p_2\leftrightarrow p_3$, respectively. They correspond to the symmetry of the triangle integrals
$I_{\bigtriangleup} $ and $ I_{\bigtriangledown}$. In the following we will not make explicit this separation.

Since the coefficients $c_{\bigtriangleup} $ 
and $c_{\bigtriangledown}$ which will be determined from $C_{3}^{\rm {{\rm (a)}}} $ and $C_{3}^{\rm {(b)}}$ multiply 
integrals whose leading small-$q$ scaling is classical, $I_{\bigtriangleup} \sim |q|^{-1} \sim I_{\bigtriangledown}$, it suffices to 
evaluate the triple cuts only to leading order in the classical limit. That is, we are free to use the leading order part of the relations \eqref{SpinEval} between Lorentz generators, polarization and spin tensors. 

For the spin-independent part of $C_3^{{\rm {\rm (a)}}}$ we find 
\begin{align}
\frac{C_3^{{\rm {\rm (a)}}, S_i = 0}}{64 \pi^2 G^2} = &  
   \frac{i}{q^2 P_2 (P_2 + q^2)} \pol_1 \cdot \pol_4 \pol_2 \cdot \pol_3 \Bigl\{
- 4 m_1^2 P_2^2 q^2 {} \bigl(P_2^2 + P_2 q^2 + m_2^2 (1 - 6 \sigma^2) q^2\bigr)
\nn\\
& \hskip .7 cm \null {} 
 +  P_2^4 q^4 - 8 m_1 m_2 P_2^3 \sigma q^4 
  + 16 m_1^3 m_2 P_2 \sigma q^2 {} \bigl(P_2^2 + P_2 q^2 + m_2^2 (1 - 2 \sigma^2) q^2\bigr)
\nn\\
& \hskip .7 cm \null {}
   +  2 m_1^4 \bigl( P_2^4 + 2 P_2^3 q^2 + 2 m_2^2 P_2 {} (1 - 4 \sigma^2) q^4
  + 2 m_2^4 (1 - 2 \sigma^2)^2 q^4 
\nn\\
& \hskip 1.7 cm \null {}
+ P_2^2 q^2 {} (m_2^2 (2 - 8 \sigma^2) + q^2) \bigr)
\Bigr\}
 \,.
\label{Cut3S=0}
\end{align}
where $P_2 \equiv - 2 p_2 \cdot k_5$. The presence of the two factors in the denominator, related by $p_2\leftrightarrow p_3$,
exposes the presence mentioned above of two distinct terms related by this transformation.

Expressing the result in terms of the covariant spin vector through Eq.~\eqref{SpinTensorVectorRelation}, the terms 
in $C_3^{{\rm {\rm (a)}}}$ that are linear in the covariant spin are:
\begin{align}
\frac{C_3^{{\rm {\rm (a)}}, {\rm SO}}}{64 \pi^2 G^2} = &
\frac{2\pol_1 \cdot \pol_4 \, \pol_2 \cdot \pol_3}{q^2 P_2 (P_2 + q^2)}
\Bigl\{S_{1}\left(k_5,q\right)
\nn \\ &
\big(- (q^4 (m_1^3 (m_2^3 (8\sigma^3-4\sigma )-4 m_2 P_2\sigma) + m_1^2 (-8 m_2^2 P_2\sigma^2+m_2^4 (4\sigma^2-1 )+P_2^2 )
\nn \\ &
+2 m_2 m_1 P_2\sigma {} (P_2-2 m_2^2 )+m_2^2 P_2^2 )-P_2 q^2 (4 m_2 m_1^3\sigma {} (m_2^2 (2-4\sigma^2 )+3 P_2 )
\nn \\ &
-2 m_1^2 P_2 (m_2^2 (1-10\sigma^2 )+2 P_2 )-8 m_2 m_1 P_2^2\sigma +P_2^3 )+m_1^2 P_2^3 (3 P_2-8 m_1 m_2\sigma ) )\big)
\nn \\ &
+S_{2}\left(k_5,q\right)\big(
 (q^4 (m_1^4 (m_2^2 (1-4\sigma^2 )+P_2 )+4 m_2 m_1^3\sigma {} (m_2^2 (1-2\sigma^2 )+2 P_2 )
\nn \\ &
+m_1^2 P_2 (2 m_2^2 (6\sigma^2-1 )-3 P_2 )-6 m_2 m_1 P_2^2\sigma+P_2^3 )
\nn \\ &
+m_1^2 P_2 q^2 (m_1^2 (m_2^2 (2-8\sigma^2 )+3 P_2 )+12 m_2 m_1 P_2\sigma-4 P_2^2 )+2 m_1^4 P_2^3 )\big)
\nn \\ &
+\big( q^2 (P_2-2 m_1 m_2\sigma) (q^2 (2 m_1^2 (-2 m_2^2\sigma^2+m_2^2+P_2 )+4 m_2 m_1 P_2\sigma-P_2^2 )+2 m_1^2 P_2^2 )\big)
\nn {}\\ &
\big(S_{1}\left(p_2,q\right)-S_{2}\left(p_1,q\right)\big)
\Bigr\}
\,.
\label{Cut3SO}
\end{align}
where 
\begin{equation}
S_{i}(a,b)\equiv a_\mu b_\nu 
S^{\mu\nu}(p_{i})
\,.
\end{equation}
Last, the terms in the triple cut which are linear in both $S_1$ and $S_2$ are 
\begin{align}
\frac{C_3^{{\rm {\rm (a)}}, S_1 S_2}}{64 \pi^2 G^2} = 
&-\frac{i}{4}  \frac{\pol_1 \cdot \pol_4 \, \pol_2 \cdot \pol_3}{q^2 P_2 \null (P_2 + q^2)}
\Bigl\{
-m_1^2 P_2^3 S_{1} (e^{\mu},e^{\nu}  ) S_{2} (e^{\alpha},e^{\beta}  )\eta_{\mu\alpha}\eta_{\beta\nu} q^4
\nn \\ &
-16 m_1^2 P_2^2 S_{1} (k_5,p_2  ) S_{2} (p_1,q  ) q^2
+2 m_1^2 (8\sigma m_1 m_2-3 P_2  ) P_2^2 S_{1} (q,e^{\mu}  ) S_{2} (q,e^{\nu}  )\eta_{\mu\nu} q^2
\nn \\ &
-4 m_1^2 P_2^2 S_{1} (p_2,e^{\mu}  ) S_{2} (k_5,e^{\nu}  )\eta_{\mu\nu} q^4
+8 m_1^2 P_2^2 (P_2-2\sigma m_1 m_2  ) S_{1} (k_5,e^{\mu}  ) S_{2} (q,e^{\nu}  )\eta_{\mu\nu} q^2
\nn \\ &
-16\sigma m_1^3 m_2 P_2^2 S_{1} (q,e^{\mu}  ) S_{2} (k_5,e^{\nu}  )\eta_{\mu\nu} q^2
-16 m_1^2 P_2^2 S_{1} (k_5,p_2  ) S_{2} (k_5,q  ) q^2
\nn \\ &
-4 m_1^2 (8\sigma m_1 m_2-5 P_2  ) P_2  S_{1} (p_2,q  ) S_{2} (k_5,q  ) q^2
+16 m_1^2 P_2^2 S_{1} (k_5,q  ) S_{2} (k_5,p_1  ) q^2
\nn \\ &
+4 (q^2 (8 (2\sigma^2 m_2^2+m_2^2-P_2  ) m_1^2-12\sigma m_2 P_2 m_1+3 P_2^2  )-8 m_1^2 P_2^2  ) S_{1} (k_5,p_2  ) S_{2} (k_5,p_1  ) q^2
\nn \\ &
-16 m_1^2 P_2^2 S_{1} (p_2,q  ) S_{2} (k_5,p_1  ) q^2-8 m_1^2 P_2^3 S_{1} (q,e^{\mu}  ) S_{2} (p_1,e^{\nu}  )\eta_{\mu\nu} q^2
\nn \\ &
+2 P_2 ( (4 (3 P_2-4\sigma^2 m_2^2  ) m_1^2+12\sigma m_2 P_2 m_1-3 P_2^2  ) q^2+8 m_1^2 P_2^2  ) S_{1} (k_5,e^{\mu}  ) S_{2} (p_1,e^{\nu}  )\eta_{\mu\nu} q^2
\nn \\ &
-2 P_2 ( ( ( (8-64\sigma^2  ) m_2^2+20 P_2  ) m_1^2+52\sigma m_2 P_2 m_1-11 P_2^2  ) q^2+16 m_1^2 P_2^2  ) S_{1} (k_5,q  ) S_{2} (p_1,q  )
\nn \\ &
+2 P_2 \big( (32\sigma m_2 m_1^3-8 (3 P_2-7\sigma^2 m_2^2  ) m_1^2-42\sigma m_2 P_2 m_1+8 P_2^2  ) q^2
\nn \\ &
+8 m_1^2 (4\sigma m_1 m_2-3 P_2  ) P_2  \big) S_{1} (k_5,q  ) S_{2} (k_5,q  )
\nn \\ &
+4 m_1 \big( (8\sigma m_2 ( (1-2\sigma^2  ) m_2^2+P_2  ) m_1^2+2 P_2 ( (6\sigma^2-2  ) m_2^2+P_2  ) m_1-3\sigma m_2 P_2^2  ) q^2
\nn \\ &
+8\sigma m_1^2 m_2 P_2^2  \big) S_{1} (k_5,e^{\mu}  ) S_{2} (k_5,e^{\nu}  ) \eta_{\mu\nu} q^2 \Bigr\}
\,.
\label{Cut3S1S2} 
\end{align}
The complete triple cut $C_3^{{\rm (a)}}$ in the classical limit is is
\begin{align}
C_3^{{\rm (a)}} = C_3^{{\rm (a)}, S_i=0}+C_3^{{\rm (a)}, {\rm SO}}+C_3^{{\rm (a)}, S_1S_2} \, .
\label{fulltriple}
\end{align}
The triple cut $C^{(b)}_3$, corresponding to Fig.~\ref{TripleCutFigure}(b) and determining the coefficient 
$c_{\bigtriangledown}$  in Eq.~\eqref{Org_Scalar_Integrals} is obtained by applying the transformation
 $(m_1, m_2, p_1, p_2, p_3, p_4, k_5)\rightarrow (m_2, m_1, p_2, p_1, p_4, p_3, -k_5)$ to the Eq.~\eqref{fulltriple}.

\subsection{Extracting integral coefficients \label{GetCoefficients}}

Armed with the expressions for the quadruple and triple cuts, we proceed to extracting the coefficients of the scalar box and triangle 
integrals in Eq.~\eqref{Org_Scalar_Integrals}. The construction \cite{Forde,Kilgore:2007qr} 
begins with solving the triple and quadruple cut conditions. They determine the loop momentum in terms of one free parameter while the latter, which may be obtained from the former for a special 
value of that parameter, give a discrete set of solutions. For a suitable parametrization of the loop momentum, the coefficient of triangle integrals are then obtained as the term in the evaluation of the triple cut on the solution of the cut condition that is independent of the free parameter. The coefficient of the box integrals is given by the sum over the solutions of the quadruple cut conditions of the evaluation 
of the quadruple cuts on these solutions. 

We will begin by solving the triple cut conditions in the appropriate parametrization~\cite{Forde,Kilgore:2007qr}. From here we will extract 
the loop momentum that solves the quadruple cuts and subsequently use them to extract the integral coefficients. Finally, 
we reconstruct the classical limit of the one-loop four-point amplitude of arbitrary-spin particles.

\subsubsection{The triple cuts and the coefficients of scalar triangle integrals}

Let us consider the triple cut in \fig{TripleCutFigure}(a), whose expression is found in Eqs.~\eqref{Cut3S=0}-\eqref{Cut3S1S2}.
The on-shell conditions for the cut legs are
\begin{eqnarray}
k_5^2=0,\qquad 2q\cdot k_5-q^2=0,\qquad p_1\cdot k_5=0 \, ,
\label{CutConditions}
\end{eqnarray}
where, as before, $q=p_2+p_3$. Their solution is parameterized as~\cite{Forde,Kilgore:2007qr}
\begin{eqnarray}
{k_5}(T)^\mu=x q^\mu+y p_1^\mu+T {a_1}^\mu + \frac{\alpha {a_2}^\mu}{T}\,,
\label{FordeParametrisation}
\end{eqnarray}
where $x, y$ and $\alpha$ are free parameters to be determined by \eqref{CutConditions}, $T$ 
parameterizes the component  of the loop momentum that is not fixed by the three cut conditions
and the vectors ${a_1}^\mu$ and ${a_2}^\mu$ are given by~\cite{Kilgore:2007qr}
\begin{eqnarray}
{a_1}^\mu=\left\langle\Qf|\sigma^\mu|\Pf\right]\,, \hskip 1.5 cm 
{a_2}^\mu=\left\langle\Pf|\sigma^\mu|\Qf\right] \, .
\label{a1a2definition}
\end{eqnarray}
The null momenta $\Pf$ and $\Qf$ are chosen to be
\begin{eqnarray}
\Pf^\mu\equiv p_1^\mu+\frac{m_1^2}{\gamma} q^\mu,\qquad
\Qf^\mu\equiv q^\mu+\frac{q^2}{\gamma}p_1^\mu \, ;
\label{FlattenedMomenta}
\end{eqnarray}
the parameter $\gamma$ is determined by requiring that $\Pf$ and $\Qf$ are null:
\begin{eqnarray}
\gamma= \frac{1}{2}(q^2\pm\sqrt{q^2(q^2-4m_1^2)}) \,.
\label{gammasolutions}
\end{eqnarray}
Three of these parameters are fixed by imposing the three on-shell conditions in \eqn{CutConditions}
\begin{equation}
x = -\frac{2m_1^2}{q^2-4m_1^2}\,, \hskip 1.5 cm 
y = -\frac{q^2}{q^2-4m_1^2},\hskip 1.5 cm 
\alpha=\frac{m_1^2q^2}{2(q^2-4m_1^2){a_1}\cdot{a_2}}\, .
\label{xyalpha}
\end{equation}

The construction of solutions to the on-shell conditions corresponding to the triple cut in \fig{TripleCutFigure}(b) is similar 
and may be obtained from that corresponding to  \fig{TripleCutFigure}(a) through 
\begin{equation}
m_1 \rightarrow m_2 \, , \hskip 1.5 cm
y \rightarrow -y\,, \hskip 1.5 cm
\gamma \rightarrow -\gamma \, .
\end{equation}

Evaluating the triple cut \eqref{Cut3S=0}-\eqref{Cut3S1S2} on these solutions yields rational functions of the remaining 
free parameter $T$. The singularities of these functions have different physical interpretations. 
As discussed before, Eqs.~\eqref{Cut3S=0}-\eqref{Cut3S1S2} contain propagator singularities that correspond to the contributions
of the box scalar integrals in Eq.~\eqref{Org_Scalar_Integrals} to the triple cut. They are reflected by singularities at values of $T$
solving the equations
\begin{align}
k_5(T)\cdot p_2 = 0\,, 
\qquad
\text{or}
\qquad
k_5(T)\cdot p_3 = 0 \, . 
\label{extracut}
\end{align}
Each of them is a quadratic equation for $T$ and thus has two solutions; it is not difficult to see that, away from special momentum configurations, $T$ takes some finite values. 
From the discussion above it is clear that the (sum over the) corresponding residues are closely related to the box integral coefficients,
which may indeed be extracted this way: the solutions of the first \eqref{extracut} equation lead to $d_{\rm B}$ while those of the second
 \eqref{extracut} equation lead to $d_{\overline{\rm B}}$. 
 
It has been shown in \cite{Forde,Kilgore:2007qr} 
that  the coefficient of the triangle integral is given by the $T$-independent part of the average 
 of the evaluation of the triple cut \eqref{Cut3S=0}-\eqref{Cut3S1S2} on the two solutions of the
triple-cut on-shell conditions, Eqs.~\eqref{FordeParametrisation}, \eqref{a1a2definition}, \eqref{xyalpha} and \eqref{gammasolutions}.
The relevant terms come therefore from the $T$-independent parts of of the loop momentum $k_5(T)$ as well as from terms containing
the product $a_1^\mu a_2^\nu$.  This product is given by
\begin{equation}
{a_1}^\mu {a_2}^\nu=2(\Qf^\mu \Pf^\nu+\Qf^\nu \Pf^\mu-\eta^{\mu\nu}\Qf\cdot \Pf)
-2i\epsilon^{\mu\nu\alpha\beta}{\Qf}_\alpha{\Pf}_\beta\,,
\end{equation}
which can be expressed in them of the full external momenta using \eqn{FlattenedMomenta}.

Following this procedure, the triple cut $C_3^{{\rm (a)}}$, corresponding to
\fig{TripleCutFigure}(a), yields the coefficient of the triangle
integral $I_{\bigtriangleup}$ in Eq.~\eqref{Org_Scalar_Integrals}:
\begin{align}
\label{tri_up}
c_{\bigtriangleup}  =  &
\nn -32 \pi^2 G^2 m_1^2 
\pol_1\cdot \pol_4 \pol_2\cdot \pol_3
\bigg\{
6m_1^2 m_2^2 {} (5\sigma^2-1) 
+ 
\frac{2m_1m_2 (5\sigma^2-3) \sigma} {\sigma^2-1}
\Big(3 i S_{2}(p_1,q) - 4 i S_{1}(p_2,q) \Big)
\\
& \null 
+  
\frac{2}{(\sigma^2-1)}
\Big\{3(5\sigma^2-1)S_{1}(p_2,q)S_{2}(p_1,q)
+(5\sigma^2-3)\sigma\eta_{\mu\alpha}\Big[
m_1 m_2 \big(S_{1}(e^\mu,q) S_{2}(e^\alpha,q)
\nn
\\
& \null 
-q^2 \eta_{\nu\beta}S_{1}(e^\mu,e^\nu) S_{2}(e^\alpha,e^\beta)\big)
+\frac{q^2 }{\sigma^2-1}\Big(2\sigma S_{1}(e^\mu,p_2) S_{2}(e^\alpha,p_1)
\nn
\\
& \null 
\hskip 1cm
+\frac{m_1+m_2 \sigma}{m_2}S_{1}(e^\mu,p_2) S_{2}(e^\alpha,q)-\frac{m_2+m_1 \sigma}{m_1}S_{2}(e^\mu,p_1) S_{1}(e^\alpha,q)\Big)
\Big]\Big\}  
\bigg\}  
+ \Ord(q^2 \textrm{S}_i^2)\, .
\end{align}
The coefficient $c_{\bigtriangledown}$, of the integral $I_{\bigtriangledown}$, 
can be obtained from $c_{\bigtriangleup}$ through the map 
\begin{equation}
(m_1, m_2, p_1, p_2, S_1, S_2, q, \sigma)\rightarrow
 (m_2, m_1, p_2, p_1, S_2, S_1, -q, \sigma) \,.
\end{equation}

\subsubsection{The quadruple cuts and the coefficients of scalar box integrals}

The coefficients of scalar box integrals are given by the average over the values of the quadruple cut on the 
solutions of the quadruple cut conditions~\cite{GeneralizedUnitarity, Forde,Kilgore:2007qr}.  As discussed in the previous subsection,
these solutions may be obtained from those of the triple cuts, Eqs.~\eqref{FordeParametrisation}, \eqref{a1a2definition}, 
\eqref{xyalpha} and \eqref{gammasolutions}, by further demanding that the additional propagator of the 
desired box diagram is on shell. We may, alternatively, start with a parametrization of the loop 
momentum which is slightly more convenient for this purpose,
\begin{eqnarray}
k_5^\mu=\alpha p_1^{\mu}+\beta p_2^\mu+\gamma q^\mu+\delta \eta^{\mu} \, ,
\end{eqnarray}
where $\alpha$, $\beta$, $\gamma$ and $\delta$ are free, complex parameters, and $\eta$ 
is a null reference vector, whose precise value should not affect the final answer. By choosing $\eta$ to be orthogonal 
to $p_1$ and $p_2$, we find the two solutions for the loop momentum
\begin{equation}
k_5^\mu=\frac{q^2}{2q\cdot \eta} \eta^\mu,\qquad\qquad
k_5^\mu=
\frac{N_{\alpha}p_1^\mu+N_{\beta} p_2^\mu+N_{\gamma} q^\mu}{{\cal N}} 
-\frac{q^2}{2q\cdot \eta} \eta^\mu \, ,
\label{QuadCutSolution}
\end{equation}
where
\begin{align}
N_{\alpha}&=-2m_2(m_2+m_1\sigma)q^2,\qquad
N_{\beta}=2m_1(m_1+m_2\sigma)q^2,\\
N_{\gamma}&=4m_1^2m_2^2(\sigma^2-1),
\qquad\quad\ \
{\cal N}=N_\gamma+(N_\beta-N_\alpha)/2 \, .
\end{align}

While it is possible to keep $\eta$ arbitrary (up to its properties stated above) and have it drop out of the final expressions
for $d_{\rm B}$, it is more convenient to use an explicit form that manifests its properties. We choose it to be
\begin{align}
\eta^\mu&=\left\langle p_1^{\flat}|\sigma^\mu |p_2{^\flat}\right] \, ,
\\
p_1^{\flat}&=p_1+m_1^2 \zeta p_2 \,,\qquad
p_2^{\flat}=p_2+m_2^2\zeta p_1 \,, \qquad
\zeta =-\frac{\sigma \pm \sqrt{\sigma^2-1}}{m_1m_2} \,.
\nonumber
\end{align}
The two values of $\zeta$ are determined by demanding that $p_1^{\flat}$ and $p_2^{\flat}$ are null. Both are necessary
for determining the coefficient of the box integral. 
To express the dependence on $\eta$ in terms of $p_1$ and $p_2$ it is useful to multiply and divide the $\eta$-dependent terms in 
Eq.~\eqref{QuadCutSolution} by $q\cdot {\bar \eta}$, 
\begin{eqnarray}
\frac{\eta^\mu}{q\cdot \eta}=\frac{q\cdot \bar{\eta}\eta^\mu}{q\cdot \bar{\eta}q\cdot \eta } \,,
\end{eqnarray}
and use the identity
\begin{eqnarray}
\eta^{\mu}\bar{\eta}^{\nu}=2\left(
{p_1{^\flat}}^{\mu}{p_2{^\flat}}^{\nu}+
{p_2{^\flat}}^{\mu}{p_1{^\flat}}^{\nu}-
\eta^{\mu\nu}{p_1{^\flat}}\cdot {p_2{^\flat}}
\right)-2i\epsilon^{\mu\nu\alpha\beta}{p_1{^\flat}}_{\alpha}{p_2{^\flat}}_{\beta} \,.
\end{eqnarray}

Using this procedure on the quadruple cut Eqs.~\eqref{fullquad}, we find that the coefficient $d_{\rm B}$ of the 
box integral $I_{\rm B}$ is given by
\begin{align}
d_{\rm B}=&
64G^2m_1m_2 \pi^2 \Big\{4m_1^3m_2^3(2\sigma^2-1)^2 \pol_1\cdot\pol_4 \pol_2\cdot\pol_3 \nn
\\
&+8 i {} m_1^2m_2^2\sigma {} (2\sigma^2-1)( 
      \pol_1\cdot\pol_4  M_{23}(p_1, q)  
 -  M_{14}(p_2, q)\pol_2\cdot\pol_3)
\cr
+&4m_1 m_2 (2\sigma^2-1) M_{14}(p_2, q)M_{23}(p_1, q)
\label{BoxCoefficientCovariant}\\
+&4m_1^2 m_2^2 \sigma {} (2\sigma^2-1)  \eta_{\mu\nu}M_{14}(e^\mu, q) M_{23}(e^\nu, q)
-2q^2m_1^2 m_2^2 \sigma {} (2\sigma^2-1) \eta_{\mu\nu}\eta_{\rho\sigma} M_{14}(e^\mu, e^\rho) M_{23}(e^\nu, e^\sigma)
\cr
+&\frac{2q^2 m_1 m_2} {\sigma^2-1}(4\sigma^4 -2\sigma^2-1)
\eta_{\mu\nu}M_{14}(e^\mu, p_2)M_{23}(e^\nu, p_1)
\nn \\
-&\frac{q^2 m_2} {\sigma^2-1}\left((4\sigma^4 -2\sigma^2-1)m_1 + \sigma {} (4\sigma^2-3) m_2\right) 
\eta_{\mu\nu}M_{14}(e^\mu, q)M_{23}(e^\nu, p_1)
\cr
+&\frac{q^2 m_1} {\sigma^2-1}\left((4\sigma^4 -2\sigma^2-1)m_2 + \sigma {} (4\sigma^2-3) m_1\right) 
\eta_{\mu\nu} M_{14}(e^\mu, p_2) M_{23}(e^\nu, q) \Big\} 
+{\cal O}(q^3) \,,
\nonumber
\end{align}
where $M_{ij}(a,b)$ and their counterparts with free indices are defined in Eqs.~\eqref{defMxx}.
As in the case of the quadruple cut, we kept intact the dependence on Lorentz generators and polarization tensors, 
anticipating that comparison with the effective field theory will require a careful choice of the subleading in the classical limit.

The coefficient $d_{\overline{\rm B}}$ of the second crossed-box integral $I_{\overline{\rm B}}$ is obtained from $d_{\rm B}$ 
above by interchanging $p_2$ and $p_3$.

\subsection{The one-loop amplitude in the classical limit}

We can now reconstruct the classical part of the four-point amplitude~\eqref{Org_Scalar_Integrals}. 
As we will see in detail in the next section and used in earlier literature~\cite{Donoghue, RothsteinClassical, BjerrumClassical, Vaidya,  CachazoGuevara, Spin2PM,  CliffIraMikhailClassical}, 
the new physical information in this amplitude arises from triangle integrals. It is therefore convenient to 
collect their contribution in $\mathcal{M}_{\bigtriangleup+\bigtriangledown}$ defined as
\begin{align}
i\mathcal{M}^\text{1 loop}_4 &= d_{\rm B} I_{\rm B} + d_{\overline{\rm B}} I_{\overline{\rm B}} +  
c_{\bigtriangleup}I_{\bigtriangleup}+c_{\bigtriangledown}I_{\bigtriangledown} \,.
\\
&\equiv d_{\rm B} I_{\rm B} + d_{\overline{\rm B}} I_{\overline{\rm B}} + \mathcal{M}_{\bigtriangleup+\bigtriangledown} \, .
\label{Full_Amp}
\end{align}

The two triangle integrals are related by interchanging the masses $m_1$ and $m_2$ and are well-known; in an 
expansion around the classical limit they are~\cite{Donoghue, BjerrumClassical, CliffIraMikhailClassical, 2PMDiVecchia}
\begin{eqnarray}
I_{\bigtriangleup} = \int \frac{d^4l}{(2 \pi)^2} \frac{1}{(l^2+i\epsilon)((l+q)^2+i\epsilon) ((l+p_1)^2-m_1^2+i\epsilon)} = -\frac{i}{32m_1} \frac{1}{\sqrt{-q^2}} + \cdots\,, \nn \\
I_{\bigtriangledown}=  \int \frac{d^4l}{(2 \pi)^2} \frac{1}{(l^2+i\epsilon)((l+q)^2+i\epsilon) ((l-p_2)^2-m_2^2+i\epsilon)}  = -\frac{i}{32m_2} \frac{1}{\sqrt{-q^2}} + \cdots\, .
%
\label{eq:triangleIntegrals}
\end{eqnarray}
Thus, together with their coefficients \eqref{tri_up}, their contribution to the amplitude, is
\begin{align}
i\mathcal{M}_{\bigtriangleup+\bigtriangledown}  =  
&\frac{2 \pi^2 i G^2 m_1 m_2}{\sqrt{-q^2}} 
\pol_1\cdot \pol_4 \pol_2\cdot \pol_3 
\bigg\{ 3m_1 m_2 {} (m_1+m_2) (5\sigma^2-1)\nn  \\
& \null  
+ 
\frac{(5\sigma^2-3) \sigma} {\sigma^2-1}
\Big((3m_1+4m_2) i {} S_2(p_1,q) - (3m_2+4m_1) i {} S_1(p_2,q) \Big)
\label{OneloopTriAmplLeviCivita}\\
\nn
& \null 
+\frac{(m_1+m_2)}{(\sigma^2-1)m_1 m_2}
\Big\{3(5\sigma^2-1)S_{1}(p_2,q)S_{2}(p_1,q)
+(5\sigma^2-3)\sigma\eta_{\mu\alpha}\Big[
m_1 m_2 \big(S_{1}(e^\mu,q) S_{2}(e^\alpha,q)
\nn
\\
& \null 
-q^2 \eta_{\nu\beta}S_{1}(e^\mu,e^\nu) S_{2}(e^\alpha,e^\beta)\big)
+\frac{q^2 }{\sigma^2-1}\Big(2\sigma S_{1}(e^\mu,p_2) S_{2}(e^\alpha,p_1)
\nn
\\
& \null 
\hskip 1cm
+\frac{m_1+m_2 \sigma}{m_2}S_{1}(e^\mu,p_2) S_{2}(e^\alpha,q)-\frac{m_2+m_1 \sigma}{m_1}S_{2}(e^\mu,p_1) S_{1}(e^\alpha,q)\Big)
\Big]\Big\}  
\bigg\} + \Ord(q^2 \textrm{S}_i^2)\, .
\nonumber
\end{align}
Using identities of the Levi-Civita tensor, this can be expressed in terms of the covariant spin vector:
\begin{align}
i\mathcal{M}_{\bigtriangleup+\bigtriangledown}  =  
\frac{ \pi^2 i G^2 m_1 m_2}{\sqrt{-q^2}} &
\pol_1\cdot \pol_4 \pol_2\cdot \pol_3 
\bigg\{ 
6m_1 m_2 {} (m_1+m_2) (5\sigma^2-1) \nn  \\
& \null 
+ 
\frac{2i (5\sigma^2-3) \sigma} {\sigma^2-1}\epsilon^{\mu\nu\rho\sigma}p_1{}_\mu p_2{}_\nu q_\rho
\Big((3m_1+4m_2) {} \frac{S_2{}_\sigma}{m_2} + (3m_2+4m_1) \ {} \frac{S_1{}_\sigma}{m_1} \Big)
\label{OneloopTriAmpl}\\
\nn
& \null 
+  
\frac{2(m_1+m_2) (20\sigma^4-21\sigma^2+3)}{\sigma^2-1} (q\cdot S_1 q\cdot S_2 - q^2 S_1 \cdot S_2)\\
&\qquad\qquad
+\frac{ 8 q^2\sigma^3 (m_1+m_2) (5\sigma^2-4)p_1\cdot S_2 p_2 \cdot S_1 } {m_1 m_2 (\sigma^2-1)^2}
\bigg\}  
+ \Ord(q^2 \textrm{S}_i^2)\,.
\nonumber
\end{align}
The terms containing both $S_1$ and $S_2$ are now expressed as scalar products; this property will be useful in \sect{sec:EFT} for systematically organizing the interactions of spinning particles. The box integrals $I_{\rm B}$ and $I_{\overline{\rm B}}$ expanded in the classical limit are also well-know~\cite{Iwasaki:1971vb, 
BjerrumClassical, 2PMDiVecchia}; we do not list explicitly their 
contribution to the complete amplitude~\eqref{Full_Amp} because, on the one hand, it will turn out to be physically unimportant 
and on the other it can be easily reconstructed given the $d_{\rm B}$ integral coefficient in Eq.~\eqref{BoxCoefficientCovariant}.
Later in this section we compare Eq.~\eqref{Full_Amp} with existing results in the 
literature~\cite{DamgaardSpinAmplitude, HolsteinRoss, Spin2PM_1} .

The complete classical amplitude~\eqref{Full_Amp} agrees with the spin-1/2 amplitude constructed 
in~\cite{DamgaardSpinAmplitude} after changing the orientation of external momenta in the latter to match ours 
and upon making the replacements
\begin{align}
\pol_i\cdot\pol_j\rightarrow 1 \,,
\qquad
\pol_1M^{\mu\nu}\pol_4=S^{\mu\nu}(p_1) \,,
\qquad
\pol_2M^{\mu\nu}\pol_3=S^{\mu\nu}(p_2) \,.
\end{align}
The former can be understood as a choice of normalization of the amplitude and the latter two are consistent with the leading term
in our second Eq.~\eqref{SpinEval} and are a consequence of the effective spinors used in~\cite{DamgaardSpinAmplitude}.

Refs.~\cite{HolsteinRoss} and \cite{Spin2PM_1} present the amplitude as different expansions around the zero-momentum limit:
in the former it is an expansion in the spatial momenta of external particles while in the latter it is an expansion in $(\sigma-1)$.
Accounting for the nonrelativistic normalization of~\cite{HolsteinRoss} it is not difficult to see that the small momentum expansion 
of our expression for $\mathcal{M}_{\bigtriangleup+\bigtriangledown} $ recovers the terms listed in eq.~(94) of that reference 
and the expansion in $(\sigma-1)$ of $\mathcal{M}_{\bigtriangleup+\bigtriangledown} $ recovers the terms listed in eqs.~$(7.11)$, 
$(7.13)$ and $(7.18)$ of reference of Ref.~\cite{Spin2PM_1}.

\subsection{Tree and one-loop summary}
\label{tree_and_loop_summary}

To facilitate the extraction of the two-body effective Hamiltonian in
the next section, we now summarize the one-loop and tree-level
four-higher spin amplitudes for obtained in this section and
Sec.~\ref{sec:tree}, respectively.
For this purpose we normalize the amplitudes nonrelativistically, by
dividing by a factor of $4 E_1 E_2$, and manifest the dependence on
the rest-frame spin that arises from presence of Lorentz generators in
vertices.
We do the latter in two steps: we will first expose the rest-frame
spins coming from the dependence on the covariant spin vectors in the
amplitude, but not that coming from the product of polarization
tensors. The coefficients of the various spin-dependent monomials in
the resulting expressions are decorated with a subscript ``cov'',
which emphasizes their covariant origin.
We subsequently extract the remaining spin dependence in the product of polarization tensors. 
The reason for this stepwise treatment is that the coefficients of the
various spin-dependent monomials in the final expressions are simple
linear combinations of the ``covariant'' coefficients, with additional
energy- and mass-dependent factors arising from
Eqs.~\eqref{SpinEval} and~\eqref{full_epsdoteps}.

To further facilitate comparison with EFT calculations in the next section, 
we specialize the expressions of the amplitudes and their associated particle
spins to the center-of-mass frame, defined as (recall that all momenta are outgoing)
\begin{align}
p_1=-(E_1, \bm p)
\, ,\hskip 1.5 cm 
p_2=-(E_2, -\bm p)
\, , \hskip 1.5 cm 
q=(0, \bm q)
\, , \hskip 1.5 cm 
\bm p\cdot \bm q = \frac{\bm q^2}{2}\,.
\label{COMdef}
\end{align}
Eq.~\eqref{eq:spin_covToNR_map} then gives the relation between the
covariant spin vectors and the corresponding the rest-frame ones:
\begin{align}
S_1^\mu = \left(\frac{\bm p \cdot \bm S_1}{m_1}, \bm S_1 + \frac{\bm p \cdot \bm S_1}{(E_1+m_1) m_1}\bm p \right),\hskip 1.5 cm 
S_2^\mu = \left(-\frac{\bm p \cdot \bm S_2}{m_2}, \bm S_2 + \frac{\bm p \cdot \bm S_2}{(E_2+m_2) m_2}\bm p \right) .
\label{eq:spin_boost}
\end{align}

The two covariant spin-dependent factors in the tree-level amplitude
Eq.~\eqref{treeS1S2} are $q\cdot S_i $ and ${p_1}_\mu {p_2}_\nu q_\rho
{S_i}_\sigma$. Using Eqs.~\eqref{eq:spin_boost}, it is straightforward to find
that
\begin{align}
q\cdot S_i = \bm q\cdot \bm S_i- \frac{\bm q^2 \bm p\cdot \bm S_i}{2m_i(E_i+m_i)}
\, , \hskip 1.5 cm 
\epsilon^{\mu\nu\rho\sigma}
{p_1}_\mu {p_2}_\nu q_\rho {S_i}_\sigma = E\, (\bm p\times\bm q)\cdot \bm S_i \, ,
\label{VdotS}
\end{align}
where $E=E_1+E_2$ is the total energy of the incoming particles and we
neglected terms that are of a higher-order in $\bm q$, which appear
because of the fourth relation in \eqref{COMdef}. While such terms
are important at loop level, they can be safely ignored at tree level
because they cancel the graviton propagator and thus cannot contribute
to the long-range potential.

With these preliminaries, the tree-level \eqref{treeS1S2} with a
nonrelativistic normalization, can be written as
\begin{align}
\frac{\mathcal{M}_4^\text{tree}}{4E_1E_2} & =
\frac{4 \pi  G }{\bm q^2} \pol_4\cdot\pol_1 \pol_3\cdot \pol_2
\bigg[
a^{(0)}_{\text{cov},1} 
+i\sum_{j=1}^2 a^{(1,j)}_{\text{cov},1}(\bm p\times\bm q)\cdot \bm S_j 
+ a^{(2,1)}_{\text{cov},1} \bm q\cdot \bm S_{1} \, \bm q \cdot \bm S_{2}
\bigg]
\, .
\label{eq:M_1PM_full}
\end{align}
The ``cov'' subscript decorating the coefficients reflects the fact
that they originate from terms with covariant dependence on the spin
vector; the second subscript `1' reflects that the are tree-level
coefficients. The first superscript in the $a$ coefficients represents the number of spin
vectors multiplying this coefficient while the second superscript
denotes the spin-dependent monomials with the given number of spins. For
monomials linear in spin we identify it with the spin label. While
here we encounter a single two-spin monomial, we chose to nevertheless
index it in anticipation of the fact that more monomials will appear
in the one-loop amplitude.
The explicit expressions for the coefficients can be easily read off from the amplitude \eqref{treeS1S2}; accounting for 
the nonrelativistic normalization, they are
\begin{align}
a_{\textrm{cov},1}^{(0)} =
-\frac{m^2\nu^2 }{\xi \gamma^2}(1-2\sigma^2)
\,,\quad
a_{\textrm{cov},1}^{(1,i)} = 
\frac{\nu}{\xi \gamma^2}\frac{2\sigma E}{m_i}
\,,\quad
a_{\textrm{cov},1}^{(2,1)} =
-\frac{\nu}{\xi \gamma^2}(1-2\sigma^2)
\, ,
\label{eq:1PMCoefficients}
\end{align}
where we used the variables
\begin{align}
m=m_1+m_2
\ , \quad
E=E_1+E_2
\ , \quad
\gamma = \frac{E}{m}
\ , \quad
\nu = \frac{m_1m_2}{m^2}
\ , \quad
\xi = \frac{E_1E_2}{E^2} \ .
\label{var_defs}
\end{align}
It is
interesting to note that a second two-spin monomial appears, $\bm q^2
\bm S_1\cdot \bm S_2$, with coefficient equal to
$a_{\textrm{cov},1}^{(2,2)}=-a_{\textrm{cov},1}^{(2,1)} $. Because of
the factor of $\bm q^2$ however, it cannot contribute to a long-range
interaction at tree level, so it is dropped. This structure will
appear again at one loop, where the factor of $\bm q^2$ is no longer 
implies that this monomial can be dropped.

The one-loop amplitude is given in Eqs.~\eqref{Full_Amp},
\eqref{BoxCoefficientCovariant} and \eqref{OneloopTriAmpl}; it
contains two additional covariant spin-dependent monomials, $p_2\cdot
S_1$ and $p_1\cdot S_2$ apart from those already appearing at tree
level.  They can be easily expressed in terms of the rest-frame spin
vectors,
\begin{align}
p_2\cdot S_1 = -\frac{E}{m_1}\bm p\cdot \bm S_1
\,, \qquad
p_1\cdot S_2 = +\frac{E}{m_2}\bm p\cdot \bm S_2 \,,
\label{new_PdotS}
\end{align}
using Eq.~\eqref{eq:spin_boost}.  Together with Eqs.~\eqref{VdotS}
they can be used to write the one-loop amplitude as
\begin{align}
\frac{\mathcal{M}_4^\text{1 loop}}{4E_1E_2} & =
\frac{2\pi^2  G^2 }{|\bm q |} \pol_4\cdot\pol_1 \pol_3\cdot \pol_2
\bigg[
a^{(0)}_{\text{cov},2} 
+i\sum_{j=1}^2 a^{(1,j)}_{\text{cov},2}(\bm p\times\bm q)\cdot \bm S_j 
 \nn \\
& \hspace{1.5cm}
+ a^{(2,1)}_{\text{cov},2}  \bm q\cdot \bm S_{1} \, \bm q \cdot \bm S_{2}
+ a^{(2,2)}_{\text{cov},2}  \bm q^2 \, \bm S_1\cdot \bm S_2 
+ a^{(2,3)}_{\text{cov},2} \bm q^2 \, \bm p\cdot \bm S_1 \, \bm p\cdot \bm S_2
\bigg] \nn \\
& \hskip 1 cm \null 
-
i a_{\rm B} I_{\rm B} -i a_{\bar{\rm B}} I_{\bar{\rm B}}\, .
\label{eq:M_2PM_full}
\end{align}
For the terms on the first two lines we used the same labeling scheme for the coefficients as at tree level. The two subscripts indicate
that the coefficients originate from covariant dependence on the spin vectors and that they are appear at one loop, respectively. 
The new superscripts compared to those already appearing for tree-level coefficients indicate that they multiply bilinears in spin
which are labeled as $2$ and $3$, continuing the list of monomials bilinears in spin started at tree level.
They are:
\begin{align}
a^{(0)}_{\textrm{cov}, 2} & =
\frac{3\nu^2 m^3}{4 \xi \gamma^2} (5\sigma^2-1) 
 \,,&
\nn \\[0.5ex]
a^{(1,1)}_{\textrm{cov},2} & 
= \frac{\nu}{4\xi \gamma^2} \frac{\sigma(5\sigma^2-3)}{\sigma^2-1} \frac{(4m_1+3m_2)}{m_1} E 
\,,&
a^{(1,2)}_{\textrm{cov},2} & = 
\frac{\nu}{4\xi\gamma^2} \frac{\sigma(5\sigma^2-3)}{\sigma^2-1} \frac{(3m_1+4m_2)}{m_2} E
\,,\nn \\[0.5ex] 
a^{(2,1)}_{\textrm{cov},2} & = 
-a^{(2,2)}_{\textrm{cov},2}= 
\frac{m \nu }{4 \xi \gamma^2}\left(
\frac{20\sigma^4-21\sigma^2+3}{\sigma^2-1} \right) 
\,,&
a^{(2,3)}_{\textrm{cov},2} & =
\frac{m \nu}{4\xi \gamma^2} \left(\frac{20\sigma^3-15\sigma^2-6\sigma+3}{(\sigma-1)\bm p^2}\right)
\,.
\label{eq:2PMCoefficients}
\end{align}
We point out here the appearance, as in the tree-level amplitude, of
the monomial $\bm q^2 \bm S_1\cdot \bm S_2$ with coefficient
$a^{(2,2)}_{\textrm{cov},2} = -a^{(2,1)}_{\textrm{cov},2}$. It would
be interesting to understand whether this equality persists to higher
orders in perturbation theory.
The box integral coefficients,
\begin{equation}
a_{\rm B} = \frac{d_{\rm B}}{4 E^2 \gamma_1\gamma_2} \,, \hskip 2cm 
a_{\bar{ \rm B}} = \frac{d_{\bar{\rm B}}}{4 E^2 \gamma_1\gamma_2} \,,
\label{box_coefs_norm}
\end{equation}
where $\gamma_i = E_i/m_i$ is the usual Lorentz factor of particle $i$.
The corresponding box and cross-box integrals, $I_{\rm B} $ and
$I_{\bar{\rm B}}$, are given in Ref.~\cite{Beenakker:1988jr}
and were evaluated in the classical limit in
Ref.~\cite{CliffIraMikhailClassical}.  In \sect{sec:EFT} we we follow
the same integration scheme where the integrals are,
\begin{align}
I_{\rm B} &= \frac{i}{2E}\int \frac{d^{D-1} \bm \ell }{(2\pi)^{D-1}} \, {1 \over \bm \ell^2 ( \bm \ell + \bm q)^2 (\bm \ell^2 + 2 \bm p \bm \ell) }\, ,
\cr
I_{\bar{\rm B}} &= 0 \, .
\end{align} 
The box integral has a stronger-than-classical scaling in the
classical limit.  Thus, in taking the classical limit of the box
coefficient $d_{\rm B}$, first subleading terms should also be kept.
As we will discuss in the next section, they have no physical effects
and we will choose them such that the EFT we construct there
corresponds to the complete theory used to compute the amplitudes
summarized here. For this reason we did not express the box (and
cross-box) coefficients in terms of the rest-frame spin vectors; the
leading terms are, however, easy to obtain by replacing
Eqs.~\eqref{VdotS} and \eqref{new_PdotS} in
Eq.~\eqref{BoxCoefficientCovariant}.

It is not difficult to see that, upon expanding the products of polarization tensors in Eqs.~\eqref{eq:M_1PM_full} 
and \eqref{eq:M_2PM_full} using Eqs.~\eqref{SpinEval}, no further spin-dependent spin-bilinear monomials are 
generated. The coefficients of each spin-dependent monomial is slightly modified because of the additional spin 
dependence coming from $\pol_4\cdot\pol_1 \pol_3\cdot \pol_2$: each one becomes a linear combinations of $a_\text{cov}$. 
This mixing is the same at every order in $G$.
Thus, the tree-level and one-loop amplitudes fully expanded to second order in spin are
\begin{align}
\frac{\mathcal{M}_4^\text{tree}}{4E_1E_2} & =
\frac{4\pi  G }{\bm q ^2} 
\bigg[
a^{(0)}_{1} 
+i\sum_{j=1}^2 a^{(1,j)}_{1}(\bm p\times\bm q)\cdot \bm S_j 
+ a^{(2,1)}_{1}  \bm q\cdot \bm S_{1} \, \bm q \cdot \bm S_{2}
\bigg]
\, ,
\label{eq:M_1PM_full_exp}
\\
\frac{\mathcal{M}_4^\text{1 loop}}{4E_1E_2} & =
\frac{2\pi^2  G^2 }{|\bm q |} 
\bigg[
a^{(0)}_{2} 
+i\sum_{j=1}^2 a^{(1,j)}_{2}(\bm p\times\bm q)\cdot \bm S_j 
 \nn \\
& \hspace{1.5cm}
+ a^{(2,1)}_{2}  \bm q\cdot \bm S_{1} \, \bm q \cdot \bm S_{2}
+ a^{(2,2)}_{2}  \bm q^2 \, \bm S_1\cdot \bm S_2 
+ a^{(2,3)}_{2} \bm q^2 \, \bm p\cdot \bm S_1 \, \bm p\cdot \bm S_2
\bigg] \nn \\
& \hskip 1 cm \null 
-
ia_{\rm B} I_{\rm B} -i a_{\bar{\rm B}} I_{\bar{\rm B}}\, ,
\label{eq:M_2PM_full_exp}
\end{align}
with the $a_{i}^A$ given by:
\begin{align}
a^{(0)}_i & =  
a^{(0)}_{\textrm{cov},i}
\,, \nn \\
a^{(1,1)}_i & = 
a^{(1,1)}_{\textrm{cov},i}-\frac{1}{m_1^2(\gamma_1+1)} a_{\textrm{cov},i}^{(0)}
\,,  \nn \\
a^{(1,2)}_i & =
a^{(1,2)}_{\textrm{cov},i}-\frac{1}{m_2^2(\gamma_2+1)} a_{\textrm{cov},i}^{(0)} 
\,, \nn \\
a^{(2,1)}_i & = 
a^{(2,1)}_{\textrm{cov},i}
-\frac{\bm p^2}{m_2^2(\gamma_2+1)}a^{(1,1)}_{\textrm{cov},i}
-\frac{\bm p^2}{m_1^2(\gamma_1+1)}a^{(1,2)}_{\textrm{cov},i}
+\frac{\bm p^2}{m_1^2 m_2^2 (\gamma_1+1)(\gamma_2+1)} a_{\textrm{cov},i}^{(0)}
\,, \nn \\
a^{(2,2)}_i & = 
a^{(2,2)}_{\textrm{cov},i}
+\frac{\bm p^2}{m_2^2(\gamma_2+1)}a^{(1,1)}_{\textrm{cov},i}
+\frac{\bm p^2}{m_1^2(\gamma_1+1)}a^{(1,2)}_{\textrm{cov},i}
-\frac{\bm p^2}{m_1^2 m_2^2 (\gamma_1+1)(\gamma_2+1)} a_{\textrm{cov},i}^{(0)}
\,, \nn \\
a^{(2,3)}_i & = 
a^{(2,3)}_{\textrm{cov},i}
-\frac{1}{m_2^2(\gamma_2+1)}a^{(1,1)}_{\textrm{cov},i}
-\frac{1}{m_1^2(\gamma_1+1)}a^{(1,2)}_{\textrm{cov},i}
+\frac{1}{m_1^2 m_2^2 (\gamma_1+1)(\gamma_2+1)} a_{\textrm{cov},i}^{(0)}
\, ,
\label{eq:a_covRelations}
\end{align}
with $i=1, 2$.
We are not decorating them with a ``cov'' index
because they are no longer associated with a definite combination of covariant spin vectors.  
We note that, while the expansion of $\pol_4\cdot\pol_1 \pol_3\cdot \pol_2$ generates the structures 
$ \bm q^2 \, \bm S_1\cdot \bm S_2 $ and $ \bm q^2 \, \bm p\cdot \bm S_1 \, \bm p\cdot \bm S_2$ in the tree level amplitude,
they can be ignored, as done before,  because  they do not correspond to long-range interaction terms.
The  $a_{\text{cov}, i}$  with $i=1, 2$ are given in Eqs.~\eqref{eq:1PMCoefficients} 
and \eqref{eq:2PMCoefficients}, respectively,  and $a^{(2,3)}_{\textrm{cov},1}=0$.

In the following section we will use these expressions to fix the
effective interaction Hamiltonian of two spinning bodies with
arbitrarily-oriented spins, through ${\cal O}(S_1 S_2)$.

\section{Effective field theory and derived Hamiltonian}
\label{sec:EFT}

Having found scattering amplitudes of General Relativity coupled with higher-spin fields
of the type described in Sec.~\ref{sec:basics}, we will now describe their translation to a 
two-body spin-dependent conservative Hamiltonian with complete velocity dependence.
We will extract it from the two-to-two interaction of an effective field theory of the positive-energy 
modes of higher-spin fields, thus generalizing the construction of Ref.~\cite{CliffIraMikhailClassical} 
to include spin degrees of freedom. 
The matching procedure with spins was discussed at $\Ord(G)$~\cite{Vaidya,YutinEFT}, and also at $\Ord(G^2)$ for spin-orbit potential expanded in velocity~\cite{Vaidya}.
Here we will establish a general spinning formalism for higher orders in $G$ and all-order in velocity;
it has the distinct advantage of being relatively straightforward,  while simultaneously 
producing results that allow physical observables to be obtained through standard
Hamiltonian mechanics methods.

\subsection{Spin formalism}
\label{generalspin}

Unlike the Lorentz-invariant setup of earlier sections, we will parametrize the spin degrees of freedom in terms of
the rest-frame spin of the two fields, $\xi_1$ and $\xi_2$. Since the rest frames of the two particles are not necessarily 
identical (i.e. the two particles need not be simultaneously at rest), there are two copies of the little-group generators, each 
acting on only one of the two fields; thus, the two fields $\xi_1$ and $\xi_2$ carry little-group indices\footnote{This is to be 
contrasted with the full theory, where the higher-spin fields carry $SU(2)_L\times SU(2)_R$ spinor indices, cf. e.g. Eqs.~\eqref{spinorindex} and \eqref{spinorindex2}. }, which we suppress throughout. 
We will denote the two copies of the little-group generators by $\pS_1$ and $\pS_2$, respectively, and their 
components by $\hat{\mathrm{S}}^{i}_{a}$ with $a=1, 2$. Apart from generating the $SO(3)$, 
\begin{align}
[\hat{\mathrm{S}}^{i}_{a} ,\, \hat{\mathrm{S}}^{j}_{b}] = i \delta_{ab}\,\epsilon^{ijk}\, \hat{\mathrm{S}}^k_a \ ,
\label{eq:spin_SO3}
\end{align}
they also are, as in nonrelativistic quantum mechanics, the spin operators of the two particles.

We take the action of the effective field theory for the fields $\xi_{1}$ and $\xi_2$ to be
\begin{align}
	S =& \int_{\bm k}   \, 
\sum_{a=1,2} \xi_a^\dagger(-\bm k) \left(i\partial_t - \sqrt{\bm k^2 + m_i^2}\right) \xi_a(\bm k)
-\int_{\bm k, \bm k'} \, 
 \xi_1^\dagger(\bm k') \xi_2^\dagger(-\bm k') \,
  \Vmom(\bm k' ,\bm k, \pS_a) \,\xi_1(\bm k)\xi_2(-\bm k) \, ,
\label{eq:eftL}
\end{align}
where we wrote the interaction term in the center of mass frame and $\int_{\bm k} = \int \frac {d^{D-1} \bm k}{(2\pi)^{D-1}}$. 
As mentioned, all little-group (spin) indices are suppressed. While the field $\xi(\bm k)$ describes a particle with momentum $\bm k$, 
its spin is always defined respect to its own rest-frame, such that the algebra in Eq.~\eqref{eq:spin_SO3} is satisfied. This is analogous 
to the treatment of spinning particles in quantum mechanics, where commutation relations of spin operators are the same whether
of not  the particle is at rest~\cite{NRQCD}.
The hat on the potential $\Vmom(\bm k' ,\bm k, \pS_a)$ indicates that it is a momentum space quantity; it is a function 
of the incoming momentum $\bm k$, momentum transfer $\bm q = \bm k-\bm k'$, as well as spin operators 
$\pS_1$ and $\pS_2$ of the two particles. 

To connect these operators with the classical rest-frame spin vectors, which is a necessary in the matching of the EFT amplitudes
with those of the complete theory, we take the asymptotic states of $\xi_1$ and $\xi_2$ to be spin coherent states in Eq.~\eqref{CoherentState}.
Similarly, the classical two-body Hamiltonian is given by the expectation value on classical on-shell states, which satisfy $\bm p^2 = (\bm p-\bm q)^2$,
\begin{align}
H(\bm q,\bm p)  = \sqrt{\bm p^2+m_1^2}+\sqrt{\bm p^2+m_2^2} + \langle \bm n_1\bm n_2| \Vmom(\bm p-\bm q, \bm p , \pS_a) 
|\bm n_1\bm n_2 \rangle \ ,
\label{HEFT}
\end{align}
where we only keep the classical part based on the counting in Eqs.~\eqref{eq:classical_limit} and \eqref{eq:classicalSpin_counting}.
The spin-independent part, obtained by formally setting $\pS_a=0$, clearly reproduces the two-body spinless Hamiltonian.
Here the momentum transfer $\bm q$ is the conjugate of the separation $\bm r$ between the two particles.
Therefore the position-space classical potential follows from taking the Fourier transform with respect to $\bm q$
\begin{align}
V(\bm r, \bm p, \bm S_a)  = \int_{\bm q} e^{-i\bm q\cdot \bm r}\,\langle \bm n_1\bm n_2| \Vmom(\bm p-\bm q, \bm p , \pS_a)|\bm n_1\bm n_2 \rangle \ ;
\label{eq:RealVfromMomV}
\end{align}
the expectation value effectively replaces the symmetric product of spin operators $\pS_a$ with their classical expectation values $\bm S_a$ via Eq.~\eqref{CoherentState}.

The ansatz for the interaction $\Vmom(\bm k' ,\bm k, \pS_a)$, which is subsequently fixed by matching the EFT amplitudes with
those of the full theory in the classical limit, is constructed such that on the one hand it contains only long-range interactions 
between the fields $\xi_1$ and $\xi_2$ and on the other by requiring that none of the terms vanishes in the classical limit.
As discussed in Ref.~\cite{CliffIraMikhailClassical, 3PMLong}, this fixes the dependence on the momentum transfer at $\Ord(G^n)$ 
to be $|\bm q|^{n-3}$ while the dependence on the incoming momentum is arbitrary.  
The classical scaling, described in Eqs.~\eqref{eq:classical_limit} and \eqref{eq:classicalSpin_counting}, is the main tool for 
including the spin dependence. It implies that, at each $\Ord(G^n |\bm q|^{n-3})$, we may include any number of spin operators
as long as each of them is accompanied by one factor of the momentum transfer, i.e.
\begin{align}
\Vmom(\bm k' ,\bm k, \pS_a) \supset 
d^{j_1\dots j_{n_1+n_2}, i_1,\dots, i_{n_1+n_2}}(\bm p)
\,  q^{j_1}\dots  q^{j_{n_1+n_2}} \, 
\hat{\mathrm{S}}^{i_1}_{1}\dots \hat{\mathrm{S}}^{i_{n_1}}_{1} \, 
\hat{\mathrm{S}}^{i_{n_1+1}}_{2}\dots \hat{\mathrm{S}}^{i_{n_1+n_2}}_{2}
\frac{G^n}{|\bm q|^{3-n}} \ .
\label{Vgeneralterm}
\end{align}
The coefficients $d^{j_1\dots j_{n_1+n_2}, i_1,\dots, i_{n_1+n_2}}(\bm p)$ can be further expanded in independent tensor structures,
which are constrained both by the desired/expected symmetries (such as parity) and by the classical limit. The former implies that
in a parity-invariant theory an even $n_1+n_2$ requires a parity-invariant coefficient and an odd $n_1+n_2$ requires a parity-odd
coefficient (i.e. one containing a Levi-Civita tensor). 

The ansatz for the coefficients $d^{j_1\dots j_{n_1+n_2}, i_1,\dots, i_{n_1+n_2}}(\bm p)$ is also constrained by the classical limit 
and the desired long-range nature of interactions. Momentum conservation implies that $\bm p\cdot \bm q = \bm q^2/2$. Therefore, 
if such a contraction occurs, the corresponding term becomes subleading in the classical limit and therefore needs not be included. 
Similarly, if $d^{j_1\dots j_{n_1+n_2}, i_1,\dots, i_{n_1+n_2}}(\bm p)$ contains a term that leads to a contraction of two of the 
momentum transfer factors, then that term needs not be included at $\Ord(G)$ (because it is a contact interaction) but must be 
included at all higher orders.
Each independent tensor structure has a scalar coefficient, which depends only on the square of the 
center of mass momentum; as in \cite{CliffIraMikhailClassical, 3PMLong}, we take it to be $\bm p^2=(\bm k^2+\bm k'{}^2)/2$.

It is not difficult to see that at $\Ord(G)$ and in a parity-invariant theory,  the only types of operators with an even and odd number 
of spin operators of each particle are $(\bm q \cdot \pS_i)^a$ and $\FTLq \cdot \pS_i \,(\bm q \cdot \pS_i)^a$, where $a$ is an even 
integer and 
\begin{equation}
\FTLq = i (\bm k\times \bm q)
\label{Lq}
\end{equation}
is the momentum space version of the orbital angular momentum. The complete set of operators that can appear in the interaction 
potential is the tensor product of the above sets of single-particle spin operators.
All other combinations, which are proportional to $\bm q^2$ and therefore ignored at this order, must be included at 
$\Ord(G^{n\ge 2})$;  the construction of operators proceeds as described above.

Using the construction outlined above, we will next set up in detail the EFT to quadratic order in spin operators and 
through $\Ord(G^2)$ and determine the free coefficients of the interaction potential by matching its amplitudes with 
those of the full theory, summarized in Sec.~\ref{tree_and_loop_summary}.


\subsection{Potential bilinear in spin}

Following the framework described above, we now build the  the most general classical potential up to quadratic order in spins.
The classical counting in Eq.~\eqref{eq:classicalSpin_counting} implies that, in momentum space, the possible building blocks are 
\begin{align}
\textrm{linear in spin:}&\quad \FTLq \cdot \pS_i \,,\nn\\
\textrm{quadratic in spin:}&\quad \bm q\cdot  \pS_i \, \bm q\cdot\pS_j\,,
\hskip .5 cm \bm q^2\, \pS_i \cdot \pS_j\,,
\hskip .5 cm 
\bm q^2\, \bm k\cdot  \pS_i \, \bm k\cdot\pS_j,\nn \\
&\quad
\bm q\cdot \bm k \, \bm k\cdot  \pS_i \, \bm q\cdot\pS_j, \hskip .5 cm \bm q\cdot \bm k \, \bm q\cdot  \pS_i \, \bm k\cdot\pS_j\,.
\label{eq:spin_operators_raw}
\end{align}
where the subscripts $i,j=1,2$ is the particle label, the prefactors are chosen such that each operator is $\Ord(1)$
under the classical counting and  $\FTLq$, defined in \eqref{Lq}, is the momentum-space version of the orbital angular momentum.
Parity requires that an operator with an odd number of spins must contain a factor of $\FTLq$.
Note that the vectors $\bm q, \, \bm k,$ and $\FTLq$  span a complete basis in three dimension. 
The operator $\FTLq \cdot \pS_i \, \FTLq \cdot \pS_j$ can be written in terms of the
above building blocks.

If the momenta of the two particles, $\bm k$ and $\bm k'$, are on shell then $\bm k\cdot \bm q \rightarrow
\bm q^2/2$; consequently, the operators $\bm q\cdot \bm k \, \bm k\cdot \bm
S_i \, \bm q\cdot\pS_j$ and $\bm q\cdot \bm k \, \bm q\cdot \pS_i \, \bm k\cdot\pS_j$
are subleading in the classical limit.  
This observation is similar to the removal of the products $\bm k\cdot \bm q$ in favor of $\bm q^2$ in
the case of spinless particles~\cite{CliffIraMikhailClassical}. 
We can see that it is even more advantageous in the presence of spin operators, because even the 
number of independent interactions is reduced.
This choice is analogous to gauge choices in more standard derivations of two-body Hamiltonians 
from General Relativity; as we will see shortly, it corresponds 
to the so-called isotropic gauge.

Thus, a minimal basis of spin-dependent interactions in the on-shell scheme,  up to quadratic order in spin 
and linear in the spin of each particle, consists of the six operators:
\begin{alignat}{4}
& \hat \Operator^{(0)} = \mathbb{I} \,,  
&&\hat \Operator^{(1,1)} = \FTLq\cdot \pS_1 \,,
&&\hat \Operator^{(1,2)} = \FTLq\cdot \pS_2\,, 
\label{OperatorList} \\
& \hat \Operator^{(2,1)} = \bm q\cdot \pS_1 \, \bm q\cdot \pS_2 \,,\qquad
&&\hat \Operator^{(2,2)} = \bm q^2 \, \pS_1\cdot \pS_2\,,\qquad
&&\hat \Operator^{(2,3)}  = \bm q^2 \, \bm k \cdot \pS_1\,\bm k \cdot \pS_2\,. \nn
\end{alignat}
Their expectation values in the spin coherent states, as in Eq.~\eqref{HEFT}, are in one to one correspondence with the 
various spin-dependent monomials in the tree-level and one-loop amplitude in the full theory to this order in spin, see Sec.~\ref{tree_and_loop_summary}. We labeled them following the same scheme as there. 
The ansatz for the EFT potential operator $\hat V(\bm k', \bm k, \pS_i)$ to quadratic order in spin operators is
\begin{align}
\hat V(\bm k', \bm k, \pS_i) =& 
\sum_A \hat V^{A}(\bm k', \bm k)\, \hat \Operator^A \,,
\label{eq:V_mom}
\end{align}
where $A$ runs over the superscripts of the operators in \eqn{OperatorList}, 
and $\hat V^{A}(\bm k',\bm k)$ are free momentum-dependent coefficients 
can be expanded as
\begin{align}
\hat V^{A}(\bm k',\bm k) =& \frac{4\pi G}{\bm q^2} d^A_1\left(\bm p^2 \right)
+ \frac{2\pi^2 G^2}{|\bm q|}  d^A_2\left(\bm p^2 \right)+  \Ord(G^3) \,.
\label{eq:V_p}
\end{align}
The coefficients $d_i^A$ are closely related to the $d$ coefficients in Eq.~\eqref{Vgeneralterm} and may be 
interpreted as the scalars multiplying the independent tensor structures in the latter.
As mentioned in Sec.~\ref{generalspin} and following Ref.~\cite{CliffIraMikhailClassical}, we choose the
off-shell continuation $\bm p^2 = (\bm k^2 +\bm k^{'2})/2$. While this is not important at tree level, it becomes essential 
for higher-order amplitudes.

It is not difficult to see that, as discussed in Sec.~\ref{generalspin},  in the $\Ord(G)$ potential any operator $\hat \Operator^A$ which 
contains a factor of $\bm q^2$ can be ignored. Indeed, the $\bm q^2$ in such an operator it cancels the $\bm q^{-2}$ 
in the first term in \eqref{eq:V_p} and thus leads to a contact term  upon Fourier-transform to position space.
Such are  $\hat \Operator^{(2,2)}$ and $\hat \Operator^{(2,3)}$, so we may therefore choose
\begin{equation}
d^{(2,2)}_1 = d^{(2,3)}_1 =0 \,.
\end{equation}
Starting at $\Ord(G^2)$ however, the $\bm q^2$ in these operators does not cancel the non-analytic $\bm q$ 
dependence, and therefore yields relevant long-distance effects and, in general, should not be ignored.

The position-space classical potential follows straightforwardly from the Fourier transform of $\bm q$ as in Eq.~\eqref{eq:RealVfromMomV}. (Here we strip off the coherent states.)
To carry out the Fourier transform it is necessary to identify the complete dependence on $\bm q$.  This amounts 
to expressing all $\bm k$ and $\bm k'$ in terms $\bm p$ and $\bm q$; the latter may be ignored in the classical limit.
This gives rise to the position-space potential
\begin{align}
V(\bm r, \bm p, \pS_i) &= 
\sum_A V^{A}(\bm r, \bm p)\, \Operator^A \, .
\label{eq:H_r}
\end{align}
The six independent position-space operators are,
\begin{alignat}{4}
& \Operator^{(0)} = \mathbb{I} \,, 
&&\Operator^{(1,1)} = \frac{1}{\bm r^2}\,\bm L\cdot \pS_1 \,, 
&&\Operator^{(1,2)} = \frac{1}{\bm r^2}\,\bm L\cdot \pS_2 \,, \nn \\
& \Operator^{(2,1)} = \frac{1}{\bm r^4}\,\bm r\cdot \pS_1 \, \bm r\cdot \pS_2\,, \qquad
&&\Operator^{(2,2)} = \frac{1}{\bm r^2}\, \pS_1\cdot \pS_2\,,\qquad
&&\Operator^{(2,3)} = \frac{1}{\bm r^2}\, \bm p\cdot \pS_1\,\bm p\cdot \pS_2 \, ,
\label{RealSpaceOperators}
\end{alignat}
where $\bm L = \bm r\times \bm p$ is the orbital angular momentum, and the prefactors 
are expanded in $G$,
\begin{align}
V^{A}(\bm r,\bm p) & = 
\frac{G}{|\bm r|} c^A_1(\bm p^2) 
+ \left(\frac{G}{|\bm r|}\right)^2  c^A_2(\bm p^2)+\Ord(G^3) \,.
\label{eq:V_r}
\end{align}
The earlier choice to trade the products $\bm k\cdot \bm q$ for $\bm q^2$ using momentum conservation 
translates in position space to the absence of the product $\bm p\cdot \bm r$ from the expression of the Hamiltonian.
Thus, it corresponds to the so-called isotropic gauge.

The relation between the coefficients of the momentum space and position space identity operator is 
trivial, $c^{(0)}_i = d^{(0)}_i$. However,  some of the spin operators $\hat\Operator^A$ include non-trivial 
(tensor-like) $\bm q$ dependence. Thus, the Fourier-transform  of $\hat V(\bm k', \bm k, \pS_i)$ in 
Eqs.~\eqref{eq:V_mom} and \eqref{eq:V_p} leads, general, to nontrivial linear relations between the  $d^A$ in 
Eq.~\eqref{eq:V_p} and the $c^A$ coefficients in Eq.~\eqref{eq:V_r}.
They are summarized in \tab{table:operator_summary}.
We note that, while at $\Ord(G)$ the momentum space potential depends only on $\hat\Operator^{(2,1)}$, 
its Fourier-transform depends on both $\Operator^{(2,1)}$ and $\Operator^{(2,2)}$; they enter only in the combination
$3\Operator^{(2,1)} -  \Operator^{(2,2)}$, which may be identified as (proportional to) the quadrupole of the system of two 
particles which is not inherited from the quadrupole of either one.

\def\spacer{$\frac{1}{X_{|_{|_X}}}$}
\begin{table}[t]
	\centering 
	\def\arraystretch{2.3}
	\begin{tabular*}{0.97 \textwidth}{c  @{\extracolsep{\fill}} l @{\extracolsep{\fill}} l @{\extracolsep{\fill}} l @{\extracolsep{\fill}} l}
		\hline\hline
		Label $A$ & $ {\hbox{mom. space} \atop  \hbox{operator  \hskip .42 cm }}$ & $ {\hbox{real space} \atop  \hbox{operator \hskip .18 cm }}$ &
                  $\Ord(G)$   &  $\Ord(G^2)$   \\
		\hline
		(0) & $\mathbb{I}$ & ~$\;\mathbb{I}$ & $c^{(0)}_1=d^{(0)}_1$ &  $c^{(0)}_2=d^{(0)}_2$ \\
		\hline
		(1,1) & $\FTLq\cdot \pS_1$ & \vphantom{\spacer} $\dfrac{1}{\bm r^2}\,\bm L\cdot \pS_1$ & 
		$c^{(1,1)}_1 = - d^{(1,1)}_1$ \hskip .8 cm \null & $c^{(1,1)}_2 = -2d^{(1,1)}_2$  \\
		\hline
		(1,2) & $\FTLq\cdot \pS_2$ & \vphantom{\spacer} $\dfrac{1}{\bm r^2}\,\bm L\cdot \pS_2$ & $c^{(1,2)}_1 = - d^{(1,2)}_1$ &$c^{(1,2)}_2 = -2d^{(1,2)}_2$ \\
		\hline
		(2,1) & $\bm q\cdot \pS_1 \, \bm q\cdot \pS_2$ & \vphantom{\spacer} $\dfrac{1}{ \bm r^4}\,\bm r\cdot \pS_1 \, \bm r\cdot \pS_2$ & $c^{(2,1)}_1 = -3 d^{(2,1)}_1$ &$c^{(2,1)}_2 = -8 d^{(2,1)}_2$ \\
		\hline
		(2,2) & $\bm q^2 \, \pS_1\cdot \pS_2$ & \vphantom{\spacer} $\dfrac{1}{\bm r^2}\, \pS_1\cdot \pS_2$ & $c^{(2,2)}_1 = d^{(2,1)}_1$ &$c^{(2,2)}_2 = 2 d^{(2,1)}_2-2d^{(2,2)}_2$ \\
		\hline
		(2,3) & $\bm q^2 \, \bm k\cdot \pS_1\,\bm k\cdot \pS_2 \quad$ &\vphantom{\spacer}  $\dfrac{1}{\bm r^2}\, \bm k\cdot \pS_1\,\bm k\cdot \pS_2 \quad$ & $c_1^{(2,3)} = 0$ &$c^{(2,3)}_2 = -2 d^{(2,3)}_2$ \\
		\hline \hline
	\end{tabular*}
	\caption{\small Summary of momentum- and real-space operators for spin
          interactions and relations between their coefficients, through
          bilinear order in spins. The first column lists operator
          labels, $A$.  The operators in momentum space and in position space, $\hat\Operator^A$  and $\Operator^A$, 
          are given in column two and three, respectively.  In the
          last column we give the relations between the coefficients
          of momentum-space operators $d^{A}_i$ and position-space counterpart $c^{A}_i$ at $\Ord(G)$ and $\Ord(G^2)$.  The
          momentum-space and position-space potentials are defined in
          \eqns{eq:V_p}{eq:V_r}. }
	\label{table:operator_summary}
\end{table}

\subsection{EFT four-point scattering amplitude}

To guarantee that the EFT described above corresponds to the full theory set up and used in 
earlier sections and free coefficients of the EFT Lagrangian, we compare its tree-level and one-loop 
four-point scattering  amplitude with the tree-level and one-loop amplitudes are summarized in 
Sec.~\ref{tree_and_loop_summary}.
To this end, in this section we evaluate the EFT two-to-two scattering amplitude. Before
proceeding in the next section to determine its free coefficients, we use the state-of-the-art 
spin$_1$-spin$_2$ Hamiltonians \cite{JaranowskiSchafer4PN, SteinhoffADMforSpin} to 
verify that the EFT amplitude reproduces the suitable expansion of the full theory amplitudes. 


Given the simple structure of the EFT Lagrangian, it is straightforward to derive the Feynman rules. 
The propagator and vertices are:
\begin{equation}
	\includegraphics[scale=.55,trim={0 0.2cm 0 0}, clip]{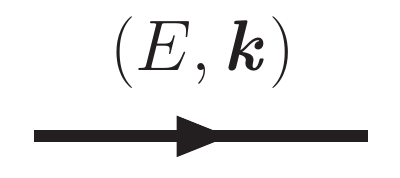} = \frac{i\, \mathbb{I}}{E-\sqrt{\bm k^2+m^2}+i\epsilon}\,, \qquad
	\vcenter{\hbox{\includegraphics[scale=.55,trim={0 0.5cm 1cm 0}, clip]{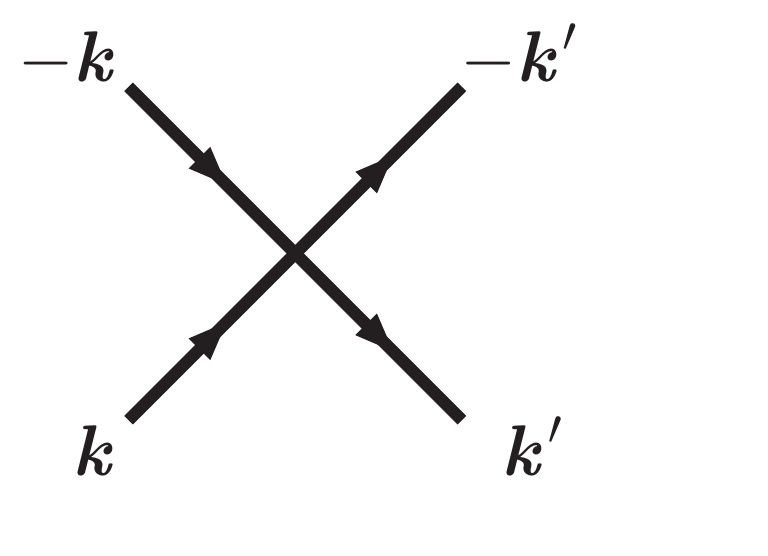}}} =
	-i \hat{V}(\bm k', \bm k, \pS_{i})\,,
\label{eq:EFT_FeynmanRules}
\end{equation}
where $\mathbb{I}$ in the numerator of propagator is an identity operator.
As emphasized above, the vertices should be viewed as operators whose ordering is important.
As in the spinless case, off-shell continuation of the potential needs to be defined in order to have consistent amplitude. 
We use $\bm p^2 = (\bm k^2+\bm k^{'2})/2$ in the coefficients $d^{(A)}_i(\bm p^2)$, and also choose Eq.~\eqref{OperatorList} as the off-shell definitions for the operators.

\begin{figure}[tb]
	\begin{center}
		\includegraphics[scale=.6,trim={0 1.5cm 0 0},clip]{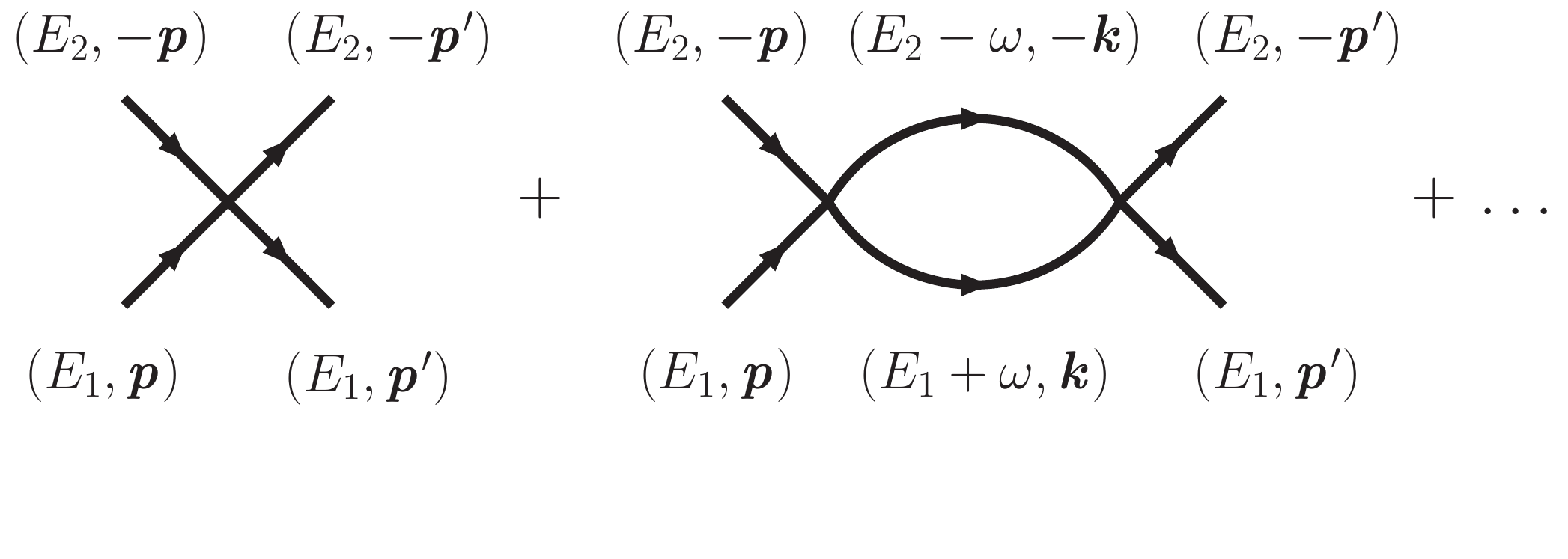}
	\end{center}
	\vspace*{-0.5cm}
	\caption{\small The EFT scattering amplitude is given by the
          sum of bubble diagrams. We use center of mass frame for
          external kinematics. The loop momentum $\bm k = \bm p + \bm
          l$ where $\bm l$ is the momentum transfer flowing downward in the left-most
          vertex of the second diagram. }
	\label{fig:EFT_amp}
\end{figure}

To illustrate the calculation, consider the amplitude up to
$\Ord(G^2)$. The two relevant Feynman diagrams are given in
\Fig{fig:EFT_amp}.  It is not difficult to see that the Feynman rules give thew following expression for the 
two-to-two scattering amplitude stripped of external-state spinors:
\begin{align}
\widehat{\mathcal{M}}^{\textrm{EFT}}
=& -\hat V(\bm p',\bm p) 
- \int_{\bm k} \int \frac{d \omega}{2\pi} \frac{\hat V(\bm p',\bm k)\, \hat V(\bm k,\bm p)}{\left(E_1+\omega- \sqrt{\bm k^2+m_1^2}\right) \left(E_2-\omega- \sqrt{\bm k^2+m_2^2} \right)} \nn \\[3pt]
=& -\hat V(\bm p',\bm p) - \int_{\bm k} \frac{\hat V(\bm p',\bm k)\, \hat V(\bm k,\bm p)}{E_1+E_2- \sqrt{\bm k^2+m_1^2}- \sqrt{\bm k^2+m_2^2}} \ .
\label{eq:eft_FeynmanDiagram}
\end{align}
The second line is obtained by carrying out the $\omega$ integral, using the standard $i\epsilon$ prescription.

Unlike the case of spinless particles, the potential $V$ entering each vertex is an operator and therefore the order 
of $\hat V(\bm p',\bm k)$ and $\hat V(\bm k,\bm p)$ in the numerator is essential 
As typically done in quantum mechanics and quantum field theory, they are ordered  from left to right, beginning with the 
vertex adjacent to the final state, followed by the one adjacent to the initial state. In terms of the one-loop Feynman graph in 
\Fig{fig:EFT_amp}, vertices are read against the arrows denoting the momentum (and charge) flow.
Each vertex brings spin operators of the same particle. In close similarity with the full theory, to consistently take 
the classical limit it is necessary to decompose such products in irreducible representations of the rotation group\footnote{In the full theory,
products of Lorentz generators were decomposed in irreducible representations of the Lorentz group.}; this is done by repeated use of 
the $SO(3)$ algebra in Eq.\eqref{eq:spin_SO3} . For example, a product of two spin operators is organized as
\begin{align}
\hat{\mathrm{S}}^{i}_{a} \hat{\mathrm{S}}^{j}_{a}
= \frac{1}{2}\left\lbrace \hat{\mathrm{S}}^{i}_{a}, \hat{\mathrm{S}}^{j}_{a}\right\rbrace 
+\frac{1}{2} \left[ \hat{\mathrm{S}}^{i}_{a}, \hat{\mathrm{S}}^{j}_{a} \right]
= \frac{1}{2}\left\lbrace \hat{\mathrm{S}}^{i}_{a}, \hat{\mathrm{S}}^{j}_{a}\right\rbrace
+\frac{i}{2}\, \epsilon^{ijk}\, \hat{\mathrm{S}}^{k}_{a} \ ;
\label{eq:spin_decomposition}
\end{align}
this is similar with Eq.~\eqref{eq:cov_symmetrization} in the full theory, written there for the generators of the Lorentz group. 
Although the commutator may appear to be subleading  in classical counting, it can still yield relevant contributions 
when it appears in a loop diagram.  

The propagator in Eq.~\eqref{eq:eft_FeynmanDiagram} simplifies when expanded around the
classical limit~\cite{CliffIraMikhailClassical},
\begin{align}
\frac{1}{E_1+E_2- \sqrt{\bm k^2+m_1^2}- \sqrt{\bm k^2+m_2^2}} 
=  -\frac{2\xi E}{\bm k^2 -\bm p^2} - \frac{1-3\xi}{2\xi E}+\cdots \ ,
\end{align}
where $E=E_1+E_2$, $\xi = E_1 E_2/(E_1+E_2)^2$, and the ellipsis stands for higher order 
in classical counting which are irrelevant for $\Ord(G^2)$.  As
indicated in \fig{fig:EFT_amp}, $\bm k = \bm p + \bm l$, where $\bm
p$ is the center-of-mass external momentum and $\bm l$ is the momentum
transfer in the left-most vertex of the send diagram.  We can see that
the first term is $\Ord(|\bm l|^{-1}) \sim \Ord(|\bm q|^{-1})$ using
$\bm k = \bm p +\bm l$ and expanding in $\bm l$. Therefore, as stated above, the
commutator term in Eq.~\eqref{eq:spin_decomposition} can be relevant for 
the classical limit of the EFT amplitude when it interferes with the
propagator. We will judiciously keep such contributions when
evaluating the second term in Eq.~\eqref{eq:eft_FeynmanDiagram}.

The final two-to-two scattering amplitude $\mathcal{M}^{\rm EFT}$ in EFT is obtained 
by contracting 
$\widehat{\mathcal{M}}^{\textrm{EFT}}$
with suitable external-state states.  
As discussed in Sec.~\ref{generalspin} and Sec.~\ref{sec:basics}, 
the relevant ones for the classical limit are the spin coherent states defined in Eq.~\eqref{CoherentState}.
Thus, 
\begin{align}
\mathcal{M}^{\rm EFT}  \equiv \langle \bm n_1,\bm n_2|\, \widehat{\mathcal{M}}^{\rm EFT} \,| \bm n_1,\bm n_2\rangle \, .
\end{align}
Since these states are momentum-independent, the net effect is expectation value is to simply turn 
the spin operator $\pS_i$ into classical expectation values $\bm S_i$; moreover, this expectation value does 
not lead to any terms that are subleading in the classical limit.

At $\Ord(G)$, the  EFT amplitude follows simply from evaluating the tree-level diagram in \Fig{fig:EFT_amp}, 
which corresponds to the first term in Eq.~\eqref{eq:eft_FeynmanDiagram}; it is directly given by the $\Ord(G)$ 
potential. Keeping only terms up to bilinear order in spin that contribute to long-range interactions, the amplitude is
\begin{align}
\mathcal{M}^{\rm EFT}_{\rm 1PM} =
\frac{4\pi G }{\bm q^2}
\left[
a^{(0)}_1 + a^{(1,1)}_1 \FTLq\cdot \bm S_1 + a^{(1,2)}_1 \FTLq\cdot \bm S_2
+ a^{(2,1)}_1 \, \bm q\cdot \bm S_1 \, \bm q\cdot \bm S_2 \right]
.
\label{eq:M_1PM}
\end{align}
The $a_1^A$  coefficients~\footnote{These EFT amplitude coefficients are formally distinct from 
the full theory coefficients. However, for the EFT to correspond to the full theory, the scattering amplitudes of the two theories must 
be the same. Enforcing this condition, referred to as ``EFT matching'' which we will do in Sec.~\ref{matching}, leads to amplitudes' coefficients being equal, so we use the same notation for both of them.}
 are directly given by the coefficients in the momentum-space potential~\eqref{eq:V_p},
\begin{equation}
a^A_1 = -d^A_1 \,.
\label{eq:a1}
\end{equation}
As discussed in \sect{sec:basics}, a simple rule for tracking relevance in
the classical limit, is that each power of spin comes with a single power of $\bm
q$ relative to the spinless case; terms with higher powers are irrelevant.
If the required powers of $\bm q$ appears as ${\bm q}^2$ at this order
we can drop the contributions, because they cancel the $1/\bm q^2$ pole
and will not yield long-distance contributions to the potential.  

The EFT amplitude at $\mathcal{O}(G^2)$ order receives contributions
from both terms in Eq.~\eqref{eq:eft_FeynmanDiagram} and can be
written as
\begin{align}
\mathcal{M}^{\rm EFT}_{\rm 2PM} & =
\frac{2\pi^2 G^2}{|\bm q|}
\bigg[
a^{(0)}_2 + a^{(1,1)}_2 \FTLq\cdot \bm S_1 + a^{(1,2)}_2 \FTLq\cdot \bm S_2  \nn \\
& \hspace{1.5cm}
+ a^{(2,1)}_2 \bm q\cdot \bm S_{1} \, \bm q \cdot \bm S_{2}
+ a^{(2,2)}_2 \bm q^2 \, \bm S_1\cdot \bm S_2
+ a^{(2,3)}_2 \bm q^2 \, \bm p\cdot \bm S_1 \, \bm p\cdot \bm S_2
\bigg] \nn \\
& \hskip 1 cm \null 
+(4\pi G)^2\, a_{\rm iter}\int \frac{d^{D-1}\ell}{(2\pi)^{D-1}}
\frac{2\xi E}{\bm \ell^2 (\bm \ell+\bm q)^2 (\bm \ell^2+2\bm p\cdot \bm \ell)} \,,
\label{eq:M_2PM}
\end{align}
where we have expanded in $\bm q$ and kept only terms that are relevant in the 
classical limit.
The first two lines are of order $1/|\bm q|$, including the scaling of the spin vectors; 
this is the expected order of the classical potential at $\Ord(G^2)$.
The integral in the last term originates from the one-loop diagram in \fig{fig:EFT_amp}; 
since the two vertices are identical and given by the $\Ord(G)$ potential, we refer to this term 
as an ``iteration''. It is not difficult to see that the integral is of order $1/\bm q^2$.
Thus, for the amplitude to be accurate to $\Ord(1/|\bm q|)$, an extra power of 
$\bm q$ beyond the leading contribution must be kept in the coefficient $a_{\rm iter}$.
While this part of the amplitude does not contain information beyond the one in the 
tree-level potential,  it is nevertheless very important.
Similarly to the spinless EFT, the iteration part of the EFT one-loop amplitude is IR-divergent. 
Since the EFT is constructed to match the low-energy part of the desired 
complete theory, the IR divergences in EFT should be the same as in the full theory 
one-loop amplitude.\footnote{Higher-loop iteration terms have a similar interpretation.}
In doing so the first subleading terms in the $a_{\rm iter}$ coefficients are crucial and the required
match can be enforced by appropriately choosing the subleading terms in the relation between
the Lorentz generators and the spin tensor in the full theory. We emphasize that the only effect of
such a choice is to guarantee the match of IR divergences and has no consequence on the finite 
part of the amplitude and on the $\Ord(G^2)$ potential. 

To write compact expressions for the coefficients of the 2PM EFT amplitude in terms of those of the 
momentum space Hamiltonian it is convenient to define the functions
\begin{align}
A_0[X] &= 
\left[(1-3\xi) + 2 \xi^2E^2 \partial  \right] X
\,, \nn \\
A_1[X] &= 
\left[(1-3\xi) + \frac{2 \xi^2 E^2}{\bm p^2} + 2 \xi^2 E^2\partial \right] X
\,,\nn \\
A_2[X] &= 
\left[\frac{1}{4}(1-3\xi) + \frac{ \xi^2 E^2}{\bm p^2} + \frac{1}{2} \xi^2 E^2\partial \right] X
\,, \nn \\
A_3[X] &= 
\left[\frac{3}{4}(1-3\xi) + \frac{ \xi^2 E^2}{\bm p^2} + \frac{3}{2} \xi^2 E^2\partial \right] X
\,,
\end{align}
where $\partial = \partial/\partial{\bm p^2}$. 
Then, the EFT amplitude coefficients in Eq.~\eqref{eq:M_2PM} are:
\begin{align}
a^{(0)}_{2} = \null &
-d^{(0)}_2+\frac{1}{2\xi E}\, A_0\left[\left(d^{(0)}_1 \right)^2\right] 
, \nn \\
a^{(1,i)}_{2} = \null &
-d^{(1,i)}_2+\frac{1}{2\xi E}\, A_1\left[d^{(0)}_1 d^{(1,i)}_1\right]
, \nn \\
a^{(2,1)}_{2} = \null &
-d^{(2,1)}_2 
+\frac{1}{2\xi E}\, A_3\left[d_1^{(0)}d^{(2,1)}_1\right]
+\frac{\bm p^2}{2\xi E}\, A_2\left[d_1^{(1,1)}d^{(1,2)}_1\right]
+\frac{\xi E}{8} (d_1^{(1,1)}+d_1^{(1,2)}) d^{(2,1)}_1 
\,, \nn \\[4pt]
a^{(2,2)}_{2} = \null &
-d^{(2,2)}_2
-\frac{1}{2\xi E}\, A_2\left[d_1^{(0)}d^{(2,1)}_1\right]
-\frac{\bm p^2}{\xi E}\, A_2\left[d_1^{(1,1)}d^{(1,2)}_1\right]
+\frac{\xi E}{8} (d_1^{(1,1)}+d_1^{(1,2)}) d^{(2,1)}_1 
\,, \nn \\[4pt]
a^{(2,3)}_{2} = \null & 
- d^{(2,3)}_2 
+\frac{\xi E}{\bm p^4}\, \left[d_1^{(0)}d^{(2,1)}_1\right]
+\frac{1}{\xi E}\, A_2\left[d_1^{(1,1)}d^{(1,2)}_1\right]
-\frac{\xi E}{2\bm p^2} (d_1^{(1,1)}+d_1^{(1,2)}) d^{(2,1)}_1
\, .
\label{eq:EFT_2PM_S1S2}
\end{align}
Recalling that the first superscript of the Hamiltonian coefficients represent the number of spin operators in the 
corresponding operator, it is easy to infer that the combination $(d_1^{(1,1)}+d_1^{(1,2)}) d^{(2,1)}_1$ arises from 
three-spin terms in the numerator of the amplitude \eqref{eq:eft_FeynmanDiagram}. Such terms nevertheless 
contribute to the two-spin terms in the amplitude through the commutator identity Eq.~\eqref{eq:spin_decomposition}.

We note that the spin-dependent sector of the amplitude contains $1/\bm p^2$ threshold singularities through the functions 
$A_{1,2,3}[d^{A}_1 d^{B}_1]$. This singularity is physical because it appears in the amplitude. These terms arise 
from the reduction to scalar integrals of various tensor integrals in the one-loop diagram in \Fig{fig:EFT_amp}.
This is intimately connected to the spin dependence and, as pointed out in Sec.~\ref{sec:oneloop}, has a counterpart 
in the amplitude calculation in the full theory. 
The residue of this singularity is completely fixed by $\Ord(G)$ terms in the Hamiltonian, so the singular terms should be the same in the EFT and in the full theory.
The 2PM Hamiltonian obtained by demanding that the two amplitudes are identical turns out to be local in $\bm p$.

Last, the iteration coefficient $a_{\rm iter}$ in \eqn{eq:M_2PM} is fixed by the $\Ord(G)$ terms in the Hamiltonian:
\begin{align}
\label{eq:M_2PM_box}
a_{\rm iter}=&
\left(d_1^{(0)} \right)^2 + d_1^{(0)} d^{(1,1)}_1 \FTLq \cdot \bm S_1+d_1^{(0)} d^{(1,2)}_1 \FTLq \cdot \bm S_2   \\
&+d_1^{(0)}d^{(2,1)}_1 \,\bm q\cdot \bm S_1 \, \bm q\cdot \bm S_2
+\frac{1}{2}\left(d_1^{(1,1)}d^{(1,2)}_1 + \frac{d_1^{(0)}d^{(2,1)}_1}{\bm p^2} \right)
\Bigl[\FTLq \cdot \bm S_1 \, \FTLq \cdot \bm S_2
- \bm p^2 \, \bm q\cdot \bm S_1\, \bm q\cdot \bm S_2 \Bigr] 
\,.
\nn
\end{align}
This expression is through $\mathcal{O}(|\bm q|)$, where we count $\bm S_i \sim 1/|\bm q|$. 
In particular, the subleading terms appearing when mapping  $\FTLq \cdot \bm S_1 \, \FTLq \cdot \bm S_2$ 
to the basis in Eq.~\eqref{eq:spin_operators_raw} are important.

The EFT formalism described in Sec.~\ref{generalspin} and illustrated here for the spin-bilinears 
also allows us to compute scattering amplitudes starting from canonical Hamiltonians in the General 
Relativity literature.  The gauge invariance of scattering amplitudes provides a straightforward test 
of the (in)equivalence of different-looking Hamiltonians that avoids the explicit construction of canonical
transformations.
Given some Hamiltonian constructed from the outset in terms of the classical spin, the calculation 
of amplitudes follows the same steps as above. We first promote the classical spins in the potential to 
operators $\bm S_a\mapsto {\hat{\bm S}}_a$. Since the classical Hamiltonian has, by construction, 
classical scaling, this operation does not introduce an ordering ambiguity; any such ordering, arising 
through Eq.~\eqref{eq:spin_decomposition}, is subleading.
The amplitudes then follow from the Feynman rules in Eq.~\eqref{eq:EFT_FeynmanRules} and the 
expression Eq.~\eqref{eq:eft_FeynmanDiagram} for the tree-level and one-loop Feynman diagrams.
As in our calculation, the ordering of the vertices is relevant once inserted in the one-loop (and higher-loop) 
diagrams. 
While our construction exploits on-shell conditions to eliminate $\bm p\cdot \bm r/|\bm r|$ and only 
keeps operators \eqref{RealSpaceOperators} in our position-space potential \eqref{eq:H_r}, typical 
GR-derived Hamiltonians depend on $\bm p\cdot \bm r/|\bm r|$ as well as other operators like
$\bm p \cdot \pS_1 \, \bm r \cdot \pS_2$. It is essential that they all be kept in the off-shell vertex and 
that on shell conditions are enforced only for the external states of the amplitude.

Using this approach, we can evaluate the EFT amplitude for the available 
post-Newtonian Hamiltonian with spin-orbit and spin$_1$-spin$_2$ interactions. 
In the same gauge as the potential for spinless bodies in Eq.~(8.41)
of Ref.~\cite{JaranowskiSchafer4PN}, the spin-dependent
next-to-leading order Hamiltonian may be found in Eqs.~(7.26)-(7.29)
in Ref.~\cite{SteinhoffADMforSpin}, and the next-to-next-leading order
one in Eqs.~(138)-(139) of Ref.~\cite{SteinhoffNNLOSpin}.\footnote{See
  Ref.~\cite{EFTSpinHamiltonianEquiv} for the equivalence of the spin
  Hamiltonian derived using NRGR.}
This counting translates to expansions up to ${\cal O}(G v^4)$ and ${\cal O}(G^2v^2)$ at the first and second order in an expansion in Newton's constant, respectively. See \Fig{table:PNPM} for the comparison of spin$_1$-spin$_2$ potential.

The tree-level EFT amplitude following from these Hamiltonians, through $\Ord (\bm p^4)$,  has the same structure
as Eq.~\eqref{eq:M_1PM}; the coefficients are
\begin{align}
a_{1}^{(0)} &= 
m_1 m_2 + \frac{3 m_1^2 + 8 m_1 m_2 + 3 m_2^2}{
2 m_1 m_2} \bm p^2  - \frac{5 m_1^4 - 18 m_1^2 m_2^2 + 5 m_2^4}{8 m_1^3 m_2^3}\bm p^4
+\dots \,,
\label{eq:1PMCoefficientsNNLO}  \\
a_{1}^{(1,1)} &=
 \frac{4 m_1 + 3 m_2}{2 m_1} 
+\frac{18 m_1^2 + 8 m_1 m_2 - 5 m_2^2}{8 m_1^3 m_2} \bm p^2
-\frac{15 m_1^4 + 15 m_1^2 m_2^2 + 12 m_1 m_2^3 - 
7 m_2^4}{16 m_1^5 m_2^3} \bm p^4
+ \dots \,,
\nn \\
a_{1}^{(2,1)} &=
1 + \frac{2 m_1^2 + 9 m_1 m_2 + 2 m_2^2}{4 m_1^2 m_2^2}\bm p^2  - \frac{6 m_1^4 + 15 m_1^3 m_2 - 4 m_1^2 m_2^2 + 
	15 m_1 m_2^3 + 6 m_2^4}{16 m_1^4 m_2^4}\bm p^4
+\dots \,,
	\nn
\end{align}
where the ellipsis stand for $\Ord(v^{n\ge 6})$
and $a_{1}^{(1,2)}$ is obtained by exchanging $(m_1,\gamma_1)$ and $(m_2,\gamma_2)$ 
in $a_{1}^{(1,1)}$. It is not difficult to see that these expressions reproduce the coefficients of the full theory 
amplitude in Eqs.~\eqref{eq:1PMCoefficients} and \eqref{eq:a_covRelations}, through $\Ord(v^4)$.

The available Hamiltonians determine the $\Ord(G^2)$ amplitude only through  $\Ord(v^2)$. The structure of 
the amplitude  is the same as Eq.~\eqref{eq:M_2PM}. The coefficients of the various spin-dependent monomials 
are to the relevant order in velocity are
\begin{align}
a_{2}^{(0)} &= 
3 m_1 m_2 (m_1+m_2) + \frac{
3  (m_1 + m_2) (3 m_1^2 + 10 m_1 m_2 + 3 m_2^2)}{
4 m_1 m_2} \bm p^2
+\dots\,,
\nn \\[2pt]
a_{2}^{(1,1)} &= 
\frac{m_1 m_2^2 (4 m_1 + 3 m_2)}{2 (m_1 + m_2) \bm p^2} + 
\frac{20 m_1^3 + 53 m_1^2 m_2 + 41 m_1 m_2^2 + 9 m_2^3}{
4 m_1(m_1+ m_2)} \nn \\[2pt]
&\quad + \frac{3 (30 m_1^4 + 71 m_1^3 m_2 + 43 m_1^2 m_2^2 - m_1 m_2^3 - 
4 m_2^4)}{16 m_1^3 m_2 (m_1 + m_2)} \bm p^2
+\dots \,,
\nn \\[2pt]
a_{2}^{(2,1)} &=
\frac{m_1^2 m_2^2}{2 (m_1 + m_2) \bm p^2}
+ \frac{(7 m_1^2 + 15 m_1 m_2 + 7 m_2^2)}{2 (m_1 + m_2)} \nn \\[2pt]
&\quad + \frac{3 (3 m_1^4 + 39 m_1^3 m_2 + 74 m_1^2 m_2^2 + 39 m_1 m_2^3 + 3 m_2^4)}{16 m_1^2 m_2^2 (m_1 + m_2)} \bm p^2
+\dots\,, \nn \\[2pt]
a_{2}^{(2,2)} &= -a_{2}^{(2,1)} \nn \\[2pt]
a_{2}^{(2,3)} &=
\frac{m_1^2 m_2^2}{(m_1 + m_2)\bm p^4}
+ \frac{19 m_1^2 + 40 m_1 m_2 + 19 m_2^2}{4 (m_1 + m_2) \bm p^2}
\nn\\[2pt]
&\quad+ \frac{3 (3 m_1^4 + 45 m_1^3 m_2 + 86 m_1^2 m_2^2 + 
45 m_1 m_2^3 + 3 m_2^4}{16 m_1^2 m_2^2 (m_1 + m_2)}
+\dots \, .
\label{eq:2PMCoefficientsNNLO}
\end{align}
The ellipsis stand for $\Ord(v^{n\ge 4})$ and $a_{2}^{(1,2)}$ is obtained by interchanging $m_1$ and $m_2$ in $a_{2}^{(1,1)}$.  
Note that the operator associated with $a^{(2,3)}_2$ is $\bm p\cdot \pS_1 \, \bm p\cdot \pS_2$, 
which carries $\Ord (\bm p^2)$, its coefficient is determined only though $\Ord(v^0)$.
We also note the coefficients of the spin-dependent monomials exhibit  $1/\bm p^2$ singularities; 
similarly to the full theory amplitude and to the calculation of the EFT amplitude from the 
potential~\eqref{eq:V_mom},  they originate from the tensor reduction of EFT one-loop integrals 
and their residues are controlled by $\Ord(G)$ Hamiltonian terms. 
The agreement outlined here serves as a highly non-trivial test of the spin EFT formalism
formulated in this section and of the higher-spin field theory construction used in earlier sections.


\subsection{Conservative spin Hamiltonian from matching}
\label{matching}

With the amplitudes of the EFT and of the full theory in hand, we turn to constructing 
the Hamiltonian. It is fixed by demanding  that the two amplitudes are the same,
\begin{align}
{\cal M}^\text{EFT}_\text{1PM} = \frac{{\cal M}^\text{tree}_4}{4E_1 E_2}
\ , \qquad
{\cal M}^\text{EFT}_\text{2PM} = \frac{{\cal M}^\text{1 loop}_4}{4E_1 E_2}
\ .
\label{IDamplitudes}
\end{align}
The EFT amplitude is parametrized in \eqns{eq:M_1PM}{eq:M_2PM}, with coefficients
given in Eqs.~\eqref{eq:a1} and~\eqref{eq:EFT_2PM_S1S2} while the full theory amplitude
may be found, in the same parametrization, in Eqs.~\eqref{eq:1PMCoefficients}, 
\eqref{eq:2PMCoefficients} and \eqref{eq:a_covRelations}. 
The equality of amplitudes \eqref{IDamplitudes} implies that coefficients of identical spin-dependent 
monomials --- both denoted by $a_i^A$ -- are also identical. From here we extract the coefficients $d^A_i$
of the momentum-space potential.

Carrying this out at $\Ord(G)$ we find that $d^A_1$ are given by
\begin{align}
d_{1}^{(0)} =&
\frac{m^2 \nu^2}{\xi \gamma^2}(1-2\sigma^2) 
\,,\nn \\
d_{1}^{(1,i)} =& 
-\frac{\nu}{\xi \gamma^2 }\frac{2\sigma E}{m_i} -\frac{1}{m_i^2 (\gamma_i+1)} d_{1}^{(0)}
\,,\nn \\
d_{1}^{(2,1)} =&
\frac{\nu}{\xi \gamma^2 }(1-2\sigma^2)
+\left(\frac{1}{m_1 (\gamma_1+1)} + \frac{1}{m_2 (\gamma_2+1)} \right)\frac{2\sigma \bm p^2}{E \xi }
+\frac{\bm p^2}{m_1^2 m_2^2 (\gamma_1+1)(\gamma_2+1)} d_{1}^{(0)}
\,,
\label{eq:1PM_potential}
\end{align}
and $d^{(2,2)}_1= d^{(2,3)}_1=0$ because they do not mediate long-range interactions. The variables used here are
defined in Eq.~\eqref{var_defs} and below Eq.~\eqref{box_coefs_norm}.
The position-space potential follows immediately via \tab{table:operator_summary}
\begin{alignat}{6}
&  c_1^{(0)} = d_1^{(0)}\,,\qquad
&& c_1^{(1,i)} = -d_1^{(1,i)}\,,\qquad
&& c_1^{(2,1)} = -3d_1^{(2,1)}\,,\qquad
&& c_1^{(2,2)} = d_1^{(2,1)}\,,\qquad
&& c_1^{(2,3)} = 0.
\label{eq:1PM_potential_real}
\end{alignat}
At $\Ord (G)$, the amplitude and potential is directly related, so the potential retains the structure 
in Eq.~\eqref{eq:a_covRelations}. The relation between $c_{1}^{(2,1)}$ and $c_{1}^{(2,2)}$ implies that, 
at this order, the spin-bilinear part of the position-space potential depends only on the two-particle 
quadrupole $3 \bm r\cdot \bm S_1\, \bm r\cdot \bm S_2 - r^2 \bm S_1\cdot \bm S_2$.  

The coefficients \eqref{eq:1PM_potential_real} determine the $\Ord(G)$ Hamiltonian to leading order in the classical limit.
As discussed earlier in this section, we may consider keeping subleading terms in this Hamiltonian, as they may yield 
leading-order classical terms in the $\Ord(G^2)$ amplitude and thus modify the classical Hamiltonian at that order.
Fortunately, the on-shell conditions force the correction to be either suppressed by two powers of $\bm q$ compared to 
the classical terms, or proportional to the  operators $\bm q^2 \, \bm p\cdot \pS_i \, \bm q\cdot\pS_j$. The latter structure 
cancels one graviton pole, leading to a contact term which gives a vanishing contribution in the one-loop EFT amplitude. 
The former may yield at most contributions to the EFT amplitude that are suppressed by one power of $\bm q$.
We conclude therefore that the coefficients \eqref{eq:1PM_potential_real} are sufficient to generate the correct EFT 
amplitudes  through $\Ord(G^2)$.

As a non-trivial consistency check, which verifies that the EFT we constructed corresponds to
the full theory used in earlier sections,  we can compare the iteration coefficient, 
$a_\text{iter}$ in Eq.~\eqref{eq:M_2PM_box}, and the classical limit of the $a_{\rm B}$ of the energy-integrated 
box integral. They are both determined by $\Ord(G)$ data and multiply the same IR-divergent 
three-dimensional integral, which is $\Ord(|\bm q|^{-1})$.
We find
\begin{align}
\frac{d_{\rm B}}{4E_1 E_2} = (4\pi G)^2\, 4\xi E^2\, a_{\rm iter}  + \Ord(|\bm q|) \ ,
\label{box_stuff}
\end{align}
which is indeed required for the equality of the EFT and full theory amplitudes, Eq.~\eqref{IDamplitudes}, at $\Ord(\bm q^{-2})$.
The match of the IR divergent part can be extended to subleading order in two equivalent ways. On the one hand we can extend 
the $\Ord(G)$ EFT Hamiltonian by subleading ($\Ord(|\bm q|)$) terms which are adjusted for such that  the equality above holds 
to $\Ord(|\bm q|)$. They are related to the fact that the relation \eqref{SpinEval} needed to express $d_{\rm B}$ in terms of 
the classical spin vector hold only to leading order in the classical limit.
Conversely, we can include next-to-leading order terms in Eq.~\eqref{SpinEval} and verified that such terms extend the equality 
\eqref{box_stuff} to $\Ord(|\bm q|)$. 
The details of this subleading correction are not important; their only effect is to restore the equality of the IR-divergent part of 
the EFT and full-theory one-loop amplitudes without altering the classical EFT Hamiltonian through $\Ord(G^2)$.

The coefficients $d_2^A$ of the $\Ord(G^2)$ Hamiltonian are found from Eq.~\eqref{IDamplitudes} with the EFT amplitude 
coefficients  in \eqn{eq:EFT_2PM_S1S2} and full theory coefficients in Eqs.~\eqref{eq:2PMCoefficients} and \eqref{eq:a_covRelations}. 
The spin-independent term and spin-orbit interaction coefficients $d^{A}_{i}$ are
\begin{align}
d^{(0)}_{2}& =
\frac{\nu^2 m^3}{\xi  \gamma^2} \left[ \frac{3}{4} \left(1-5 \sigma^2\right)
-\frac{4 \nu  \sigma {} \left(1-2 \sigma^2\right)}{\gamma  \xi } -
\frac{\nu ^2 (1-\xi) \left(1-2 \sigma ^2\right)^2}{2 \gamma ^3  \xi ^2}
\right]
\,, \nn \\[0.5ex]
d^{(1,1)}_{2}& =
\frac{\nu E}{4 \xi \gamma^2 m_1}\biggl[
-\frac{(5\sigma^2-3) \sigma}{\sigma^2-1}(4m_1+3m_2)
+\frac{2(2\sigma^2-1)}{\sigma^2-1}
\left(2\sigma E {} \left(\gamma_1^{-2}+\gamma_2^{-2} \right)
+\gamma_1^{-1} (1-2\sigma^2)m_2
\right) \nn \\[0.5ex]
&\quad+\frac{4\nu(6\sigma^2-1)E}{\xi \gamma^2}
\biggr]-\frac{1}{m_1^2(\gamma_1+1)}d_2^{(0)}
\,.
\label{eq:2PM_potential}
\end{align}
The variables used here are defined in Eq.~\eqref{var_defs} and below Eq.~\eqref{box_coefs_norm} 
and $d_2^{(1,2)}$ is obtained by interchanging $(m_1,\gamma_1)$ and $(m_2,\gamma_2)$ in $d_2^{(1,1)}$.
The expressions the coefficients $d_2^{(2,1)},d_2^{(2,2)},d_2^{(2,3)}$ of the spin-bilinear operators 
are more complicated so we provide them in the Mathematica ancillary file \texttt{coefficients.m}.  

The coefficient functions of the position-space potential are readily obtained through the relations in \tab{table:operator_summary}
\begin{align}
c_2^{(0)} &= d_2^{(0)}\,,\qquad
c_2^{(1,i)} = -2d_2^{(1,i)}\,,\nonumber\\
c_2^{(2,1)} &= -8d_2^{(2,1)}\,,\quad
c_2^{(2,2)} = 2d_2^{(2,1)}-2d_2^{(2,2)}\,,\quad
c_2^{(2,3)} = -2d_2^{(2,3)} \,.
\label{eq:2PM_potential_real}
\end{align}

We can verify that the probe limit of our all-orders-in-velocity
Hamiltonian reproduces the results of
Ref.~\cite{TestBodyQuadraticSpin}, where the Hamiltonian was
constructed in this limit up to quadratic order in spins. (See also
Refs.~\cite{SpinTestMassAdditional,Spin2PM_testBH}.) Taking $m_1, |\bm
p| \ll m_2$, so we have a spinning probe particle 1 in a Kerr
background by particle 2 and mapping those results into isotropic
gauge\footnote{We thank Justin Vines for sharing this result.}, the
probe-limit real-space potential up to bilinear order in spin is
\begin{align}
V(\bm p, \bm r, \bm S_i) =&
\left(
\frac{Gm_2}{m_1}\frac{(2 \gamma_1 + 1)}{\gamma_1 (\gamma_1 + 1)}
-\frac{G^2m_2^2}{m_1}\frac{(9 \gamma_1^3 + 7 \gamma_1^2 + 2 \gamma_1 + 2)}{2 r \gamma_1^3 (\gamma_1 + 1)}
\right)\,\frac{\bm L\cdot \bm S_1}{r^3}
+\left(2G-\frac{6G^2m_2}{r} \right) \frac{\bm L\cdot \bm S_2}{r^3} \nn \\
&+G\frac{(2 \gamma_1 - 1)  
}{\gamma_1}\,\frac{1}{r^5}\,(3 \bm r\cdot \bm S_1\, \bm r\cdot \bm S_2 - \bm r^2 \bm S_1\cdot \bm S_2) \nn \\
&
+G^2 m_2\frac{
	(-40 \gamma_1^4 -28 \gamma_1^3 +14 \gamma_1^2 +6 \gamma_1 +6 )}{2 \gamma_1^3 (\gamma_1 + 1)}\,
\frac{\bm r\cdot \bm S_1\, \bm r\cdot \bm S_2}{r^6} \nn \\
&+G^2 m_2\frac{
	(16 \gamma_1^4 + 13 \gamma_1^3 - 5 \gamma_1^2 - 2 \gamma_1 - 2)
	}{2 \gamma_1^3 (\gamma_1 + 1)} \, \frac{\bm S_1\cdot \bm S_2}{r^4} \nn \\
&+G^2 \frac{m_2}{m_1^2}\frac{ (\gamma_1 - 1)}{2 \gamma_1 (\gamma_1 + 1)^2}\,\frac{\bm p\cdot \bm S_1\, \bm p\cdot \bm S_2}{r^4}
\, .
\label{Vrealspace}
\end{align}
Only the leading contribution in small $m_1/m_2$ is kept for each type of spin-dependent monomial.
Mapping the expression above to the form in Eq.~\eqref{eq:H_r} and
Eq.~\eqref{eq:V_r} yields the coefficient $c^A_i$.  We can see that
the $\Ord(G)$ position-space potential contains only the combination
$3 \bm r\cdot \bm S_1\, \bm r\cdot \bm S_2 - \bm r^2 \bm S_1\cdot \bm
S_2$, in agreement with our result, cf. the discussion below
Eq.~\eqref{eq:1PM_potential_real}.  Eq.~\eqref{Vrealspace} is in
complete agreement with the probe limit of our potential.

We note that the combination $3 \bm r\cdot \bm S_1\, \bm r\cdot \bm S_2 - \bm r^2 \bm S_1\cdot \bm S_2$, as a symmetric traceless tensor in $\bm r$, can be interpreted 
as the quadrupole of the two-particle system which is is not induced by the quadrupole of the individual constituents and 
thus it is entirely due to their interaction. It may therefore be natural to organize the spin dependence in terms of this operator, 
even at higher orders in $G$.  At ${\cal O}(G^n)$ it can also be identified as the traceless-symmetric structure in momentum space Hamiltonian,
$|\bm q|^{n-3}(3 \bm q\cdot \bm S_1\, \bm q\cdot \bm S_2 - \bm q^2 \bm S_1\cdot \bm S_2)$; while the second term drops out 
of the ${\cal O}(G)$ Hamiltonian (because it represents a contact interaction), the Fourier-transform of the remainder, including 
the additional $\bm q$-dependent factors, is the operator on the second line of Eq.~\eqref{Vrealspace}.

\subsection{Summary of EFT formulas}
\label{sec:EFT_summary}

Here we collect the formulae that define the EFT constructed in this section and its coefficient functions determined 
by matching its amplitudes with those of the full theory. 

Our real-space Hamiltonian is 
\begin{align}
H = \null &  H^{(0)}(r^2, p^2)+H^{(1,i)}(r^2, p^2) \bm L\cdot \bm S_i
\nn \\
&
+ H^{(2,1)}(r^2, p^2) \bm r\cdot \bm S_1 \bm r\cdot \bm S_2
+ H^{(2,2)}(r^2, p^2) \bm S_1 \cdot \bm S_2
+ H^{(2,3)}(r^2, p^2) \bm p\cdot \bm S_1 \bm p\cdot \bm S_2 +\dots \,,
\label{Hamiltonian_general}
\end{align}
where the ellipsis stand for terms quadratic in the spin of each particle as well as for terms of higher orders in spin.
This Hamiltonian corresponds to the one in
Eqs.~\eqref{eq:H_r} and \eqref{RealSpaceOperators}.
As usual, the coefficients in \eqn{Hamiltonian_general} can be expanded in Newton's constant $G$, as 
in \eqn{eq:V_r}:
\begin{eqnarray}
H^{(0)}(r^2, p^2)&=&\sqrt{\bm p ^2 + m_1^2}+\sqrt{\bm p^2 + m_2^2} + \frac{G}{r} c_1^{(0)}(\bm p^2)+ \left(\frac{G}{r} \right)^2 c_2^{(0)}(\bm p^2)+ {\cal O}(G^3) \,, \nn
\\
H^{(1,i)}(r^2, p^2)&=&  \frac{1}{r^2}\left[\frac{G}{r} c^{(1, i)}_{1}(\bm p^2)+ \left(\frac{G}{r} \right)^2 c^{(1, i)}_{2}(\bm p^2)
+{\cal O}(G^3)\right], \nn
\\
H^{(2,1)}(r^2, p^2)&=&\frac{1}{r^4}\left[\frac{G}{r} c^{(2,1)}_{1}(\bm p^2)+ \left(\frac{G}{r} \right)^2 c^{(2,1)}_{2}(\bm p^2)
+{\cal O}(G^3)\right] ,
\\
H^{(2,2)}(r^2, p^2) &=&\frac{1}{r^2}\left[\frac{G}{r} c^{(2,2)}_{1}(\bm p^2)+ \left(\frac{G}{r} \right)^2 c^{(2,2)}_{2}(\bm p^2)
+{\cal O}(G^3)\right] , \nn
\\
H^{(2,3)}(r^2, p^2) &=&\frac{1}{r^2}\left[\frac{G}{r} c^{(2,3)}_{1}(\bm p^2)+ \left(\frac{G}{r} \right)^2 c^{(2,3)}_{2}(\bm p^2)
+{\cal O}(G^3)\right] .
 \nn
\end{eqnarray}
The first coefficient function, $H^{(0)}(r^2, p^2)$, is the
Hamiltonian that describes the gravitational interaction of spinless
particles; the remaining ones give systematically spin-dependent interactions.  
The coefficients $c_i^A$, are the same as in Eq.~\eqref{eq:V_r}.

From Eqs.~\eqref{eq:1PM_potential} and~\eqref{eq:1PM_potential_real} we have the explicit values
of the $\Ord(G)$ position-space Hamiltonian coefficients:
\begin{align}
c_1^{(0)} &= \frac{m^2\nu^2}{\xi \gamma^2}(1-2\sigma^2) \,, \nn \\
c_1^{(1,i)} &= \frac{\nu}{\xi \gamma^2}\frac{2\sigma E}{m_i} +\frac{1}{m_i^2 (\gamma_i+1)} c_{1}^{(0)} \,, \nn \\
c_1^{(2,1)} & = - \frac{3\nu}{\xi \gamma^2}(1-2\sigma^2) 
-\left(\frac{3}{m_1 (\gamma_1+1)} + \frac{3}{m_2 (\gamma_2+1)} \right)\frac{2\sigma  \bm p^2}{E \xi}
-\frac{3\bm p^2}{m_1^2 m_2^2 (\gamma_1+1)(\gamma_2+1)} c_{1}^{(0)} \,, \hskip 2 cm  \nn  \\
c_1^{(2,2)} &= -\frac{1}{3} c_1^{(2,1)}\,, \nn \\
c_1^{(2,3)} & = 0 \,.
\label{eq:1PM_cCoeffs}
\end{align}
Similarly, the $\Ord(G^2)$ terms are obtained from \eqns{eq:2PM_potential}{eq:2PM_potential_real}. The spin-independent 
and spin-orbit ones are given by,
\begin{align}
c_2^{(0)} &= \frac{\nu^2 m^3}{\xi \gamma^2} \left[ \frac{3}{4} \left(1-5 \sigma^2\right)
-\frac{4 \nu  \sigma {} \left(1-2 \sigma^2\right)}{\gamma  \xi } -
\frac{\nu ^2 (1-\xi) \left(1-2 \sigma ^2\right)^2}{2 \gamma ^3  \xi ^2}
\right]
\,, \nn \\[0.5ex]
c_2^{(1,1)} & = -
\frac{\nu E}{2\xi \gamma^2 m_1}\biggl[
-\frac{(5\sigma^2-3) \sigma}{\sigma^2-1}(4m_1+3m_2)
+\frac{2(2\sigma^2-1)}{\sigma^2-1}
\left(2\sigma E {} \left(\gamma_1^{-2}+\gamma_2^{-2} \right)
+\gamma_1^{-1} (1-2\sigma^2)m_2
\right) \nn \\[0.5ex]
&\quad+\frac{4\nu (6\sigma^2-1)E}{\xi \gamma^2}
\biggr] + \frac{2}{m_1^2(\gamma_1+1)}c_2^{(0)}
\,, \nn \\
c_2^{(1,2)} & = c_2^{(1,1)} \Bigr|_{m_1 \leftrightarrow m_2, \gamma_1 \leftrightarrow \gamma_2} \ .
\label{eq:2PM_cCoeffs}
\end{align}
The remaining lengthier coefficients $c_2^{(2,1)}$, $c_2^{(2,2)}$, and $c_2^{(2,3)}$
are found in the ancillary file~\cite{AttachedFile}.
The variables used here are defined in Eq.~\eqref{var_defs} and below Eq.~\eqref{box_coefs_norm}:
\begin{align}
\gamma  &= \frac{E}{m}\, , 
\hskip 1 cm \gamma_1 = \frac{E_1}{m_1}\,, 
\hskip 1 cm  \gamma_2 = \frac{E_2}{m_2} \,, 
\hskip 1 cm 
 \xi = \frac{E_1 E_2}{E^2}\,, 
 \hskip 1 cm 
 \sigma = \frac{p_1 \cdot p_2}{m_1 m_2} \,,
 \nn\\
 E & = E_1+ E_2 
\hskip 1 cm
m  = m_1+ m_2 
\hskip 1 cm
\nu = \frac{m_1 m_2}{m^2}
\end{align}
where $E_1$ and $E_2$ are the energies of the two particles.

The EFT amplitude coefficients can be obtained from the real-space Hamiltonian coefficients from
\tab{table:operator_summary} and \eqns{eq:a1}{eq:EFT_2PM_S1S2}.  At $\Ord(G)$ we have
\begin{align}
& a^{(0)}_1= -c^{(0)}_1 \,, \hskip 3.2cm 
a^{(1,1)}_1 = c^{(1,1)}_1\,, \hskip 1.5 cm 
a^{(1,2)}_1 = c^{(1,2)}_1\,, \nn \\ 
& a^{(2,1)}_1 = \frac{1}{3} c^{(2,1)}_1 = -c^{(2,2)}_1  \,, \hskip 1.2 cm
a^{(2,2)}_1 = 0\,, \hskip 2 cm
a^{(2,3)}_1 = 0\,,
\label{eq:aEFT_1PM_Summary}
\end{align}
while $\Ord(G^2)$ we have
\begin{align}
a^{(0)}_{2} = \null &
-c^{(0)}_2+\frac{1}{2\xi E}\, A_0\left[\left(c^{(0)}_1 \right)^2\right] 
, \nn \\
a^{(1,i)}_{2} = \null &
\frac{1}{2} c^{(1,i)}_2 -\frac{1}{2\xi E}\, A_1\left[c^{(0)}_1 c^{(1,i)}_1\right]
, \nn \\
a^{(2,1)}_{2} = \null &
\frac{1}{8} c^{(2,1)}_2 
-\frac{1}{6\xi E}\, A_3\left[c_1^{(0)} c^{(2,1)}_1\right]
+\frac{\bm p^2}{2\xi E}\, A_2\left[c_1^{(1,1)} c^{(1,2)}_1\right]
+\frac{\xi E}{24} (c_1^{(1,1)} + c_1^{(1,2)}) c^{(2,1)}_1 
, \nn \\[4pt]
a^{(2,2)}_{2} = \null &
\frac{1}{8} c^{(2,1)}_2 + \frac{1}{2}c^{(2,2)}_2
+\frac{1}{6\xi E}\, A_2\left[c_1^{(0)} c^{(2,1)}_1\right]
-\frac{\bm p^2}{\xi E}\, A_2\left[c_1^{(1,1)}c^{(1,2)}_1\right]
+\frac{\xi E}{24} (c_1^{(1,1)}+c_1^{(1,2)}) c^{(2,1)}_1 
, \nn \\[4pt]
a^{(2,3)}_{2} = \null & 
\frac{1}{2} c^{(2,3)}_2 
-\frac{\xi E}{3 \bm p^4}\, \left[c_1^{(0)} c^{(2,1)}_1\right]
+\frac{1}{\xi E}\, A_2\left[c_1^{(1,1)}c^{(1,2)}_1\right]
-\frac{\xi E}{6\bm p^2} (c_1^{(1,1)}+c_1^{(1,2)}) c^{(2,1)}_1 
.
\label{eq:aEFT_2PM_Summary}
\end{align}
where
\begin{align}
A_0[X] &= 
\left[(1-3\xi) + 2 \xi^2E^2 \partial  \right] X
\,, \nn \\
A_1[X] &= 
\left[(1-3\xi) + \frac{2 \xi^2 E^2}{\bm p^2} + 2 \xi^2 E^2\partial \right] X
\,,\nn \\
A_2[X] &= 
\left[\frac{1}{4}(1-3\xi) + \frac{ \xi^2 E^2}{\bm p^2} + \frac{1}{2} \xi^2 E^2\partial \right] X
\,, \nn \\
A_3[X] &= 
\left[\frac{3}{4}(1-3\xi) + \frac{ \xi^2 E^2}{\bm p^2} + \frac{3}{2} \xi^2 E^2\partial \right] X
\,,
\end{align}
The explicit values of the amplitude coefficients $a^A_i$ summarized
in \sect{tree_and_loop_summary} are much simpler than the
corresponding $c_i^A$ Hamiltonian coefficients.  This is not surprising
given that Hamiltonians are gauge dependent, while amplitudes are gauge 
independent. It would be interesting to see if there exists a  a formulation of the EFT 
which leads to a Hamiltonian which exhibits the simplicity of the scattering amplitude.

\section{Physical Observables}
\label{sec:angle}

The two-body Hamiltonian constructed in the previous section allows us
to compute observables of scattering processes and of bound motion of spinning bodies.  
We will focus here on scattering observables, to point out a simple connection 
to the eikonal phase.
Unlike the spinless case where the motion occurs in a plane and therefore there is 
only a single scattering angle, the spinning  case has several interesting observables. 
Since the orbital angular momentum is no longer conserved, the scattering process is 
three-dimensional and thus there are two deflection angles. Moreover, since the spins 
are not separately conserved, they also change in a scattering process. 
As for the spinless case, these scattering observables are useful
stepping stones for the construction of effective one-body
Hamiltonians~\cite{EOB,EOBSpinDamour,EOBSpin} which can be used to evaluate
bound-state dynamics.  In this case, the dynamics is similarly
rich with three-dimensional motion and multiple angles, leading to
nontrivial modulation of gravitational wave signals which may be used
to determine the properties of the binary constituents.

\subsection{Classical mechanics of particles with spin}
\label{sec:classical_EoM}

\begin{figure}[tb]
\begin{center}
\includegraphics[scale=.7]{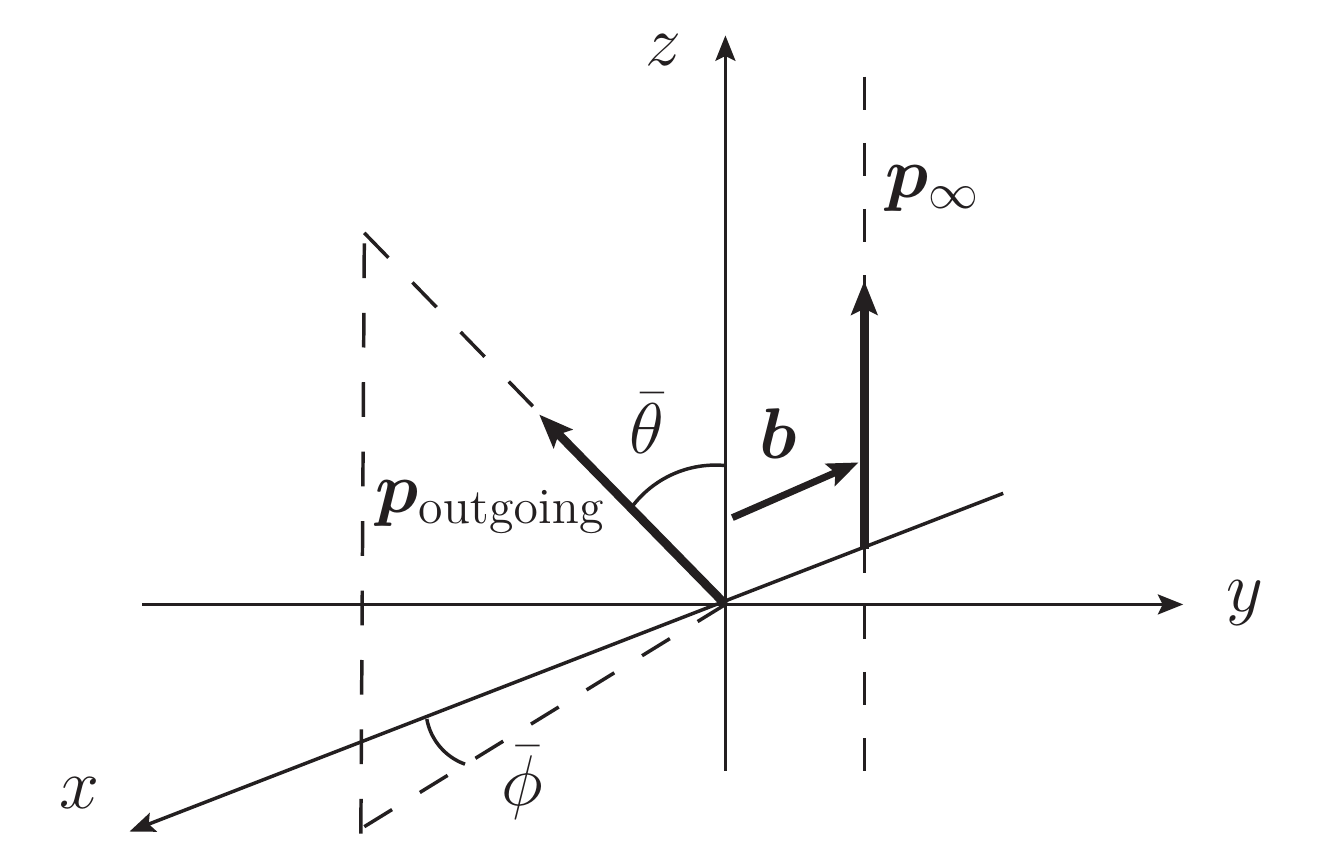}
\end{center}
\vskip -.5 cm
\caption{\small A representation of a scattering process in the presence of spin. 
The center of mass is placed at the origin; $\bm p_\infty$ is the incoming 
momentum, $\bm p_{\rm outgoing}$ is the outgoing momentum and $|\bm p_{\rm outgoing}|=|\bm p_\infty|$ . 
Its direction has been translated to pass through the origin to facilitate the depiction of the two scattering angles 
${\bar\theta}$ and ${\bar\phi}$.
}
\label{AngleFigure}
\end{figure}

Consider the general problem of an arbitrary Hamiltonian describing
the interaction of two particles with rest-frame spin three-vectors
${\bm S}_1$ and ${\bm S}_2$ in their center of mass frame, $H=H({\bm
  r}, {\bm p}, {\bm S}_1, {\bm S}_2)$.  In our case we will truncate to a
fixed number of spin vectors, namely bilinear in spin.  This is
consistent, since terms in the solutions to the equations of motion
with a certain number of spin vectors do not receive contributions
from terms in the Hamiltonian with a larger number of spin vectors.
While, as usual, $\bm r$ and $\bm p$ are canonically-conjugate to each
other, the spin variables do not have a natural canonical
conjugate. To derive the equations of motion we use the fact that they
must generate $SO(3)$, so
\begin{equation}
\{{S}^i_a,\, {S}^j_b\}=\delta_{ab} \, \epsilon^{ijk} S^k_a\,, \hskip 1.5 cm a,b=1,2 \,.
\label{PBspin}
\end{equation}
where $\{A,B\}$ is the Poisson bracket of $A$ and $B$.  One way to
understand this relation is by recalling that in the complete theory
the spin vector is given by the expectation value of the Lorentz
generator. Applying Ehrenfest's theorem to the Lorentz algebra leads
to \eqn{PBspin}.  A similar strategy for deriving the equations of
motion for the spin variables is found in, for example,
Refs.~\cite{EOBSpinDamour, PortoReview}.

The equations of motion are then,
\begin{equation}
\dot {\bm r} = \frac{\partial H}{\partial {\bm p}} \,,
\hskip 1.5 cm 
\dot {\bm p} = -\frac{\partial H}{\partial {\bm r}} \,,
\hskip 1.5 cm 
\dot {\bm S}_a = -{\bm S}_a \times \frac{\partial H}{\partial {\bm S}_a} ~,~~~a=1,2 \, ,
\label{general_eom}
\end{equation}
where in the spin equation of motion no summation is implied on the right-hand side.
One can use spherical polar coordinates but for the purpose of 
finding the impulse $\Delta \bm p$ we find it convenient to use Cartesian coordinates.
One may either solve the equations of motion for coordinates and momenta and spins 
as a function of time or one may choose the $z$ coordinate as effective time parameter.

There are two conservation laws that aid the construction of classical
solutions. These fix the energy and the total angular momentum in
terms of their asymptotic values:
\begin{align}
E & = H({\bm r}, {\bm p}, {\bm S}_1, {\bm S}_2)=
\sqrt{p_\infty^2+m_1^2}+\sqrt{p_\infty^2+m_2^2} \,,
\cr
{\bm I} &= {\bm r}\times {\bm p}+{\bm S}_1 + {\bm S}_2 =  
{\bm L}_\text{in} + {\bm S}_{1, \text{in}} + {\bm S}_{2,\text{in}} \,.
\label{conservation}
\end{align}
where ${\bm p}_\infty = p_\infty {\bm e}_z$ is the incoming momentum
at infinity\footnote{The notation $p_\infty$ is sometimes defined differently, as
  in Ref.~\cite{Damour:2019lcq}.}, as indicated in \fig{AngleFigure}.  
We take the orbital angular momentum at infinity to be
\begin{equation}
{\bm L}_\text{in} = {\bm b}\times {\bm p}_\infty  = b p_\infty {\bm e}_y \, ,
\end{equation}
where $\bm b = -b\bm e_x$ and $b$ is the impact parameter.  Note that under the conservative dynamics,
the spins cannot exchange energy with the remainder of the system because under
the equations of motion,
\begin{equation}
\frac{d |{\bm S}_a|^2}{dt} = 2 {\bm S}_a \cdot \dot {\bm S}_a 
= - 2 {\bm S}_a \cdot \left({\bm S}_a \times \frac{\partial H}{\partial {\bm S}_a}\right) = 0\,,
\label{S2Preserve}
\end{equation}
where, as in the spin equation of motion \eqref{general_eom}, the
index $a=1,2$ is not summed.

We solve the equations of motion perturbatively in Newton's constant, i.e. we search for a 
solution for coordinates, momenta and spins of the form
\begin{align}
\bm r(t) &= \bm r_0(t) + G \bm r_1(t)+ G^2 \bm r_2(t)+\dots \ , 
\nonumber\\
\bm p(t) &= \bm p_0(t) + G \bm p_1(t)+ G^2 \bm p_2(t)+\dots \ , 
\\
\bm S_a(t) &= \bm S_{a,0}(t) + G \bm S_{a,1}(t)+ G^2 \bm S_{a,2}(t)+\dots \ .
\nonumber
\end{align}
Replacing them in the equations of motion \eqref{general_eom} leads to iterative relations
between the time derivative of the $n$-th term in the expansions above and all the 
lower-order terms. The $\Ord(G^0)$ terms describe the motion of a free spinning particle in flat space,
i.e. a straight line fixed by the initial momentum, the impact parameter and initial spin.
The first-order differential equations for the higher-order terms can be integrated; the relevant 
boundary conditions are that $\bm r_{n\ge 1}$, $\bm p_{n\ge 1}$ and $ \bm S_{a,n\ge 1}$ vanish 
at $t=-\infty$. The contribution of each order in $G$ to an observable $O$, such as the 
impulse and spin kick, is then
\begin{equation}
\Delta O_n =   \int_{-\infty}^\infty dt\,\frac{dO_n}{dt} = O_n(t=+\infty) - O_n(t=-\infty) \, ,
\label{observables}
\end{equation}
with the complete result being their sum weighted with the appropriate powers of $G$.

The incoming and outgoing trajectories approach straight lines at
$t=\pm\infty$, respectively, which are along the incoming and outgoing
momenta. Thus, the polar and azimuthal scattering angles can be read
off in terms of their components or, alternatively in terms of the
incoming momentum and the impulse. Starting with an initial momentum
along some generic direction defined by the angles $\theta_0$ and
$\phi_0$,
\begin{align}
\bm p_{\textrm{incoming}} = p_{\infty}\sin{\theta}_0 \cos{\phi}_0\, \bm e_x 
                                           +p_{\infty}\sin{\theta}_0 \sin{\phi}_0 \bm e_y 
					 +p_{\infty}\cos {\theta}_0 \bm e_z \ ,
\label{pin}
\end{align}
then the outgoing momentum, expressed in terms of the scattering angles ${\bar\theta}$ and ${\bar\phi}$, is					 
\begin{align}
\bm p_{\textrm{outgoing}} &= \bm p_{\textrm{incoming}} + \Delta \bm p 
\nn\\
&= p_{\infty}\sin(\theta_0+\bar{\theta}) \cos(\phi_0+\bar{\phi})\, \bm e_x 
  +p_{\infty}\sin(\theta_0+\bar{\theta}) \sin(\phi_0+\bar{\phi}) \bm e_y 
+p_{\infty}\cos(\theta_0+\bar{\theta}) \bm e_z \ .
\label{pout}
\end{align} 
The scattering angles can then be easily extracted in terms of the components of the impulse $ \Delta \bm p $ and the incoming 
momentum. It is worth noting that, if e.g. $\theta_0=\pi$ and $\phi_0=0$, 
then the azimuthal angle evaluated at finite Newton's constant exhibits a discontinuity in the limit $G\rightarrow 0$. Indeed, for 
such a value of $\theta_0$, 
the incoming momentum has vanishing components along $\bm e_x $ and $\bm e_y $. The the components of the outgoing 
momentum  in these directions are both ${\cal O}(G)$, leading to $\tan\bar\phi = {\cal O}(1)$. This discontinuity is unphysical and 
may be easily remedied by slightly changing the initial conditions such that the incoming momentum is not parallel to a 
coordinate axis.

\subsection{Impulse, spin kick and the eikonal phase}

As we will describe in some detail in Ref.~\cite{EoMEikonalPaper}, by
solving Hamilton's equations we find a remarkably simple hidden
structure, tying the solution to the gauge-invariant amplitudes.  In
the case of spinless particles there is a direct link between the
physical observables and gauge-invariant quantities extracted from
scattering amplitudes~\cite{DamourTwoLoop, 3PMLong,
  AmplitudePotential,AmplitudePotentialBjerrum}.  Indeed,
Refs.~\cite{OConnellObservables, MOV} provide a general formalism for
systematically extracting physical observables from amplitudes and
their unitarity cuts.  For spinning particles such a relation has been
found at lowest order in $G$ found in Ref.~\cite{MOV} and further applied in Ref.~\cite{GOV_2}.  Here we show that
such relations appear to be general by rewriting the solution for the
impulse and spin kick at $\Ord(G^2)$ in terms of appropriate
derivatives acting on the eikonal phase~\cite{EikonalPapers},
showing that the notion of the eikonal phase naturally generalizes 
to the case of spin.

As for the spinless case, we define the eikonal phase $\chi = \chi_1   
+ \chi_2 + \cdots$, by Fourier
transforms of appropriate parts of amplitudes.
The $\Ord(G)$ contribution to eikonal phase, in particular,
is just the Fourier transform of the tree amplitude to impact parameter space\footnote{At 
higher orders in $G$ the $\bm b$ in the eikonal formula could differ
from the geometric impact parameter~\cite{EikonalImpact}, but we can set aside this 
distinction through $\Ord(G^2)$.} ,
\begin{equation}
\chi_1 = \frac{1}{4m_1m_2\sqrt{\sigma^2-1}}
\int \frac{d^{2-2 \eps}\bm{q}}{(2\pi)^{2- 2\eps} }e^{-i\bm{q}\cdot\bm{b}}\mathcal{M}^{\tree}(\bm{q}) \, ,
\label{FourierTree}
\end{equation}
and the $O(G^2)$ contributions to eikonal phase is given simply 
by the triangle contributions to the one-loop amplitudes,
\begin{equation}
\chi_2 =\frac{1}{4m_1m_2\sqrt{\sigma^2-1}}
\int \frac{d^2\bm{q}}{(2\pi)^2}e^{-i\bm{q}\cdot\bm{b}} 
\mathcal{M}^{\bigtriangleup+\bigtriangledown}(\bm{q})\,,
\label{FourierTriangle}
\end{equation}
where $\mathcal{M}^{\bigtriangleup+\bigtriangledown}(\bm{q})$ is given
in Eq.~\eqref{OneloopTriAmpl}.  
Since the goal is to compare with results obtained from Hamilton's equations, in the formulas above we must use 
the amplitudes in expressed in terms of the canonical  rest-frame spins.
The tree-level amplitude in this form is given in Eq.~\eqref{eq:M_1PM_full_exp} while 
the triangle part of the one-loop amplitude, $\mathcal{M}^{\bigtriangleup+\bigtriangledown}(\bm{q})$,  
is given by the first two lines of Eq.~\eqref{eq:M_2PM_full_exp}.
The coefficient of each spin-dependent structure is 
given in Eqs.~\eqref{eq:a_covRelations},~\eqref{eq:1PMCoefficients}, and~\eqref{eq:2PMCoefficients}.
Carrying out the Fourier transforms, give the
following remarkably compact expressions,
\begin{align}
\chi_{1} &= \frac{\xi E}{|\bm p|} \biggl[
-a^{(0)}_1 \ln \bm b^2 - \frac{2a^{(1,i)}_1}{\bm b^2} (\bm p \times \bm S_i) \cdot \bm b
+a^{(2,1)}_1\left(\frac{2}{\bm b^2} \bm S_{1\perp}\cdot \bm S_{2\perp} 
-4\frac{\bm S_{1\perp}\cdot \bm b \,\bm S_{2\perp} \cdot \bm b}{\bm b^4} \right)
\biggr] \,, \nonumber  \\[7pt]
\chi_{2} &= \frac{\pi  \xi E}{|\bm p|} \bigg[
\,\frac{a^{(0)}_2}{|\bm b|}
-\left(a^{(2,2)}_2 \bm S_1\cdot \bm S_2 + a^{(2,3)}_2 \bm p\cdot \bm S_1\, \bm p\cdot \bm S_2
\right)\,\frac{1}{|\bm b|^3} \label{Eikonal} \\
&\hskip 2.0 cm 
\null - \frac{a^{(1,i)}_2}{|\bm b|^3} (\bm p \times \bm S_i) \cdot \bm b
+a^{(2,1)}_2\left(\frac{1}{|\bm b|^3} \bm S_{1\perp}\cdot \bm S_{2\perp} 
-3\frac{\bm S_{1\perp}\cdot \bm b \,\bm S_{2\perp}\cdot \bm b}{|\bm b|^5} \right)
\bigg] \,, \nn
\end{align}
\vskip .2 cm 
\noindent
where we define,
\vskip .01 cm 
\begin{equation}
\bm S_{\perp i} \equiv \bm S_{i} -  \bm p \, 
\frac{\bm S_{i} \cdot \bm p} {\bm p ^2} \,.
\end{equation}
Note that the $\bm p$ here, following from the notation in our amplitudes, 
is the incoming momentum $\bm p_{\infty}$ in \Fig{AngleFigure}.~which should not be confused with the intermediate $\bm p$ in \Sec{sec:classical_EoM}.
In \eqn{Eikonal}, 
we can also use the model-independent expressions of $a^A_i$ given in terms of the
$c_i^A$ coefficients appearing in the potential via
\eqns{eq:aEFT_1PM_Summary}{eq:aEFT_2PM_Summary}.  Remarkably the simple
gauge-invariant functions in \eqn{Eikonal} encode the physical
information for classical scattering processes through $\Ord(G^2)$,
including spinless, spin-orbit and spin$_1$-spin$_2$ interactions in a
form valid to all orders in velocity.

To extract the impulse and spin kick from the eikonal phase, consider
the kinematic configuration shown in \fig{AngleFigure} where
\begin{equation}
\bm p = (0,0, p_\infty)\,, \hskip 1.5 cm
\bm b = (-b,0, 0)\,, \hskip 1.5 cm
\bm L= \bm L_{\rm in} = p_\infty (0 , b,0)\,.
\label{InitialValues}
\end{equation}
As mentioned before,  $\bm p$ here represents the incoming momentum $\bm p_\infty$.
By evaluating the eikonal phase on this kinematics and comparing to the solution of the equations of motion, we find that
impulse in the $x$-$y$ plane is~\cite{EoMEikonalPaper},
\begin{align}
\Delta \bm p_\perp   &= -\{ \bm P_\perp, \chi \}
-\frac{1}{2}\,\{\chi, \{\bm P_\perp, \chi\} \}
-\mathcal{D}_{SL}\left(\chi, \{\bm P_\perp, \chi\} \right)
+\frac{1}{2}\,\{\bm P_\perp,\mathcal{D}_{SL}\left( \chi,  \chi \right) \} \,,
\label{Deltapxy}
\end{align}
and the spin kick for all three components is
\begin{align}
\Delta \bm S_i &= -\{ \bm S_i, \chi \}
-\frac{1}{2}\,\{\chi, \{\bm S_i, \chi\} \}
-\mathcal{D}_{SL}\left(\chi, \{\bm S_i, \chi\} \right)
+\frac{1}{2}\,\{\bm S_i,\mathcal{D}_{SL}\left( \chi,  \chi \right) \} \,,
\label{SpinKick}
\end{align}
where both relations are valid up to $\Ord(G^2)$ and
\begin{align}
\{\bm P_\perp, f \} & \equiv -\nabla_{\bm b} f \,, \hskip 2 cm 
\{S_a^i, f \}  \equiv \epsilon^{ijk} \, \frac{\partial f}{\partial S^j_a} S_a^k \,, 
\hskip 1.2 cm \hbox{($a$ not summed)},\nn \\[7pt]
\mathcal{D}_{SL}\left(f, g \right) & \equiv 
 - \sum_{a=1,2} \,\epsilon^{ijk} S^k_a\, \frac{\partial f}{\partial S^i_a}\,\frac{\partial g}{\partial L^j} 
= \frac{1}{\bm p^2} \sum_{a=1,2} 
      \biggl( \frac{\partial f}{\partial S_a^j} \frac{\partial g}{\partial b^j} \, \bm S_a \cdot \bm p 
            - p^j \frac{\partial f}{\partial S_a^j} \, \bm S_a \cdot \nabla_{\bm b} g \biggr)\, ,
\label{Brackets}
\end{align}
where $f$ and $g$ depend on $\bm S$, $\bm p$ and $\bm b$.  In the brackets $\bm p$ should be taken
as inert.
To use the first form of $D_{SL}$, we replace $\bm b$ with $\bm L$ via
\begin{equation}
\bm b = \frac{1}{\bm p^2} \, \bm p \times \bm L \,,
\end{equation}
consistent with $\bm p \cdot \bm b =0$.  After evaluating the
derivatives, we substitute in the values at $t = - \infty$ given in
\eqn{InitialValues} to obtain the impulse and spin kick.

While \eqn{Deltapxy} does not directly give the impulse in the $z$ direction,
this quantity follows from energy conservation,
\begin{equation}
\Delta p_z = - \frac{1}{2 |\bm p|} (\Delta \bm p)^2\,,
\end{equation}
which can be iteratively solved as a series in $G$.  
From the structure of the brackets and $D_{SL}$ in \eqn{Brackets},
the value $\bm S_a^2$ is preserved, as required from the equation
of motion \eqref{S2Preserve}.

The expressions for the impulse and spin kick in
\eqns{Deltapxy}{SpinKick} then matches the derived results from the
equations of motion, as we have explicitly verified.  Together 
with Eqs.~\eqref{pout} and \eqref{pin}, they relate the scattering angles 
and the eikonal phase; in the limit of vanishing spin this relation reproduces 
the standard one~\cite{EikonalPapers}, implying the proportionality of the 
sine of half of the scattering  angle and the derivative of the eikonal phase 
with respect to the absolute value of the impact parameter.
An ancillary Mathematica text file~\cite{AttachedFile} contains the explicit values
of impulse and spin kick for the initial conditions in~\eqn{InitialValues}.

The expressions in \eqns{Deltapxy}{SpinKick} strongly suggest an all orders
generalization.
For example, one matches
the above expressions to the order that they are valid, by
\begin{equation}
\Delta \mathcal O =  e^{-i \chi \mathcal D} [ \mathcal O, e^{i \chi \mathcal D} ]\,,
\label{resummation}
\end{equation}
where the commutator is related to the brackets in \eqn{Brackets} by $[f,g] =
i\{f,g\}$ and
\begin{equation}
\chi\, \mathcal D\, g \equiv \chi g + i D_{SL}(\chi, g)\,,
\end{equation}
for any function $g$ appears to the right of $\mathcal D$. 
Eq.~\eqref{resummation} is interpreted as being multiplied from the right by a function independent 
of the orbital angular momentum; alternatively, one may simply define $D_{SL}$ to vanish when it is the 
right-most operator in that expression.
We defer a detailed discussion of the derivation
of these results and their implications to
Ref.~\cite{EoMEikonalPaper}.

\section{Conclusions}
\label{sec:conclusions}

In this paper we presented a systematic method for constructing the
conservative classical Hamiltonian describing the gravitational
interaction of two massive spinning bodies. Such Hamiltonians provide
crucial input towards obtaining precision gravitational-wave
predictions from binary systems that include Kerr black holes or neutron
stars.
Our formalism extends the arbitrary-spin approach of
Refs.~\cite{LandauQED, Khriplovich:1998ev} and incorporates the
world-line interactions of Refs.~\cite{SpinEFTEarlyWorldLine,
  LeviSteinhoffLagrangian} into a field-theory framework from which
scattering amplitudes can be calculated.  The tree-level and one-loop
amplitudes we find using this formalism determine the classical
two-spinning-body Hamiltonian by EFT matching along the lines of
Ref.~\cite{CliffIraMikhailClassical}.

We constructed the tree-level amplitude to all orders in spin and
velocity and show that it reproduces the $\Ord(G)$ results of
Ref.~\cite{Vines1PM} for the Kerr black hole and extend it to general objects,
such as neutron stars with generic spin-induced multipole moments.
To demonstrate the utility of our approach, we obtained new nontrivial
results for the spin$_1$-spin$_2$ Hamiltonian at $\Ord(G^2)$ valid to all
orders in velocity.
The bilinear-in-spin part of the one-loop amplitude containing the
complete velocity dependence agrees with the spin-1/2 calculation in
Ref.~\cite{DamgaardSpinAmplitude}.  This is in line with
expectations~\cite{HolsteinRoss, Vaidya, SpinSmallYutin} that this is
sufficient at one loop to capture such spin bilinears.  We gave an
argument for the lower bound on the value of the spin which is 
sufficient in our formalism to capture the complete spin dependence of 
an amplitude at a given loop order.

To extract the two-spinning-body Hamiltonian to all orders-in-velocity
we extended the EFT approach of Ref.~\cite{CliffIraMikhailClassical}
to include spin degrees of freedom.
In doing so we encountered a subtlety: for the one-loop infrared
divergences of the EFT and of the full theory to be the same, it was
necessary to have a specific treatment of the terms in the relation
between Lorentz generators and spin tensors that are subleading in the
classical limit. This procedure guarantees that the constructed EFT
corresponds to the relativistic theory we started with.
Because of the stronger-than-classical scaling of parts of the
loop-level amplitudes, we expect that the matching of infrared
divergences must be revisited at every loop order and increasingly
more subleading terms be included. Further study is needed to
determine whether this procedure is sufficient to fix the subleading
terms to all orders in Newton's constant.
It may instead be possible construct a more involved EFT that makes the
matching of infrared divergences more straightforward or even avoids it
altogether.
It would be important to explore both of these strategies towards the
obviously-interesting problem of systematically constructing the spin-dependent
two-body Hamiltonian at $\Ord(G^3)$ and beyond. 

By suitably choosing the initial conditions, the Hamiltonian derived
here can be used to describe any dynamical problem, including the
important bound-state cases.  For constructing precision gravitational-wave
templates that incorporate the new spin information, it is necessary
to import these results into models, such as the effective one
body approach~\cite{EOB, Antonelli:2019ytb}.

In this paper we summarized the results of solving the equations of
motion for our bilinear-in-spin Hamiltonian in a scattering process,
and defer a more detailed discussion to
Ref.~\cite{EoMEikonalPaper}. We obtained the impulse and spin kick in
a scattering process.  Their construction is substantially more
intricate than for the spinless case because orbital angular momentum
is not conserved and, consequently, the scattering trajectory is no
longer planar.  Despite this additional complexity, the results for the 
impulse and spin kick obtained from the solution to the equations of motion
are neatly encoded in the eikonal phase~\cite{EikonalPapers,EikonalImpact}, obtained by
Fourier transforming relevant parts of the amplitude. 
It is rather striking that the eikonal phase
determines the scattering observables, including the spin kick.  This
points to a much greater hidden simplicity than visible in the
Hamiltonian and equations of motion. Based on our results it does seem
that a general simple formalism should exist that translates the
eikonal phase into generic physical observables.  The formalism of
Refs.~\cite{OConnellObservables,MOV}, which directly expresses physical
observables in terms of scattering amplitudes and their unitarity cuts
should provide important guidance for further developments along these
lines.

We validated our results for the spin-dependent two-body Hamiltonian and the
associated observables through several nontrivial checks.  Our primary
test is that, after expanding in velocity, our result agrees with the state-of-art 
 calculations of spin-orbit and spin$_1$-spin$_2$ potential in post-Newtonian 
 framework~\cite{SteinhoffADMforSpin, SteinhoffNNLOSpin} in the overlapped region.
Truncating our Hamiltonian to spin-orbit interactions, we also
reproduce the all-orders-in-velocity scattering angle obtained in
Ref.~\cite{GOV}, whose spin-orbit part is in agreement with
Ref.~\cite{Damour2PMSpinOrbit}, for the configuration where the spins
of the two bodies are aligned with the orbital angular momentum.
An additional nontrivial test is in the test-mass limit, in which we
reproduce the all-orders-in-velocity results of
Refs.~\cite{TestBodyQuadraticSpin}.  (See also Refs.~\cite{SpinTestMassAdditional,Spin2PM_testBH}.)

While the Lagrangian for higher-spin fields we constructed here is not directly suitable for
quantum loop calculations with internal higher-spin fields, it is
sufficient for tree-level calculations, which in turn are sufficient
for constructing all unitarity cuts required in the classical
limit.
A very interesting direction, that can usefully impact the complexity
of constructing of higher order spin-dependent Hamiltonians, is to
systematically expand our understanding of the double copy including
spin.
The double copy expresses gravitational amplitudes in terms of simpler
gauge-theory ones.  Here we pointed out some tantalizing double-copy
relations.  This includes double-copy properties of the two-matter one
graviton tree-level vertices, corresponding to the energy-momentum
tensor for arbitrary spin~\cite{Vines1PM}. In addition we presented a
KLT-like factorization for the tree-level gravitational Compton
amplitude.

An obvious problem is to extend the results obtained here to high
powers of spin at $\Ord(G^2)$ and beyond. As we have argued,
amplitudes of low-spin particles are in general insufficient for this purpose
because of special  relations between Lorentz generators in fixed
representations.  We set up our arbitrary-spin formalism precisely to avoid these limitations.  The
higher-spin Lagrangian we used captures the covariantization of the
parity-even spin-induced gravitational linear response functions and
thus includes all parity-even multipole moments. 
As the number of spin operators increases, nonlinear response
functions, described by operators with two or more gravitons, also
need to be included.  Here we avoided them by focusing on terms
that, while bilinear in spins, are at most linear in the spin of each
particle.
An important problem is the complete classification of all such
operators (each containing as many Riemann tensors as the desired
number of gravitons) and the evaluation of their contribution to the
effective Hamiltonian of massive spinning bodies.  For black holes, it
may be possible to fix coefficients through purely theoretical
considerations as done for the energy-momentum tensor~\cite{Vines1PM}.
For neutron stars or other astrophysical objects, 
the coefficients carry information about its internal structure and properties,  
and should be treated as
phenomenological parameters, to be determined by observation.
The first contribution of such an operator to the two-body Hamiltonian depends on the
number of gravitons it contains. For example, the two-graviton operators first
contribute at $\Ord({G^2})$ to conservative processes and
$\Ord({G^{3/2}})$ to processes with outgoing gravitational radiation.

It would also be of crucial importance to see what further progress
can be made in developing an eikonal formalism that includes arbitrary
spin contributions at any order and to understand in detail the extent of the 
direct links between finite parts of scattering
amplitudes and physical quantities. It would also be important to see
whether appropriate analytic continuations can relate observables of the
unbound and  bound motion, as for the spin-aligned
case~\cite{AmplitudePotential}.

In summary, we expect the amplitudes-based effective-field-theory
approach advocated here to lead to further progress on the spin
dependence of gravitational interactions.  Our linkage of scattering
observables to the eikonal phase demonstrates a surprising hidden
simplicity which suggests that better methods for constructing
physical observables may exist.  This will be further discussed in
Ref.~\cite{EoMEikonalPaper}.

\subsection*{ Acknowledgments:}
We especially thank Alessandra Buonanno, Jan Steinhoff and Justin
Vines for a number of very helpful discussions and for supplying a
simpler form of the test-mass Hamitonian based on
Ref.\cite{TestBodyQuadraticSpin}.  We also thank Clifford Cheung,
Thibault Damour, David Kosower, Aneesh Manohar, Donal O'Connell, Julio
Parra-Marinez, Rafael Porto, Ira Rothstein, and Mikhail Solon for many
helpful discussions.  Z.B. and A.L. are supported by the
U.S. Department of Energy (DOE) under award number DE-SC0009937.
R.R.~is supported by the U.S. Department of Energy (DOE) under grant
no.~DE-SC0013699.  C.H.S.~is supported by the Mani L. Bhaumik
Institute for Theoretical Physics.  M.Z.~is supported by the Swiss
National Science Foundation under contract SNF200021 179016 and the
European Commission through the ERC grant pertQCD.

\end{document}